\documentclass[aps,prx,twocolumn,superscriptaddress]{revtex4-2}

\usepackage{graphicx}
\usepackage[thinc]{esdiff}
\usepackage{subcaption}
\usepackage{amsmath}

\usepackage{verbatim}
\usepackage{dcolumn}
\usepackage{bm}
\usepackage{hyperref}
\usepackage[version=4]{mhchem}    
\usepackage{csquotes}
\hypersetup{colorlinks=true,linkcolor=red,citecolor=blue}
\usepackage{float}
\usepackage{physics}
\usepackage{xspace}
\usepackage[dvipsnames]{xcolor}
\usepackage[normalem]{ulem}
\usepackage{multirow}
\usepackage[percent]{overpic}

\usepackage{lipsum}

\newcommand{\pbesol}{PBEsol+rVV10}
\newcommand{\pbesolu}{PBEsol+$U_\mathrm{Co}$+rVV10}
\newcommand{\pbesoluv}{PBEsol+$U_\mathrm{Co}$+$V$+rVV10}
\newcommand{\pbesoluo}{PBEsol+$U_\mathrm{Co}$+$U_\mathrm{O}$+rVV10}
\newcommand{\pbesoluov}{PBEsol+$U_\mathrm{Co}$+$U_\mathrm{O}$+$V$+rVV10}

\newcommand{\pbesoluno}{PBEsol+$U_\mathrm{Co}$}
\newcommand{\pbesoluvno}{PBEsol+$U_\mathrm{Co}$+$V$}
\newcommand{\pbesoluono}{PBEsol+$U_\mathrm{Co}$+$U_\mathrm{O}$}
\newcommand{\pbesoluovno}{PBEsol+$U_\mathrm{Co}$+$U_\mathrm{O}$+$V$}

\newcommand{\editor}[2]{%
  \expandafter\newcommand\csname #1note\endcsname[1]{%
    \textcolor{#2}{(\textbf{#1:} ##1)}}%
  \expandafter\newcommand\csname #1\endcsname[1]{%
    \textcolor{#2}{##1}}%
  \expandafter\newcommand\csname #1cancel\endcsname[1]{%
    \textcolor{#2}{\sout{##1}}}%
  \expandafter\newcommand\csname #1change\endcsname[2]{%
    \textcolor{#2}{\sout{##1} ##2}}%
  \newenvironment{#1text}{\color{#2}}{\color{black}}
}

\editor{IT}{red}
\editor{VS}{magenta}
\editor{AC}{ForestGreen}

\begin{document}


\title{Assessing the magnetic states and the accuracy of first-principles Hubbard corrections for the battery cathode Li$_x$CoO$_2$ ($x=0,1$)}

\author{Valentina Sanella}
\altaffiliation[Current address: ]{PSI Center for Scientific Computing,
Theory, and Data, Paul Scherrer Institute, 5232 Villigen PSI, Switzerland}
\email{valentina.sanella@psi.ch}
\affiliation{Theory and Simulation of Materials (THEOS), and National Centre for Computational Design and Discovery of Novel Materials (MARVEL), \'Ecole Polytechnique F\'ed\'erale de Lausanne (EPFL), CH-1015 Lausanne, Switzerland}
\affiliation{Department of Materials Science, Università degli Studi di Milano-Bicocca, Italy}

\author{Cristiano Malica}
\email{cmalica@uni-bremen.de}
\affiliation{U Bremen Excellence Chair, Bremen Center for Computational Materials Science, and MAPEX Center for Materials and Processes, University of Bremen, Germany}

\author{Alberto Carta}
\affiliation{PSI Center for Scientific Computing,
Theory, and Data, Paul Scherrer Institute, 5232 Villigen PSI, Switzerland}

\author{Maria Andolfatto}
\affiliation{PSI Center for Scientific Computing,
Theory, and Data, Paul Scherrer Institute, 5232 Villigen PSI, Switzerland}

\author{Nicola Marzari}
\affiliation{Theory and Simulation of Materials (THEOS), and National Centre for Computational Design and Discovery of Novel Materials (MARVEL), \'Ecole Polytechnique F\'ed\'erale de Lausanne (EPFL), CH-1015 Lausanne, Switzerland}
\affiliation{PSI Center for Scientific Computing,
Theory, and Data, Paul Scherrer Institute, 5232 Villigen PSI, Switzerland}
\affiliation{U Bremen Excellence Chair, Bremen Center for Computational Materials Science, and MAPEX Center for Materials and Processes, University of Bremen, Germany}

\author{Livia Giordano}
\affiliation{Department of Materials Science, Università degli Studi di Milano-Bicocca, Italy}

\author{Iurii Timrov}
\email{iurii.timrov@psi.ch}
\affiliation{PSI Center for Scientific Computing,
Theory, and Data, Paul Scherrer Institute, 5232 Villigen PSI, Switzerland}

\date{\today}

\begin{abstract}
Li$_x$CoO$_2$ is a prototypical layered cathode material for Li-ion batteries, yet its accurate description from first principles remains challenging because of self-interaction errors, weak interlayer interactions, and a complex magnetic energy landscape. Here, we present a systematic investigation of the structural, electronic, magnetic, and electrochemical properties of Li$_x$CoO$_2$ ($x=0,1$) using density-functional theory augmented with self-consistent Hubbard corrections and long-range van der Waals interactions, together with a systematic exploration of possible magnetic states. The on-site interactions on Co-$3d$ and O-$2p$ states, as well as inter-site Co-O interactions, are determined from first principles using linear-response theory in the framework of density-functional perturbation theory, with L\"owdin-orthogonalized atomic orbitals employed as Hubbard projectors. For LiCoO$_2$, the inclusion of Hubbard corrections provides an accurate description of the structural properties, while the electronic structure is very sensitive to the choice of Hubbard projectors. In particular, frontier Wannier-function projectors substantially improve the description of the occupied electronic states compared with localized atomic orbitals. For CoO$_2$, we demonstrate that a systematic exploration of the magnetic energy landscape is essential to identify the lowest-energy low-spin ground state. However, the resulting Hubbard-corrected electronic structure is insulating, consistent with the prediction of the HSE06 hybrid functional, but in contrast to the experimentally observed metallic behavior. Structural relaxation further drives the system toward a different metallic solution with an electronic configuration inconsistent with low-spin Co$^{4+}$ character. Despite these limitations, the calculated intercalation voltages agree well with experiment, with deviations as small as 2\%. Importantly, this agreement does not imply an accurate description of the underlying ground state, highlighting the challenges in simultaneously describing the structural, electronic and electrochemical properties of this system. Our results demonstrate that the predictive accuracy of first-principles approaches for battery materials cannot be reliably assessed from electrochemical observables alone.
\end{abstract}

\maketitle

\section{Introduction}
\label{sec1}

The global demand for efficient and high-performance energy storage solutions has led to extensive research in battery technology in the past decades, with Li-ion batteries (LIBs) emerging as a leading technology for a wide range of applications, from portable electronics~\cite{Scrosati2005} to electric vehicles~\cite{Chen2012,Thackeray2012}. 
Among the cathode materials used in LIBs, lithium cobalt oxide (LiCoO$_2$) is one of the most commercially successful and widely used materials since its introduction by Goodenough and coworkers in 1980~\cite{Goodenough1980}. 
Despite the rapid development of alternative cathode materials, LiCoO$_2$ remains a benchmark in LIB cathodes due to its high volumetric energy density and stable performance. The layered structural framework of LiCoO$_2$ has further inspired a broad family of related cathode materials, including both Li-ion and Na-ion analogues with varied transition-metal (TM) compositions~\cite{Nayak2018}. The composition of the current layered LIB cathodes is more complex, for example cobalt has been partially substituted by nickel and manganese in the widely used class of Ni$-$Mn$-$Co (NMC) cathodes, led by NMC111~\cite{Yabuuchi2003} and NMC811~\cite{Noh2013,Xuan2019}. The doping is also extended to metals that do not come from the first transition row, like aluminum in the Ni-Co-Al system~\cite{Chen2004}.

Modeling Li$_x$CoO$_2$ from first principles using density-functional theory (DFT)~\cite{Hohenberg1964,Kohn1965} is very challenging primarily for two reasons: the localized nature of Co-$3d$ electrons and the presence of weak interlayer van der Waals (vdW) interactions. Standard exchange-correlation (xc) functionals such as local-density approximation (LDA)~\cite{Kohn1965} and generalized-gradient approximation (GGA)~\cite{Perdew1992} are known to perform poorly for TM compounds because of large self-interaction errors (SIEs)~\cite{Perdew1981, MoriSanchez2006}. These errors are particularly severe for localized $d$ and $f$ electrons, leading to their unphysical delocalization and, consequently, inaccuracies in the prediction of structural, electronic, and electrochemical properties. To mitigate these deficiencies, a variety of advanced xc functionals and corrective approaches have been developed, including DFT+$U$~\cite{Anisimov1991, Liechtenstein1995, Dudarev1998} and DFT+$U$+$V$~\cite{Campo2010, Lee2020, TancogneDejean2020}, meta-GGA functionals such as SCAN~\cite{Sun2015}, rSCAN~\cite{Bartok2019}, and r$^2$SCAN~\cite{Furness2020}, as well as hybrid functionals like PBE0~\cite{Adamo1999} and HSE06~\cite{Heyd2003,Heyd2006}. Each of these methods has its strengths and limitations~\cite{Timrov2022} Although SCAN-based functionals generally offer improved accuracy over LDA and GGA, they often still require Hubbard corrections for TM compounds, as demonstrated in SCAN+$U$~\cite{Long2020} and r$^2$SCAN+$U$~\cite{Swathilakshmi2023} studies. Beyond DFT, many-body approaches such as DFT combined with dynamical mean-field theory (DFT+DMFT)~\cite{Georges1996} have also been successfully employed to describe TM compounds with strong electronic correlations.

Among the available corrective approaches, DFT+$U$ and DFT+$U$+$V$ offer an attractive compromise between computational cost and accuracy~\cite{Himmetoglu2013, Kulik2008, Kulik2011}. However, the main challenge of these methods is that the Hubbard $U$ and $V$ parameters for a given material are not known. For this reason, in many DFT+$U$ studies of battery materials, the $U$ parameter is calibrated empirically against selected experimental observables, such as oxidation enthalpies, band gaps, or magnetic moments~\cite{Hautier2011, Aykol2014, Urban2016, Chakraborty2018, Isaacs2020}. However, this strategy depends on the availability of reliable experimental data, is not generally transferable across different properties, and introduces an undesirable degree of arbitrariness that undermines the \textit{ab initio} framework of the DFT+$U$(+$V$) method. A more appealing alternative is to determine the Hubbard parameters from first principles, thereby eliminating empirical input. Several methodologies have been developed for this purpose, including constrained DFT~\cite{Dederichs1984, Mcmahan1988, Gunnarsson1989, Hybertsen1989, Gunnarsson1990, Pickett1998, Solovyev2005, Nakamura2006, Shishkin2016}, constrained random phase approximation (cRPA)~\cite{Springer1998, Kotani2000, Aryasetiawan2004, Aryasetiawan2006}, and Hartree–Fock-based approaches~\cite{Mosey2007, Mosey2008, Andriotis2010, Agapito2015, TancogneDejean2020, Lee2020}. Among these, the linear-response formulation of constrained DFT (LR-cDFT)~\cite{Cococcioni2005} has emerged as particularly effective, and its reformulation within density-functional perturbation theory (DFPT)~\cite{Timrov2018, Timrov2021} greatly improves computational efficiency and facilitates its application to a wide range of materials.
Notably, the accuracy and efficiency of the DFT+$U$+$V$ framework, with interaction parameters $U$ and $V$ computed from DFPT, have been demonstrated in the calculation of voltages for various LIBs~\cite{Cococcioni2019, Timrov2022, Timrov2023}. These advances have been further extended to first-principles molecular dynamics simulations in combination with neural network potentials~\cite{Malica2024}.

However, none of the aforementioned xc functionals capture the long-range vdW interactions that are essential for an accurate description of layered cathode materials such as Li$_x$CoO$_2$. Meta-GGA functionals like SCAN account only for short-range vdW interactions~\cite{Grimme2016}, and thus could be insufficient for systems where interlayer binding plays a key role. Several approaches have been developed to include long-range vdW interactions. Among them, Grimme’s DFT-D3 scheme~\cite{Grimme2010} is widely used due to its simplicity and low computational cost; it adds an empirical, position-dependent energy correction that must be parametrized for each xc functional. Alternatively, a number of xc functionals incorporate vdW interactions from first principles by introducing a non-local correlation term, as implemented in the vdW-DF family~\cite{Dion2004, Thonhauser2015, Lee2010, Chakraborty2020}, rVV10~\cite{Sabatini2013}, or optPBE-vdW~\cite{Klimes2009, Klimes2011}. Therefore, achieving an accurate description of Li$_x$CoO$_2$ across different Li concentrations requires an xc treatment that simultaneously addresses SIEs and properly accounts for long-range vdW interactions.

In this respect, it is important to briefly review previous efforts to model Li$_x$CoO$_2$ using the xc functionals mentioned above. In early studies based solely on LDA~\cite{Wolverton1998, Carlier2002}, and neglecting both Hubbard and vdW corrections, the predicted voltages were systematically underestimated relative to experiment. More recent work employing advanced xc functionals has achieved higher accuracy. For example, the authors of Refs.~\cite{Kim2021, Chakraborty2018} used the Perdew–Burke–Ernzerhof (PBE)~\cite{Perdew1996} parametrization of GGA and the SCAN functional to investigate various layered Li-ion and Na-ion cathode materials. In Ref.~\cite{Kim2021}, cRPA was used to compute the Hubbard $U$ parameter, while empirical $U$ values were also explored in Refs.~\cite{Kim2021, Chakraborty2018}. For PBE and PBE+$U$, vdW interactions were incorporated via the DFT-D3 method. Both studies reported that, for Li$_x$CoO$_2$, the combination of PBE+$U$ with an empirical $U \approx 3$~eV applied to Co-$3d$ states together with DFT-D3 yields voltage predictions comparable to those obtained using SCAN (without $U$). In another systematic comparison, Ref.~\cite{Isaacs2020} considered PBE, optPBE-vdW, SCAN, and SCAN+rVV10~\cite{Peng2016}, each combined with an empirical Hubbard $U$ correction for Co-$3d$ states, across several LIB cathodes. These authors found that optPBE-vdW with empirical $U \approx 5$~eV provides the closest agreement with the experimental voltage for Li$_x$CoO$_2$. Ref.~\cite{Aykol2015} further examined multiple vdW corrections (e.g., optB88~\cite{Klimes2009}, optB86b~\cite{Klimes2011}) again combined with empirical $U$ values, as well as HSE06. It was shown that HSE06 tends to overestimate the voltage, while PBE augmented with vdW corrections and empirical $U$ in the range $2 - 5$~eV (depending on the vdW type) yields good agreement with experiments. On the other hand, Refs.~\cite{Shishkin2016, Shishkin2021} employed PBE+$U$ with Hubbard parameters determined self-consistently from LR-cDFT. The resulting $U$ values for Co-$3d$ states ranged from 3 to 5~eV, depending on computational details such as whether semicore states were included in the pseudopotentials. These works also highlighted that the computed $U$ values differ between LiCoO$_2$ and CoO$_2$, consistent with earlier LR-cDFT studies~\cite{Zhou2004}. Similarly, Ref.~\cite{Aykol2014} found that $U$ values optimized from experimental formation enthalpies differ between LiCoO$_2$ and CoO$_2$, reflecting the change in Co oxidation state and local environment. Using respective LR-cDFT Hubbard $U$ values for intercalated and deintercalated structures, Refs.~\cite{Shishkin2016, Shishkin2021} reported PBE+$U$ voltages in good agreement with experimental data. However, these studies neglected vdW interactions entirely, despite evidence from other works that vdW effects are important for Li$_x$CoO$_2$. It is important to note that both computed and optimal empirical Hubbard $U$ values depend sensitively on the choice of Hubbard projector functions. Consequently, the $U$ values reported in all of the aforementioned studies (each performed using VASP~\cite{Kresse1993}) are specific to the projector augmented wave (PAW) formalism~\cite{Blochl1994} implemented in that code, and such values are generally not transferable to other electronic-structure codes. Furthermore, the role of dynamical interactions in Li$_x$CoO$_2$ has been a subject of significant scrutiny. Ref.~\cite{Isaacs2020b} utilized the DFT+DMFT~\cite{Georges1996} formalism to demonstrate that CoO$_2$ acts as a moderately correlated system, suggesting that static Hubbard corrections often incorrectly predict insulating states or spurious charge ordering. While DFT+DMFT provides a robust framework for capturing these dynamical effects, it remains computationally complex. 
Therefore, despite extensive efforts using various xc functionals (and combinations thereof) and advanced methods like DFT+DMFT, no clear consensus has emerged regarding the most reliable first-principles approach for modeling Li$_x$CoO$_2$, particularly when aiming for a fully \textit{ab initio} description without any empirical parameters.

While all of the aforementioned DFT+$U$ studies have focused primarily on the on-site Hubbard $U$ correction for the TM ions, the role of Hubbard corrections on the ligand orbitals (i.e. the O-$2p$ states) has been entirely disregarded. Yet, it has been shown that the Hubbard parameters for O-$2p$ states are comparable to, and in some cases even larger than, those for TM-$3d$ electrons~\cite{Nekrasov2000, KirchnerHall2021, Xiong2021}, and can be essential for achieving an accurate description in certain materials~\cite{Lechermann2024}. This indicates that applying a Hubbard $U$ correction to ligand states may also be necessary. Such an approach is already used in both DFT+$U$~\cite{May2020, Lee2020, TancogneDejean2020, Berman2023, Orhan2020, Lambert2023, Kam2024} and DFT+DMFT~\cite{Georges1996, Carta2026} studies, where it has proven important for correctly capturing the electronic structure and physical properties of various TM compounds. A key effect of including $U$ on oxygen is the modification of the hybridization between TM-$3d$ and O-$2p$ states, which can significantly influence the distribution of occupied and unoccupied states. Thus the combination of fully first-principles self-consistent on-site Hubbard $U$ corrections for Co-$3d$ and O-$2p$ states, inter-site $V$ interactions between Co-$3d$ and O-$2p$ states, and long-range vdW interactions in Li$_x$CoO$_2$ needs to be investigated.

In this work, we present a first-principles study of the layered cathode material Li$_x$CoO$_2$ ($x=0,1$), aimed at assessing the accuracy and limitations of extended Hubbard functionals for describing its structural, electronic, magnetic, and electrochemical properties. We systematically investigate the impact of on-site Hubbard corrections applied to Co-$3d$ and O-$2p$ states, as well as inter-site Co--O Hubbard interactions, with all parameters computed fully from first principles and self-consistently, using the LR-cDFT approach~\cite{Cococcioni2005}, reformulated within DFPT~\cite{Timrov2018, Timrov2021} and applied in a basis of L\"owdin-orthogonalized atomic orbitals (Hubbard projectors)~\cite{Lowdin1950, Mayer2002}. Long-range vdW interactions are explicitly accounted for, and rVV10 is identified as the most accurate correction for these layered systems. Our results reveal that the predictive accuracy of an exchange-correlation functional cannot be assessed from a single physical observable. For LiCoO$_2$, we show that Hubbard-corrected calculations can accurately reproduce the crystal structure, while the electronic structure remains sensitive to the choice of Hubbard projectors. In particular, the use of frontier Wannier-function projectors leads to a quantitatively improved description of the electronic states compared with localized atomic projectors, highlighting the importance of the projector choice in DFT+$U$ calculations. For CoO$_2$, we demonstrate that the presence of multiple metastable magnetic solutions requires a systematic exploration of the magnetic energy landscape to identify the lowest-energy configuration. Although the global minimum corresponding to a low-spin Co$^{4+}$ state can be recovered, its electronic structure remains inconsistent with the experimentally observed metallic character. Furthermore, structural relaxation leads to a strongly distorted geometry and a different metallic solution characterized by charge redistribution between Co-$3d$ and O-$2p$ states, resulting in a Co electronic configuration that cannot be described within the low-spin or high-spin Co$^{4+}$ pictures. These findings highlight the challenges of describing metallic paramagnetic oxides using static mean-field approaches such as DFT+$U$ and hybrid functionals.
Finally, we find that the calculated intercalation voltages agree well with experiment, despite substantial inaccuracies in the underlying structural and electronic properties. In particular, the inclusion of first-principles on-site and inter-site Hubbard interactions yields promising energetics, as reflected in the calculated voltages, but this agreement does not guarantee an accurate description of the underlying crystal, electronic, and magnetic states. Thus, while energetics remains a key target of LR-cDFT-based Hubbard functionals, their predictive performance should be assessed across structural, electronic, magnetic, and electrochemical properties.

The remainder of this paper is organized as follows.  Section~\ref{sec:results} presents the results of this work, including a detailed analysis of the crystal and electronic structure properties, such as the projected densities of states (PDOS), atomic occupations, magnetic moments, as well as  intercalation voltages. Section~\ref{sec:conclusions} summarizes the main findings and conclusions. Section~\ref{sec:methods} provides a brief overview of the DFT+$U$+$V$ methodology, the treatment of van der Waals interactions, and describes the computational setup.

\section{Results and Discussion}
\label{sec:results}

\subsection{Crystal structure of Li$_x$CoO$_2$ ($x=0,1$)}
\label{sec:Crystal_structure}

\begin{figure}[t]
    \centering
    \includegraphics[width=0.85\linewidth]{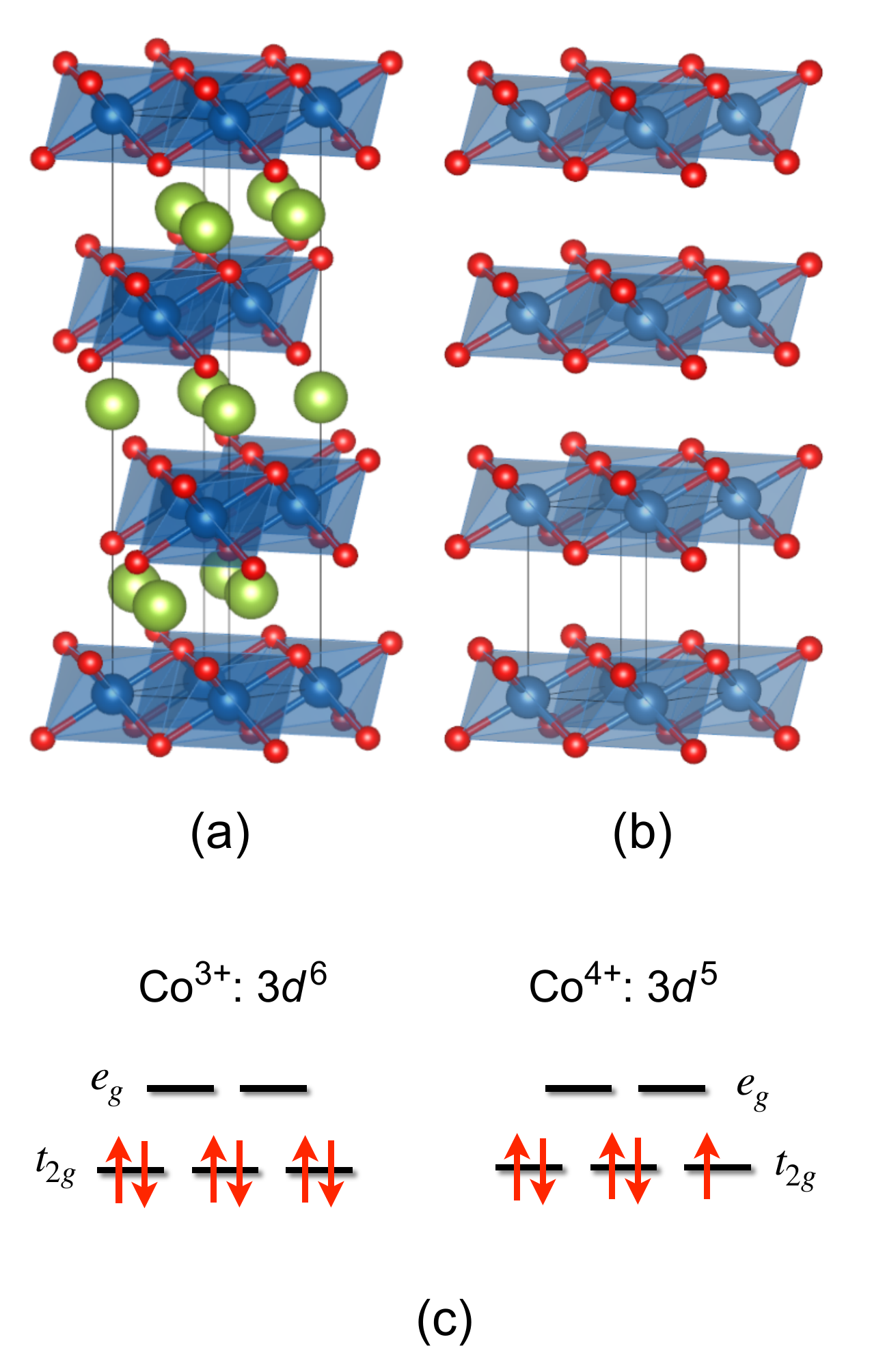}
    \caption{(a)~The O3 layered structure of LiCoO$_2$ (\textit{R$\overline{3}$m} space group), and (b)~the O1 layered structure of CoO$_2$ (\textit{P$\overline{3}$m1} space group). Lithium, cobalt, and oxygen atoms are shown in green, blue, and red, respectively. Black solid lines denote the hexagonal simulation unit cell, defined by the in-plane lattice parameter $a$ and the out-of-plane lattice parameter $c$. Structural visualizations were generated using VESTA~\cite{Momma2011}. (c)~Nominal occupations of the $3d$ manifold of a Co ion (not hybridized with ligands) in a low-spin undistorted octahedral complex with diﬀerent oxidation states ($O_h$ point group). The $t_{2g}$ and $e_g$ levels are indicated with black horizontal lines and are nondegenerate due to the crystal-field splitting; up and down red arrows correspond to spin-up and spin-down electrons, respectively.}
    \label{fig:materials}
\end{figure}

The crystal structures of LiCoO$_2$ and CoO$_2$ are shown in Fig.~\ref{fig:materials}(a) and \ref{fig:materials}(b), respectively. 
LiCoO$_2$ crystallizes in the O3 layered structure, which belongs to the \textit{R$\overline{3}$m} space group and corresponds to the G-type NaFeO$_2$ hexagonal structure~\cite{Garcia1995,Antolini2004}, containing three formula units per unit cell. 
Co$^{3+}$ adopts a low-spin configuration (see Fig.~\ref{fig:materials}(c)), and LiCoO$_2$ is experimentally observed to be a non-magnetic insulator with an indirect band gap of 2.1--2.7~eV~\cite{vanElp1991}. 
During charging, Li$^+$ ions are progressively removed from the structure, leading to several phase transitions accompanied by substantial volume changes and oxygen loss at low lithium concentrations~\cite{Garcia1995,Antolini2004}. As a consequence, the practical reversible delithiation limit in commercial cathodes is typically restricted to approximately Li$_{0.5}$CoO$_2$~\cite{Ohzuku1993}. Nevertheless, full and reversible delithiation to CoO$_2$ can be achieved under controlled conditions while retaining electrochemical activity~\cite{Amatucci1996,Motohashi2007}. Complete delithiation transforms the O3 structure into the O1 phase, in which the CoO$_2$ layers stack directly along the $c$ axis without lateral displacement. This phase crystallizes in the \textit{P$\overline{3}$m1} space group and is isostructural with CdI$_2$~\cite{Amatucci1996}. CoO$_2$ is a magnetic metal with nominal Co$^{4+}$ ions also in a low-spin configuration (see Fig.~\ref{fig:materials}(c)). Its magnetic behavior is nontrivial and has been reported either as that of a conventional paramagnetic metal~\cite{Gerken2011} or as a Pauli-paramagnetic metal with itinerant electrons~\cite{deVaulx2007,Motohashi2007}.

\subsection{LiCoO$_2$}

\subsubsection{Structural properties}

\begin{figure}[t]
    \centering
    \includegraphics[width=0.99\linewidth]{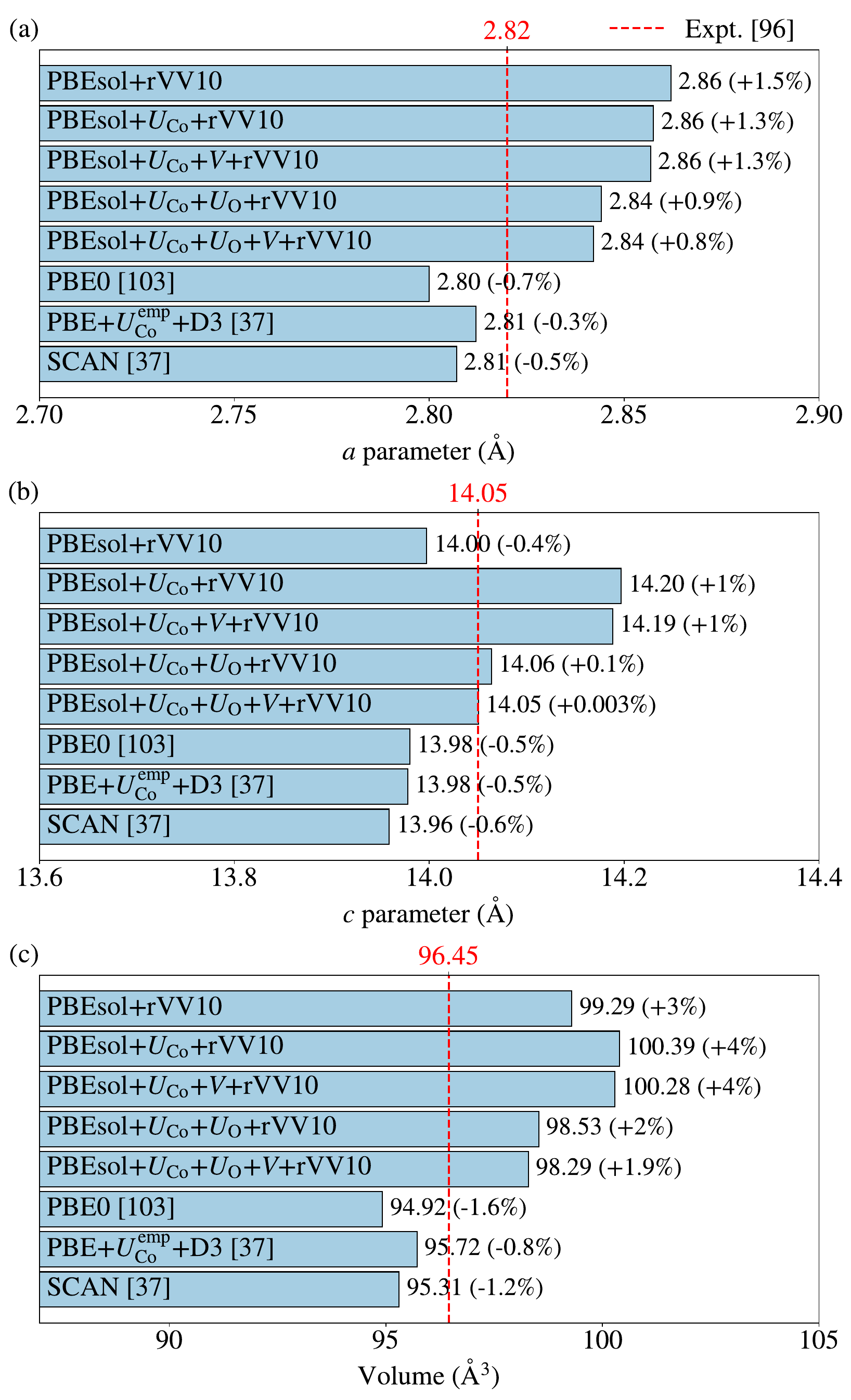}
    \caption{Structural parameters of LiCoO$_2$: (a) lattice parameter $a$, (b) lattice parameter $c$, and (c) unit-cell volume. Experimental values (shown as vertical dashed lines) are taken from Ref.~\cite{Antolini2004}. Computational results obtained with PBE0, SCAN, and PBE+$U_\mathrm{Co}^\mathrm{emp}$+D3 (using an empirical $U_\mathrm{Co}^\mathrm{emp} \approx 3$~eV for the Co-$3d$ states) are taken from Refs.~\cite{Mattila2022, Chakraborty2018}.}
    \label{fig:LCO_lat_param}
\end{figure}

Figure~\ref{fig:LCO_lat_param} compares the lattice parameters $a$ and $c$, as well as the volume of the hexagonal unit cell of LiCoO$_2$, obtained using PBEsol+rVV10 combined with various first-principles Hubbard corrections (see Table~\ref{tab:UOV}). The results are contrasted with previous computational studies~\cite{Mattila2022, Chakraborty2018} and experimental measurements~\cite{Antolini2004}. Additionally, Figs.~S1 and S2 in the Supplementary Information (SI)~\cite{Supplementary_Information} present the optimized structural parameters computed with various other vdW functionals and corrections.

Focusing first on the lattice parameter $a$ [Fig.~\ref{fig:LCO_lat_param}(a)], PBEsol+rVV10 overestimates this parameter by 1.5\%. All first-principles Hubbard-corrected functionals also yield a slight overestimation; however, they show good overall agreement with experiment, with deviations in the range of $0.8-1.3$\%. More specifically, the inclusion of $U_\mathrm{Co}$ marginally reduces the lattice parameter $a$, while the additional application of $U_\mathrm{O}$ leads to a further reduction, bringing $a$ closer to the experimental value. The inter-site interaction $V$ provides very small additional improvement. Overall, the simultaneous inclusion of all Hubbard corrections ($U_\mathrm{Co}$, $U_\mathrm{O}$, and $V$) yields the best agreement with experiment among the first-principles approaches considered in this study. By contrast, previous computational studies based on PBE0, SCAN, and PBE+$U_\mathrm{Co}^\mathrm{emp}$+D3 underestimate the lattice parameter $a$~\cite{Mattila2022, Chakraborty2018}.

Turning now to the lattice parameter $c$, a different trend emerges compared to the behavior of the lattice parameter $a$. PBEsol+rVV10 underestimates $c$ by 0.4\%. The inclusion of the $U_\mathrm{Co}$ correction leads to an overestimation of $c$, thereby worsening the agreement with experiment and resulting in a deviation of 1.0\%, while the addition of the inter-site $V$ has a negligible effect on this trend. In contrast, the inclusion of the $U_\mathrm{O}$ correction counteracts the effect of $U_\mathrm{Co}$, reducing $c$ and bringing it into excellent agreement with the experimental value, particularly when the small contribution from $V$ is also taken into account. Overall, $U_\mathrm{Co}$ and $U_\mathrm{O}$ act in opposite, partially compensating manner; however, the slightly stronger effect of $U_\mathrm{Co}$ dominates, resulting in a theoretical value of $c$ that closely matches the experimental measurement. By contrast, the aforementioned previous computational studies systematically underestimate $c$ by approximately $0.5-0.6$\%~\cite{Mattila2022, Chakraborty2018}.

Combining the trends observed for the lattice parameters $a$ and $c$, we now consider the volume of the hexagonal unit cell, given by $V_\mathrm{cell} = (\sqrt{3}/2)\,a^2 c$. As a consequence of the quadratic dependence on $a$, deviations in this lattice parameter are amplified in the resulting cell volume. PBEsol+rVV10 overestimates the experimental volume by 3.0\%. The inclusion of the $U_\mathrm{Co}$ correction further increases this overestimation to 4.0\%, with the inter-site interaction $V$ again playing only a minor role. In contrast, the addition of the $U_\mathrm{O}$ correction reduces the cell volume, yielding a smaller but still positive deviation of 2.0\% from experiment. As in the case of the lattice parameter $c$, $U_\mathrm{Co}$ and $U_\mathrm{O}$ act in opposite and partially compensating manner. The simultaneous inclusion of all Hubbard corrections ($U_\mathrm{Co}$, $U_\mathrm{O}$, and $V$) leads to a more accurate cell volume than that obtained without Hubbard corrections. By contrast, all previous computational studies systematically underestimate the cell volume, with average deviations in the range of $0.8-1.6$\%.

Overall, previous studies based on PBE0, SCAN, and PBE+$U_\mathrm{Co}^\mathrm{emp}$+D3 tend to predict slightly more accurate structural properties of LiCoO$_2$ than our most accurate results obtained with the fully first-principles Hubbard-corrected functional PBEsol+$U_\mathrm{Co}$+$U_\mathrm{O}$+$V$+rVV10.
This is due to the fact that our first-principles Hubbard parameters from DFPT are significantly larger than the empirical $U_\mathrm{Co}^\mathrm{emp}$ values (see Table~\ref{tab:UOV}). Nevertheless, our predictions for the structural properties remain very satisfactory, with deviations from experiment less than $2\%$. Importantly, when all Hubbard corrections are omitted (i.e., PBEsol+rVV10), the deviations are larger compared to the case when all Hubbard corrections are included. This demonstrates that the combined effect of the Hubbard corrections (which partially compensate each other, for example in the case of the $c$ lattice parameter and the cell volume $V_\mathrm{cell}$) results in a clear overall improvement in the prediction of the structural properties of LiCoO$_2$.

\subsubsection{Electronic structure properties}
\label{sec:LCO_electronic_structure}

\begin{figure*}[t]
    \centering
    \includegraphics[width=0.95\linewidth]{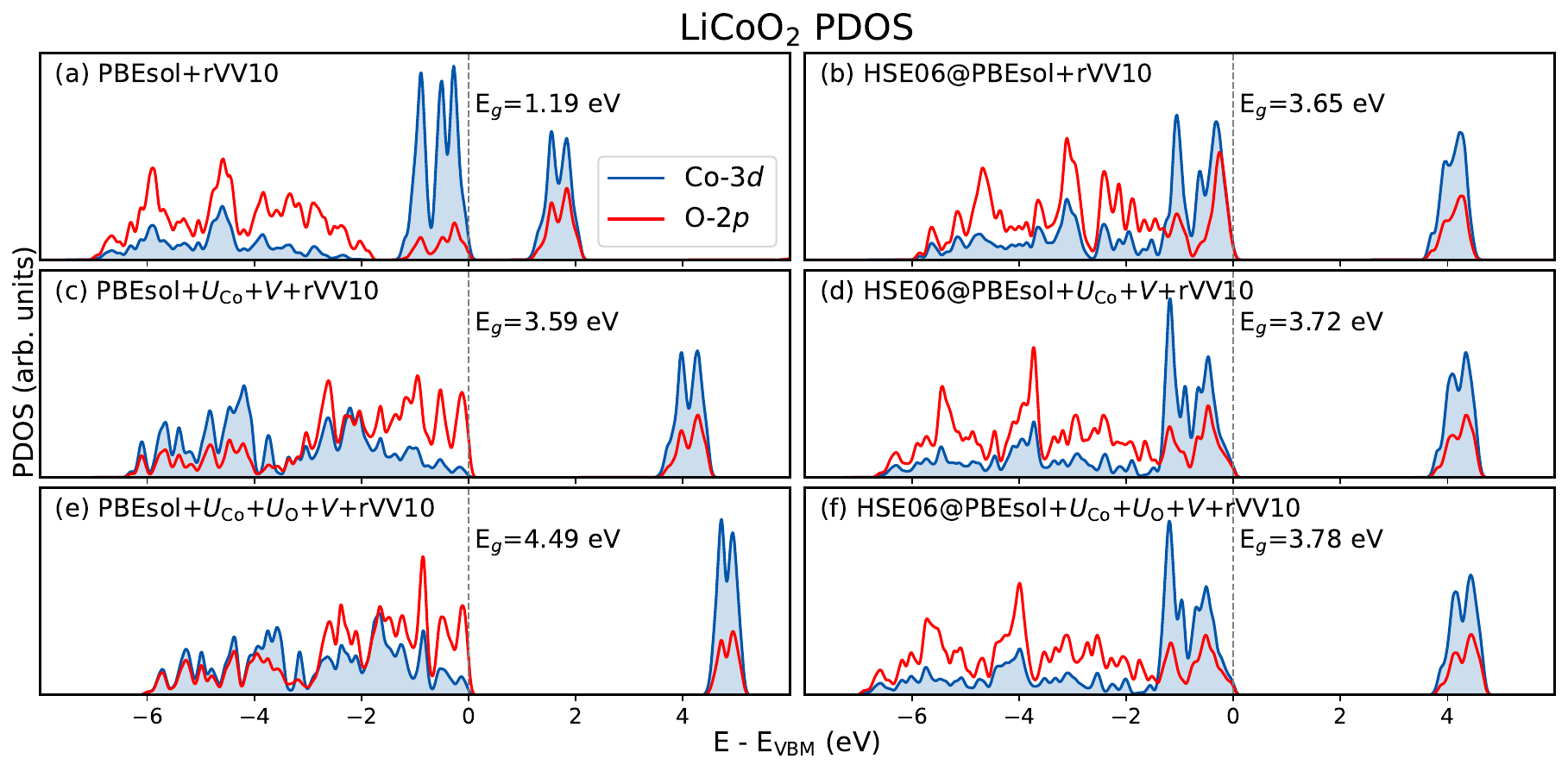}
    \caption{PDOS of LiCoO$_2$ computed using respective optimized geometries: (a) PBEsol+rVV10, (c) PBEsol+$U_\mathrm{Co}$+$V$+rVV10, and (e) PBEsol+$U_\mathrm{Co}$+$U_\mathrm{O}$+$V$+rVV10. Panels (b), (d), and (f) show the PDOS obtained with HSE06 on top of the corresponding geometries from panels (a), (c), and (e), respectively. The Co-$3d$ and O-$2p$ states are shown in blue (filled) and red (unfilled). The band gap values, $\mathrm{E}_g$, are indicated in each panel, and the energy reference is set to the valence band maximum (VBM), $\mathrm{E}_\mathrm{VBM}$. L\"owdin-orthogonalized atomic orbitals are used as Hubbard projectors for all Hubbard corrections.}
    \label{fig:licoo2dos}
\end{figure*}

We now analyze the effect of the various Hubbard corrections on the electronic structure of LiCoO$_2$. Figure~\ref{fig:licoo2dos} shows the PDOS, while Table~\ref{tab:magnlicoo2} reports the eigenvalues $\lambda^\sigma_i$ ($i=\overline{1,5}$; $\sigma=\uparrow, \downarrow$) of the diagonal occupation matrix~\eqref{eq:occ_matrix_0} (with $I=J$) and the total occupation of the Co-$3d$ states. Figure~\ref{fig:licoo2dos}(a) presents the PDOS obtained using PBEsol+rVV10 on top of its corresponding optimized geometry (see Fig.~\ref{fig:LCO_lat_param}). The valence-band maximum (VBM) is dominated by Co-$3d$ states of $t_{2g}$ character, whereas the conduction-band minimum (CBM) is primarily composed of Co-$3d$ states of $e_g$ character. Both the VBM and CBM also exhibit smaller contributions from O-$2p$ states. At lower binding energies, the occupied states are predominantly of O-$2p$ character, with a smaller admixture of Co-$3d$ states. This picture is further corroborated by the eigenvalues of the diagonal occupation matrix for the Co-$3d$ states, reported in Table~\ref{tab:magnlicoo2}. Three eigenvalues ($\lambda_3$, $\lambda_4$, and $\lambda_5$) are close to unity for both spin channels, owing to the Co$^{3+}$ oxidation state (see Fig.~\ref{fig:materials}(c)). This indicates that, in each spin channel, three electrons fully occupy the Co-$3d$ $t_{2g}$ manifold, following the analysis proposed in Ref.~\cite{Sit2011}. In contrast, the eigenvalues $\lambda_1$ and $\lambda_2$, associated with the $e_g$ manifold, are fractional, indicating that the corresponding states are only partially occupied. Although the $e_g$ states are nominally unoccupied in a free Co$^{3+}$ ion, they become partially occupied in LiCoO$_2$ due to hybridization with the O-$2p$ states~\cite{Sit2011}.

Next, we examine the electronic-structure modifications induced by the Hubbard corrections in LiCoO$_2$, using the corresponding optimized geometries. Figure~\ref{fig:licoo2dos}(c) shows the PDOS obtained upon applying the on-site $U_\mathrm{Co}$ correction together with the inter-site interaction $V$. The PDOS computed with $U_\mathrm{Co}$ alone is very similar and is shown in Fig.~S3(a)~\cite{Supplementary_Information}. The application of $U_\mathrm{Co}$ leads to significant modifications of the PDOS, most notably a pronounced redistribution of the occupied states. In particular, the three sharp peaks at the top of the valence band (originating from the Co-$3d$ $t_{2g}$ manifold in the Hubbard-uncorrected case) are no longer present. Instead, the Co-$3d$ states, comprising all $t_{2g}$ and part of the $e_g$ character, are spread over a broad energy window extending from the VBM down to approximately $-6.5$~eV. As a result, the VBM becomes predominantly O-$2p$ in character. Therefore, applying $U_\mathrm{Co}$ using L\"owdin-orthogonalized atomic orbitals as Hubbard projectors shifts the localized $t_{2g}$ states to lower energies, enhancing their hybridization with the O-$2p$ manifold. Consequently, the $t_{2g}$ states become more delocalized in energy, while remaining fully occupied. In contrast, the $e_g$ states remain partially occupied and delocalized, as confirmed by the eigenvalues of the diagonal occupation matrix reported in Table~\ref{tab:magnlicoo2}. However, this PDOS is inconsistent with experimental observations (see Fig.~19 in Ref.~\cite{Ensling2010}), which show that the VBM has predominantly Co-$3d$ $t_{2g}$ character. This discrepancy originates from the large first-principles value of $U_\mathrm{Co}$ obtained in our calculations when using L\"owdin-orthogonalized atomic orbitals as Hubbard projectors (see Table~\ref{tab:UOV}), whereas most previous DFT+$U$ studies employed substantially smaller empirical Hubbard $U$ values for the Co-$3d$ states with other types of Hubbard projectors. Turning to the conduction-band region in Fig.~\ref{fig:licoo2dos}(c), the CBM retains its overall shape upon inclusion of $U_\mathrm{Co}$ but is shifted to higher energies, leading to an increased band gap of 3.59~eV. This value overestimates the experimental band gap of 2.1--2.7~eV~\cite{vanElp1991}. The additional inclusion of the $U_\mathrm{O}$ correction further increases the band gap to 4.49~eV, without inducing qualitative changes in the overall PDOS (see Fig.~\ref{fig:licoo2dos}(e)). The case in which both $U_\mathrm{Co}$ and $U_\mathrm{O}$ are included while neglecting $V$ is shown in Fig.~S3(b) in the SI~\cite{Supplementary_Information}, again yielding a qualitatively similar PDOS. For completeness, Fig.~S4 in the SI~\cite{Supplementary_Information} reports the band gaps calculated using various other vdW functionals and corrections. Consistent with these observations, the eigenvalues of the diagonal occupation matrix reported in Table~\ref{tab:magnlicoo2} remain almost unchanged upon inclusion of $U_\mathrm{O}$. The $t_{2g}$ manifold remains fully occupied, while the $e_g$ states become slightly more depleted, as reflected in the modest reduction of $\lambda_1$ and $\lambda_2$.

\begin{table*}[t]
    \centering
    \caption{Eigenvalues of the diagonal ($I=J$) occupation matrix (see Eq.~\eqref{eq:occ_matrix_0}) for the Co-$3d$ states in LiCoO$_2$, shown separately for the spin-up ($\lambda_i^\uparrow$) and spin-down ($\lambda_i^\downarrow$) channels, along with the total L\"owdin occupation $n = \sum_{i=1}^5 (\lambda_i^\uparrow + \lambda_i^\downarrow)$~\cite{Sit2011}. 
    L\"owdin-orthogonalized atomic orbitals are used as Hubbard projectors for all cases, except PBEsol+$U$+rVV10 (WF) which corresponds to calculations using frontier Wannier functions (WF) with respective $U$. Eigenvalues highlighted in bold correspond to fully occupied states.}
    \begin{tabular}{l|ccccc|cccccc|cc} 
    \hline\hline
    Method&$\lambda_1^\uparrow$ & $\lambda_2^\uparrow$ & $\lambda_3^\uparrow$ & $\lambda_4^\uparrow$ & $\lambda_5^\uparrow$ & & $\lambda_1^\downarrow$ & $\lambda_2^\downarrow$ & $\lambda_3^\downarrow$ & $\lambda_4^\downarrow$ & $\lambda_5^\downarrow$ & & $n$ \\
    \hline
    \pbesol\ & 0.38 & 0.38 & \textbf{0.98} & \textbf{0.98} & \textbf{0.98} & & 0.38 & 0.38 & \textbf{0.98} & \textbf{0.98} & \textbf{0.98} &  & 7.39 \\
    \pbesolu\ &0.32 & 0.32 & \textbf{0.99} & \textbf{0.99} & \textbf{0.99} & & 0.32 & 0.32 & \textbf{0.99} & \textbf{0.99} & \textbf{0.99} & & 7.25 \\
    \pbesoluv\ &  0.33 & 0.33 & \textbf{0.99} & \textbf{0.99} & \textbf{0.99} & & 0.33 & 0.33 & \textbf{0.99} & \textbf{0.99} & \textbf{0.99} & & 7.27 \\
    \pbesoluo\ & 0.27 & 0.27 & \textbf{1.00} & \textbf{1.00} & \textbf{1.00} & &  0.27 & 0.27 & \textbf{1.00} & \textbf{1.00} & \textbf{1.00} & & 7.07 \\
    \pbesoluov\ & 0.28 & 0.28 & \textbf{1.00} & \textbf{1.00} & \textbf{1.00} & &  0.28 & 0.28 & \textbf{1.00} & \textbf{1.00} & \textbf{1.00} & & 7.10 \\ 
     PBEsol+$U$+rVV10 (WF) &0.01 & 0.01 & \textbf{0.99} & \textbf{0.99} & \textbf{0.99} & & 0.01 & 0.01 & \textbf{0.99} & \textbf{0.99} & \textbf{0.99} & & 6.00 \\
    \hline
    HSE06@\pbesol\ & 0.32 & 0.32 & \textbf{1.00} & \textbf{1.00} & \textbf{1.00} &&   0.32 & 0.32 & \textbf{1.00} & \textbf{1.00} & \textbf{1.00} &  & 7.27 \\
    HSE06@\pbesoluv\ & 0.33 & 0.33 & \textbf{1.00} & \textbf{1.00} & \textbf{1.00} &&   0.33 & 0.33 & \textbf{1.00} & \textbf{1.00} & \textbf{1.00} &  & 7.28 \\
    HSE06@\pbesoluov\  & 0.33 & 0.33 & \textbf{1.00} & \textbf{1.00} & \textbf{1.00} &&   0.33 & 0.33 & \textbf{1.00} & \textbf{1.00} & \textbf{1.00} &  & 7.28 \\
    \hline
    Nominal (low spin)&  0.00 & 0.00 & \textbf{1.00} & \textbf{1.00} & \textbf{1.00} & & 0.00 & 0.00 & \textbf{1.00} & \textbf{1.00} & \textbf{1.00} & & 6.00  \\
    \hline\hline
    \end{tabular}
    \label{tab:magnlicoo2}
\end{table*}

To assess the accuracy of the Hubbard-corrected PDOS discussed above, we compare these results with PDOS calculations performed using the HSE06 hybrid functional. In order to isolate the effect of the xc functional from that of the underlying geometry, the HSE06 calculations are carried out on exactly the same optimized structures used in Figs.~\ref{fig:licoo2dos}(a), (c), and (e). The resulting PDOS are shown in Figs.~\ref{fig:licoo2dos}(b), (d), and (f). The HSE06 PDOS obtained for the different geometries are qualitatively very similar, with only minor variations in the band gap values. However, a notable qualitative difference emerges when comparing the HSE06 PDOS with those obtained from the Hubbard-corrected calculations, particularly in the occupied-state region. In the HSE06 results, the VBM retains a predominantly Co-$3d$ $t_{2g}$ character, but with a significantly enhanced hybridization with O-$2p$ states. While the presence of Co-$3d$ states at the VBM is correctly captured, the strong O-$2p$ contribution appears overestimated. Experimentally, the VBM is found to be almost purely of Co-$3d$ character with only a minor O-$2p$ contribution (see Fig.~19 in Ref.~\cite{Ensling2010}). Moreover, within HSE06 the deeper-lying occupied states do not undergo substantial redistribution compared to the PBEsol+rVV10 reference, aside from a rigid shift toward the VBM (i.e., an effective bandwidth reduction). The unoccupied states are shifted to higher energies, leading to an increased band gap that nevertheless overestimates the experimental value. It is important to emphasize that these results correspond to the standard HSE06 functional, which employs a fixed fraction (0.25) of exact exchange. However, this fraction is known to be material dependent (like $U$ in DFT+$U$), since it is closely related to the electronic dielectric screening in solids. In fact, the fraction of exact exchange is inversely proportional to the dielectric constant~\cite{Skone2014}. Using the theoretical average high-frequency dielectric constant of LiCoO$_2$ ($\varepsilon_\infty = 6.1$)~\cite{Takahashi:2020}, one obtains a fraction of exact exchange of about 0.16. We verified that using this reduced exact-exchange fraction decreases the band gap to 2.7~eV, which is in excellent agreement with the upper limit of the experimental values~\cite{vanElp1991}. Finally, as shown in Table~\ref{tab:magnlicoo2}, the eigenvalues of the diagonal occupation matrix obtained with HSE06 remain very similar to those obtained in the other cases discussed above. This suggests that an analysis based solely on occupation-matrix eigenvalues in Table~\ref{tab:magnlicoo2} is not sufficient to capture the qualitative differences in the PDOS of LiCoO$_2$ arising from different xc functionals. 

Therefore, as discussed above, the PDOS obtained with Hubbard corrections based on L\"owdin-orthogonalized atomic orbitals differs substantially from both the HSE06 PDOS and the experimental spectra, particularly in the occupied-state region. This discrepancy arises from the combined effect of the relatively large first-principles value of $U_\mathrm{Co}$ obtained from DFPT and the choice of Hubbard projectors. These two factors are closely intertwined, as the computed value of $U$ depends strongly on the projector functions. Regarding the choice of Hubbard projectors, it has been shown that defining them using Wannier functions constructed as ``frontier'' (molecular) orbitals~\cite{Bajaj2021, Dabo2007} significantly improves the description of LiCoO$_2$. In particular, the resulting occupation matrices have eigenvalues that are much closer to their nominal integer values~\cite{Ting2023}. In this approach, the wannierization is restricted to the $t_{2g}$ and $e_g$ states in the vicinity of the band gap [see Figs.~S5(a) in the SI], yielding molecular orbitals with contributions from both TM and ligand ions [see Figs.~S5(c) in the SI]. This contrasts with the use of broader energy windows (including the full Co-$3d$ and O-$2p$ manifold), which produce more localized, atomic-like Wannier functions centered on the TM sites only~\cite{Isaacs2020b}. As a result, the eigenvalues of the occupation matrix in the frontier Wannier basis are significantly closer to their nominal integer values (1 and 0)~\cite{Ting2023}. We verified that replacing L\"owdin-orthogonalized atomic orbitals with frontier Wannier functions [see Figs.~S5(c) in the SI] yields occupation matrices with nearly integer eigenvalues (see Table~\ref{tab:magnlicoo2}). Moreover, we computed the PDOS using these frontier Wannier functions as Hubbard projectors and found that, with the respective ``one-shot'' (see Sec.~\ref{comp_details}) first-principles $U=3.29$~eV from DFPT~\cite{Andolfatto2026}, the overall PDOS profile agrees more closely with the HSE06 results discussed above while the band gap is significantly larger [see Fig.~S5(b) in the SI]. In particular, this approach preserves the dominant Co-$3d$ $t_{2g}$ character of the VBM of the initial PBEsol calculation, in much better agreement with experimental observations than the case shown in Fig.~\ref{fig:licoo2dos}(c). These findings highlight that accurate electronic-structure predictions for LiCoO$_2$ strongly depend not only on the value of the Hubbard $U$, but also on the nature and spatial character of the Hubbard projector functions used to apply the correction to the Kohn-Sham states. While the use of frontier Wannier functions appears to be a very promising direction, this approach is not yet at the same level of maturity as the standard self-consistent protocol based on DFPT and L\"owdin-orthogonalized atomic orbitals that we employ here~\cite{Timrov2021}. In particular, the calculation of Pulay forces and stresses in the Wannier basis is not yet available, preventing structural optimizations. Moreover, a fully self-consistent calculation of $U$ within the DFPT framework using Wannier-based Hubbard projectors is not yet readily accessible. Hence, the remainder of this paper is based on our standard self-consistent protocol employing L\"owdin-orthogonalized atomic orbitals. A more detailed study of LiCoO$_2$ using Wannier-based Hubbard projectors will be presented elsewhere~\cite{Andolfatto2026}. 

Overall, we find that applying the Hubbard corrections $U_\mathrm{Co}$, $U_\mathrm{O}$, and $V$ using L\"owdin-orthogonalized atomic orbitals as Hubbard projectors leads to an electronic structure that differs qualitatively from the HSE06 results, particularly in the distribution of occupied states and in the orbital character of the VBM. Motivated by Ref.~\cite{Ting2023}, we have shown that replacing L\"owdin-orthogonalized atomic orbitals with frontier Wannier functions, together with the corresponding Hubbard $U$ parameter, significantly improves the agreement of the calculated PDOS with both HSE06 results and experimental data compared to the standard self-consistent DFT+$U$ protocol based on L\"owdin-orthogonalized atomic orbitals~\cite{Timrov2021}. These results suggest that the use of frontier Wannier functions as Hubbard projectors~\cite{Bajaj2021, Ting2023} is a promising and powerful route toward obtaining electronic structures that are more consistent with both hybrid-functional calculations and experiments. An alternative strategy is to adopt a more targeted correction scheme, acting only on selected localized orbitals (e.g., the $t_{2g}$ manifold) while leaving other orbitals (e.g., the $e_g$ manifold) unaffected, as enabled by the orbital-resolved DFT+$U$ approach~\cite{Macke2024}.

\subsection{CoO$_2$}

As stated in Sec.~\ref{sec:Crystal_structure}, CoO$_2$ is experimentally a paramagnetic metal. Within DFT, however, an explicit treatment of paramagnetism requires computationally demanding calculations on large supercells, for example using the special quasirandom structures (SQS) approach~\cite{Zunger1990, Alling2010}. In the present work, we therefore approximate CoO$_2$ using the primitive unit cell containing one Co atom and two O atoms in a ferromagnetic configuration. We also examined an antiferromagnetic configuration by constructing a supercell containing two formula units, with the two Co atoms initialized with opposite spin orientations (not shown). The resulting antiferromagnetic solution exhibits behavior similar to the ferromagnetic state and was found to be 0.2~eV (per cell) higher in energy than the corresponding ferromagnetic solution computed in the same doubled cell. Although the ferromagnetic unit-cell description of CoO$_2$ is a simplified approximation of the true paramagnetic state, this approach is commonly adopted in the literature, particularly in high-throughput studies. Since the objective of the present work is to assess the relative accuracy of different xc functionals by comparing their predictions with each other and with available experimental data, we consider this approximation sufficient for the purpose of the present study.

\subsubsection{Magnetic energy landscape at experimental geometry}
\label{sec:Magnetic_energy_landscape}

Modeling magnetic materials within DFT+$U$ is challenging because multiple (meta-)stable electronic solutions often coexist~\cite{Ponet2024, Haddadi2026}. As a result, a DFT+$U$ calculation may converge to a local minimum rather than the true ground state. Identifying the lowest-energy solution therefore requires a systematic exploration of the magnetic energy landscape, for example using the procedure described in Ref.~\cite{Ponet2024}. This approach consists of a series of independent constrained DFT+$U$ calculations in which the occupation matrices of the Hubbard subspaces are initially constrained toward predefined target configurations. These targets correspond to different distributions of spin-up and spin-down electrons among the Co $3d$ orbitals and are generated randomly while satisfying the symmetry constraints on the occupation matrices. After the electronic density converges sufficiently close to the target occupation, the constraint is removed and the system is allowed to relax freely. Depending on the stability of the electronic state, the calculation either converges to the nearest local minimum or evolves toward a different solution. Repeating this procedure for many target occupations enables systematic sampling of the magnetic energy landscape and identification of the global minimum.

\begin{figure}[t]
    \centering
    \includegraphics[width=\linewidth]{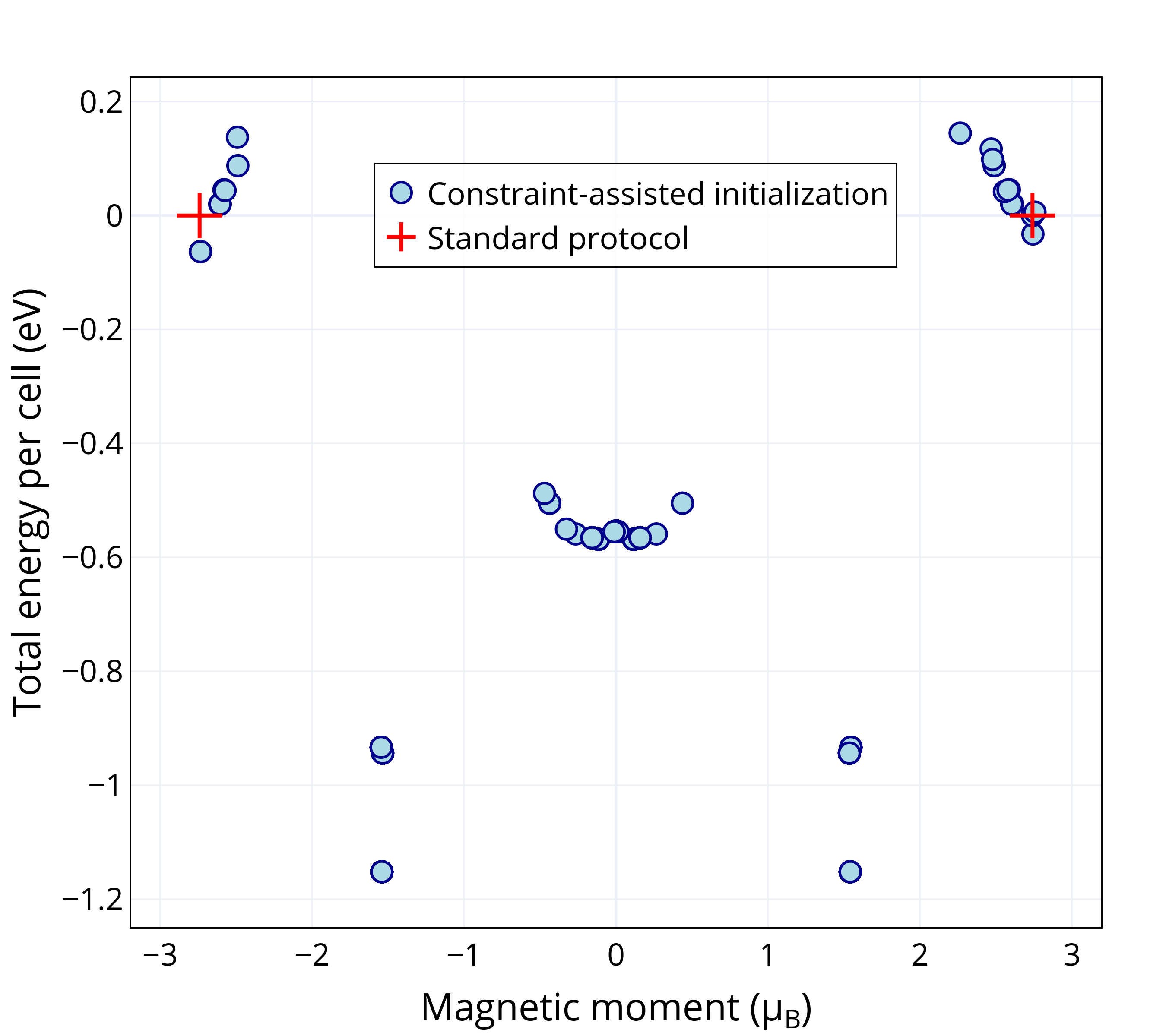}
    \caption{Magnetic energy landscape of CoO$_2$ at experimental geometry. All calculations are performed using \pbesolu\ with the value of $U_\mathrm{Co}=6.69$~eV. Blue circles denote the self-consistent solutions obtained by initially constraining the Co $3d$ occupation matrices toward randomly generated target configurations and subsequently removing the constraint during the self-consistent procedure. Red crosses represent the solutions obtained from conventional DFT+$U$ calculations without applying any occupation constraints. Energies are referenced to the conventional DFT+$U$ solution. The two lowest-energy blue circles correspond to the symmetry-equivalent global minima.}
    \label{fig:capulet}
\end{figure}

\begin{table*}[t]
    \centering
    \caption{Eigenvalues of the diagonal ($I=J$) occupation matrix for the Co-$3d$ states of CoO$_2$ using the experimental geometry and the \pbesolu\ functional with the value of $U_\mathrm{Co}=6.69$~eV, shown separately for the spin-up ($\lambda_i^\uparrow$) and spin-down ($\lambda_i^\downarrow$) channels. Also reported are the total L\"owdin occupation, $n = \sum_{i=1}^5 (\lambda_i^\uparrow + \lambda_i^\downarrow)$~\cite{Sit2011}, and the local magnetic moment, $m = \sum_{i=1}^5 (\lambda_i^\uparrow - \lambda_i^\downarrow)$, of the Co ion, as well as the total magnetization of the simulation cell, $M$. Eigenvalues highlighted in bold correspond to fully occupied states.}  
    \begin{tabular}{l|ccccc|cccccc|cccc} 
    \hline\hline
     Method &$\lambda_1^\uparrow$ & $\lambda_2^\uparrow$ & $\lambda_3^\uparrow$ & $\lambda_4^\uparrow$ & $\lambda_5^\uparrow$ & & $\lambda_1^\downarrow$ & $\lambda_2^\downarrow$ & $\lambda_3^\downarrow$ & $\lambda_4^\downarrow$ & $\lambda_5^\downarrow$ & & $n$ & $m$ ($\mu_\mathrm{B}$)& $M$ ($\mu_\mathrm{B})$\\
    \hline
     Conventional DFT+$U$ & \textbf{0.90} & \textbf{0.91} & \textbf{1.00} & \textbf{1.00} & \textbf{1.00} & & 0.09 & 0.09 & 0.44 & 0.44 & \textbf{0.99} & & 6.86 & 2.74 & 3.00\\
     Constraint-assisted DFT+$U$ &   0.66 & 0.66 & \textbf{0.99} & \textbf{0.99} & \textbf{0.99} && 0.08 & 0.35 & 0.35 & \textbf{1.00} & \textbf{1.00} &  & 7.07 & 1.54 & 1.00  \\
    \hline\hline
    \end{tabular}
    \label{tab:capulet}
\end{table*}

We applied this methodology to CoO$_2$. The magnetic energy landscape was explored using 500 independent \pbesolu\ calculations performed for the experimental crystal structure (the influence of structural relaxation is discussed in Sec.~\ref{ssec:cryst_struc_CoO2}), and the resulting solutions are shown in Fig.~\ref{fig:capulet}. In the initial scan, we employed $U_\mathrm{Co}=6.69$~eV, corresponding to the self-consistent Hubbard parameter obtained from a conventional unconstrained DFT+$U$ calculation. Because the magnetic energy landscape, including the number of accessible solutions, depends on the value of the Hubbard $U$ parameter~\cite{Ponet2024}, we performed an additional \emph{a posteriori} scan using $U_\mathrm{Co}=7.99$~eV, the self-consistent value obtained for the lowest-energy solution identified in the first scan (see Fig.~\ref{fig:capulet}). The corresponding results, reported in Fig.~S6 of the SI, confirm that the same lowest-energy solution is recovered, demonstrating that the identified ground state is robust with respect to the update of $U$. We also verified that 500 calculations are sufficient to identify the lowest-energy configuration of CoO$_2$.

We now analyze the magnetic energy landscape shown in Fig.~\ref{fig:capulet} and compare the global minimum identified by the constrained DFT+$U$ workflow with the solution obtained from a conventional unconstrained DFT+$U$ calculation. The red crosses denote solutions obtained from unconstrained DFT+$U$, which may converge to different electronic states depending on the initial conditions~\cite{Meredig:2010}, whereas the blue circles represent solutions sampled using the constrained DFT+$U$ workflow. The global minimum is identified as the lowest-energy state among all sampled solutions. The energy landscape is nearly symmetric with respect to the Co magnetic moment, reflecting the energetic equivalence of solutions with opposite spin polarization. Remarkably, the two symmetry-equivalent solutions obtained from unconstrained DFT+$U$ lie approximately 1.3~eV above the global minimum. Such a large energy difference demonstrates that a conventional DFT+$U$ calculation can become trapped in a metastable state, leading to qualitatively different electronic and magnetic properties. 

The nature of these two solutions is elucidated by the eigenvalues of the Co $3d$ occupation matrices reported in Table~\ref{tab:capulet}. The unconstrained DFT+$U$ solution exhibits five nearly fully occupied spin-up orbitals and one nearly fully occupied spin-down orbital, while the remaining orbitals have fractional occupations. This corresponds to an unphysical electronic configuration with six occupation-matrix eigenvalues close to unity, instead of the five electrons expected for the high-spin Co$^{4+}$ ion. In contrast, the global minimum identified by the constrained DFT+$U$ workflow exhibits three fully occupied spin-up orbitals and two fully occupied spin-down orbitals, in agreement with the expected low-spin Co$^{4+}$ ($t_{2g}^{5}$) configuration shown schematically in Fig.~\ref{fig:capulet}. Consistently, the total magnetic moment of the unit cell is reduced from 3~$\mu_\mathrm{B}$ for the metastable unconstrained solution to the expected value of around 1~$\mu_\mathrm{B}$ for the low-spin ground state. These findings demonstrate that identifying the global minimum is essential for obtaining the physically relevant electronic and magnetic ground state of CoO$_2$. 

\subsubsection{Structural properties}
\label{ssec:cryst_struc_CoO2}

\begin{figure}[t]
    \centering
    \includegraphics[width=0.99\linewidth]{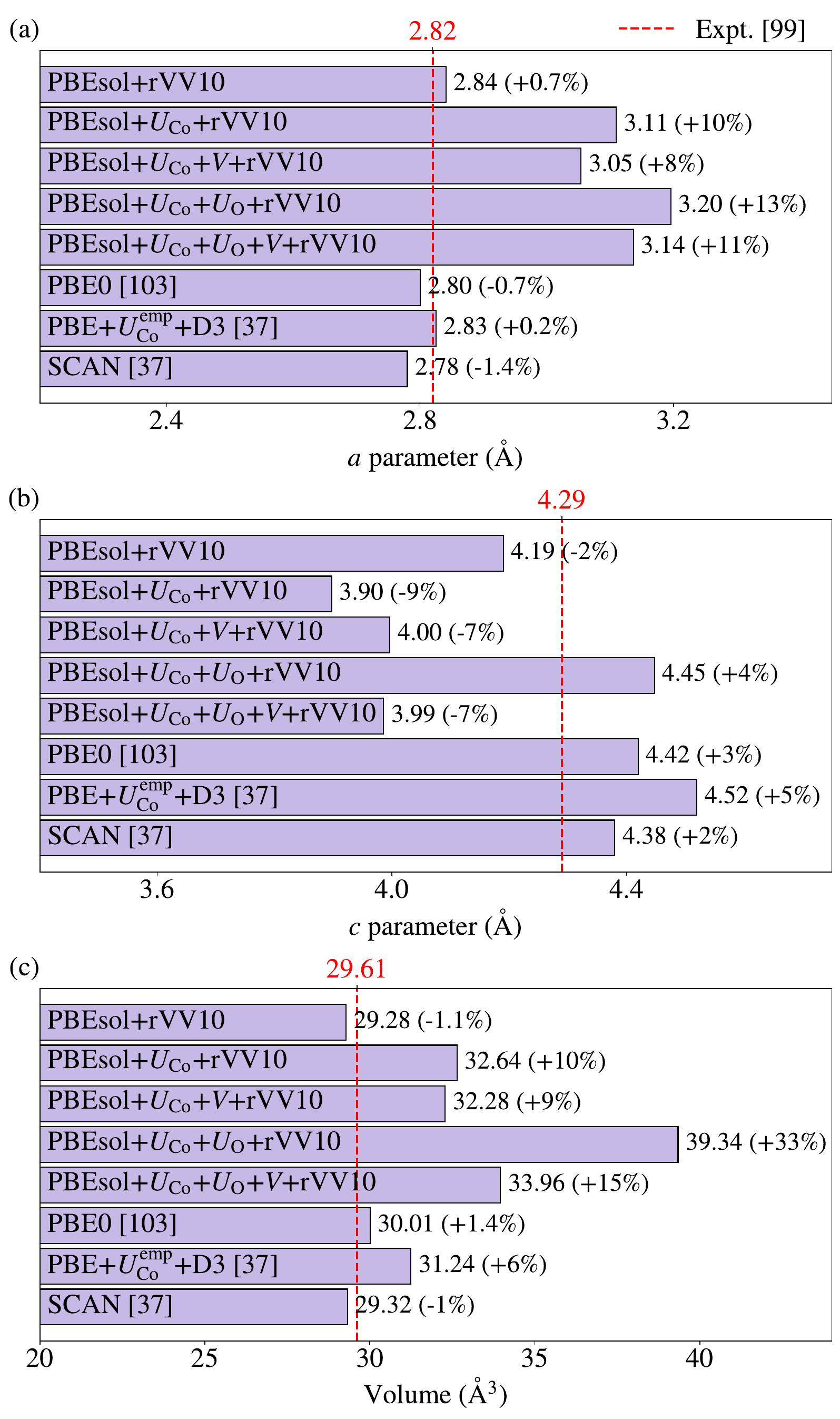}
    \caption{Structural parameters of CoO$_2$: (a) lattice parameter $a$, (b) lattice parameter $c$, and (c) unit-cell volume. Experimental values (shown as vertical dashed lines) are taken from Ref.~\cite{Amatucci1996}. Computational results obtained with PBE0, SCAN, and PBE+$U_\mathrm{Co}^\mathrm{emp}$+D3 (using an empirical $U_\mathrm{Co}^\mathrm{emp} \approx 3$~eV for the Co-$3d$ states) are taken from Refs.~\cite{Mattila2022, Chakraborty2018}}
    \label{fig:CO_lat_param}
\end{figure}

Similarly to LiCoO$_2$, we analyze here the optimized lattice parameters and cell volume of CoO$_2$. Upon complete removal of Li ions from LiCoO$_2$, the experimental cell volume decreases from 96.45~\AA$^3$ to 88.83~\AA$^3$ \cite{Amatucci1996}. This change is accompanied by a phase transition from the O3 to the O1 layered structure (see Fig.~\ref{fig:materials}). The O1 structure is described using a single formula unit, resulting in a volume reduced by a factor of three compared to LiCoO$_2$, for which three formula units were considered. Accordingly, the experimental reference volume for CoO$_2$ is 29.61~\AA$^3$ \cite{Amatucci1996}. The removal of Li ions leaves the CoO$_2$ layers separated by only weak vdW interactions, making the explicit inclusion of vdW forces essential in simulations.

Since CoO$_2$ is experimentally a paramagnetic metal, we perform spin-polarized DFT calculations, including structural optimization. Because the optimized crystal structure is coupled to the underlying electronic structure (see Sec.~\ref{sec:CoO2_electronic_structure}), the distinction between the global minimum and metastable electronic solutions discussed above can have a profound impact on the relaxed geometry. This is precisely the case for CoO$_2$. As shown below, the low-spin global minimum cannot be retained during structural optimization when Hubbard corrections are included. Instead, the system relaxes toward a metastable high-spin-like electronic state, leading to a crystal structure that deviates substantially from experiment.

Figure~\ref{fig:CO_lat_param} summarizes the lattice parameters $a$ and $c$ and the cell volume. Focusing first on the $a$ lattice parameter shown in Fig.~\ref{fig:CO_lat_param}(a), we note that experimentally $a$ remains unchanged when going from LiCoO$_2$ to CoO$_2$, with a value of 2.82~\AA~\cite{Antolini2004, Amatucci1996}. In contrast, our calculations predict a variation in $a$. Specifically, PBEsol+rVV10 overestimates $a$ by 0.7\%, which represents an improvement compared to the LiCoO$_2$ case (see Fig.~\ref{fig:LCO_lat_param}(a)). Inclusion of the $U_\mathrm{Co}$ correction significantly increases $a$ (by $10\%$), while adding $V$ reduces it to 8\%. Adding the $U_\mathrm{O}$ correction (without $V$) leads to a larger increase, yielding a deviation of 13\%, while including $V$ reduces this deviation to 11\%. Thus, in contrast to LiCoO$_2$, the optimized lattice parameter $a$ of CoO$_2$ is substantially overestimated when Hubbard corrections are applied, with the magnitude of the overestimation depending on the specific Hubbard correction employed. This sensitivity is partially related to the variations in the first-principles Hubbard parameters (see Table~\ref{tab:UOV}) and is also influenced by changes in the electronic structure. Conversely, previous computational studies using PBE0 and SCAN provide accurate predictions of $a$, underestimating the experimental value by only 0.7--1.4\%, while PBE+$U_\mathrm{Co}^\mathrm{emp}$+D3 yields nearly perfect agreement with experiment~\cite{Mattila2022, Chakraborty2018}.

The behavior of the lattice parameter $c$ in CoO$_2$ differs significantly. The PBEsol+rVV10 functional underestimates $c$ by 2\%, which is a larger deviation than observed for LiCoO$_2$ (compare Fig.~\ref{fig:CO_lat_param}(b) with Fig.~\ref{fig:LCO_lat_param}(b)). Inclusion of the $U_\mathrm{Co}$ correction further reduces $c$, leading to an underestimation of $7 - 9$\%, depending on whether the inter-site $V$ correction is applied. This represents a much larger deviation than in the LiCoO$_2$ case. The addition of $U_\mathrm{O}$ to \pbesoluv\ does not modify the prediction, while in contrast, the inclusion of $U_\mathrm{Co}$, $U_\mathrm{O}$, without $V$, overestimates $c$ by 4\%. These results indicate that the predicted $c$ parameter is highly sensitive to the choice of Hubbard corrections, with variations spanning a wide range. Previous computational studies overestimate $c$ by $2 - 5$\%. Interestingly, our \pbesoluo\ results yield a lattice parameter $c$ in good agreement with predictions obtained using PBE0, SCAN, and PBE+$U_\mathrm{Co}^\mathrm{emp}$+D3.

Considering the cell volume, we recall that it depends quadratically on $a$ and linearly on $c$, so errors in the lattice parameters are amplified accordingly in the volume prediction. As shown in Fig.~\ref{fig:CO_lat_param}(c), \pbesol\ provides very good agreement with the experimental volume, underestimating it by only 1.1\%, which is remarkable and represents an improvement over LiCoO$_2$. Inclusion of $U_\mathrm{Co}$ and inter-site $V$ leads to an overestimation of the volume by $9 - 10$\%, resulting from partial cancellation between the overestimated $a$ and underestimated $c$. This level of deviation is twice as large as that observed for LiCoO$_2$. In contrast, adding $U_\mathrm{O}$ substantially worsens the prediction, leading to an overestimation of approximately 33\% (without $V$), which is considerably larger than for LiCoO$_2$. The inclusion of $V$ reduces the deviation to $15\%$. Previous computational studies, in comparison, predict the volume in close agreement with experiment, with deviations of $1 - 6$\% ~\cite{Mattila2022, Chakraborty2018}.  

Overall, for CoO$_2$ we find that PBEsol+$U_\mathrm{Co}$+$U_\mathrm{O}$+$V$+rVV10, which provided the most accurate structural properties for LiCoO$_2$, performs poorly relative to previous computational studies based on PBE0, SCAN, and PBE+$U_\mathrm{Co}^\mathrm{emp}$+D3. As discussed below, this is both caused by and has significant consequences on the predicted electronic structure of CoO$_2$, while also highlighting the reciprocal influence between the electronic structure and the optimized crystal structure.

\subsubsection{Electronic structure properties}
\label{sec:CoO2_electronic_structure} 

\begin{figure}[t]
    \centering
    \includegraphics[width=0.98\linewidth]{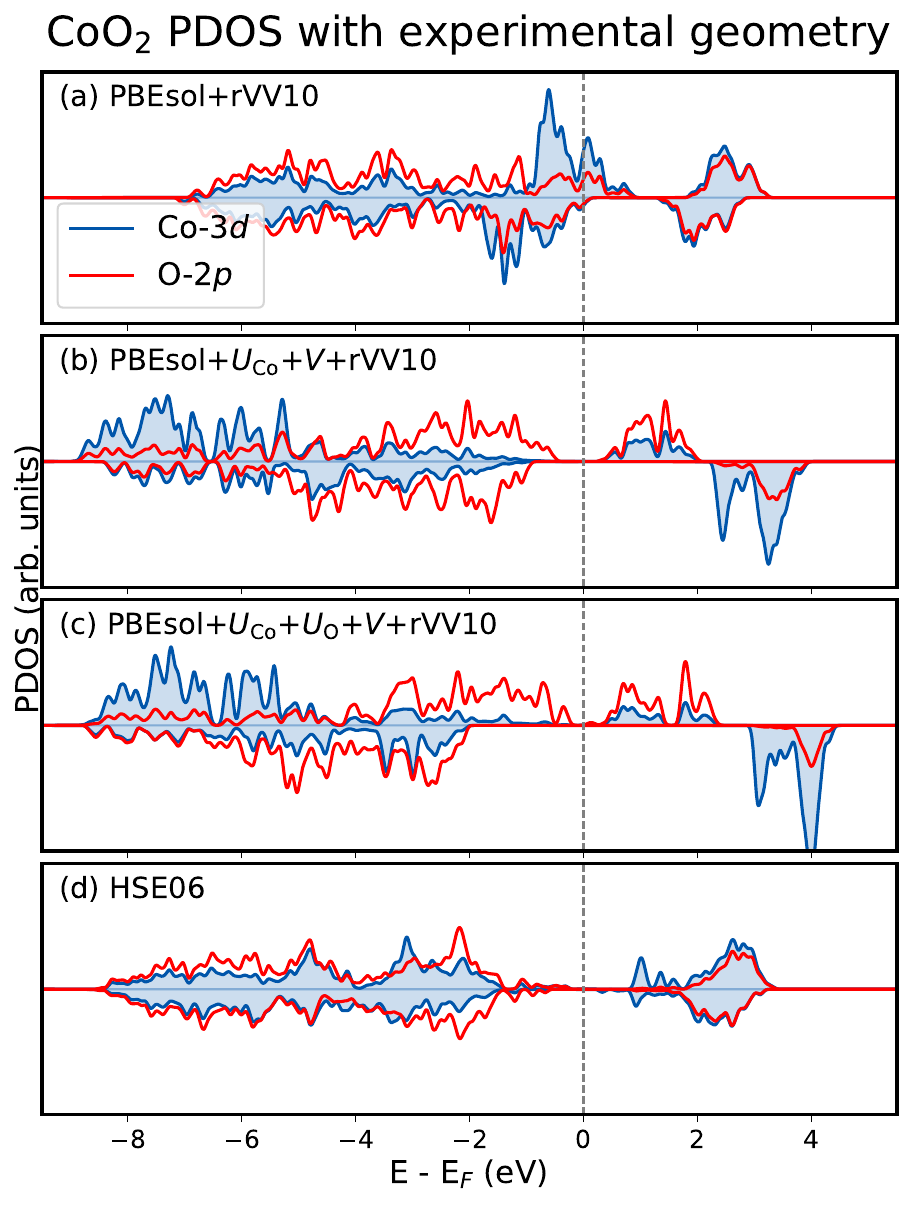}
    \caption{PDOS of CoO$_2$ computed using the experimental geometry: (a) PBEsol+rVV10, (b) PBEsol+$U_\mathrm{Co}$+$V$+rVV10, (c) PBEsol+$U_\mathrm{Co}$+$U_\mathrm{O}$+$V$+rVV10, and (d) HSE06. The Co-$3d$ and O-$2p$ states are shown in blue (filled) and red (unfilled). In each panel, the upper and lower portions of the plot correspond to the spin-up (majority) and spin-down (minority) channels, respectively. The energy reference is set to the Fermi energy, $\mathrm{E}_F$.}
	\label{fig:doscoo2_exp}    
\end{figure}

\begin{figure}[t]
    \centering
    \includegraphics[width=0.95\linewidth]{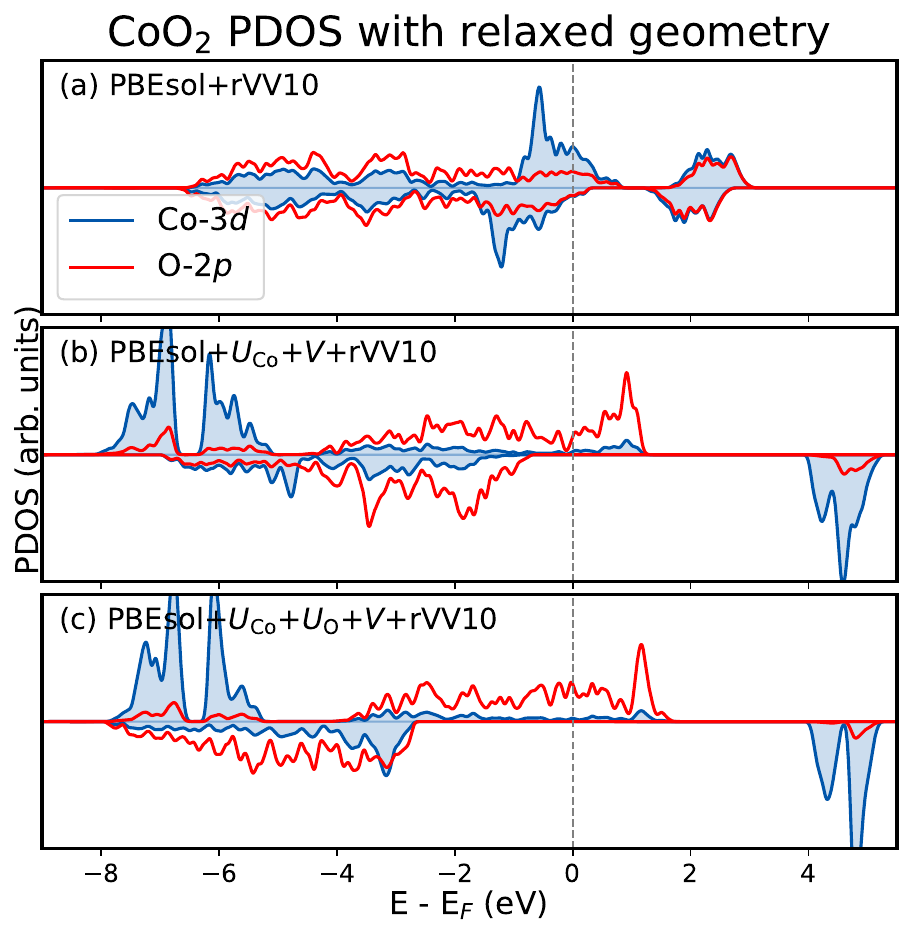}
    \caption{PDOS of CoO$_2$ computed using the respective optimized geometry: (a) \pbesol, (b) \pbesoluv, and (c) \pbesoluov. The Co-$3d$ and O-$2p$ states are shown in blue (filled) and red (unfilled). In each panel, the upper and lower portions of the plot correspond to the spin-up (majority) and spin-down (minority) channels, respectively. The energy reference is set to the Fermi energy, $\mathrm{E}_F$.}
	\label{fig:doscoo2}
\end{figure}

We now examine the effect of the different Hubbard corrections on the electronic structure of CoO$_2$, considering two cases: the experimental and relaxed crystal structure (see Sec.~\ref{ssec:cryst_struc_CoO2}). For both cases, the Hubbard parameters are computed self-consistently and are reported in Table~\ref{tab:UOV}. As discussed below, using the experimental geometry together with the magnetic energy landscape exploration described in Sec.~\ref{sec:Magnetic_energy_landscape} allows us to identify and retain the true low-spin ground state of Co$^{4+}$. In contrast, when structural relaxation is included, the electronic structure no longer remains in this low-spin state and instead converges to a high-spin-like solution. The PDOS for the experimental and relaxed structures is shown in Figs.~\ref{fig:doscoo2_exp} and \ref{fig:doscoo2}, respectively. The corresponding eigenvalues of the Co-$3d$ occupation matrices, together with the total Co-$3d$ occupations and magnetic moments, are summarized in Tables~\ref{tab:magncoo2_expt} and \ref{tab:magncoo2}.
As illustrated in Fig.~\ref{fig:materials}(c), complete delithiation of LiCoO$_2$ formally changes the oxidation state of Co from $+3$ to $+4$, corresponding to the removal of one electron per formula unit. However, because CoO$_2$ is metallic, this formal oxidation-state picture should not be interpreted as the removal of a fully localized electron from the Co-$3d$ manifold, as would be expected in an insulating system. Instead, the removed electron leads to the redistribution of the remaining electrons over the available electronic states of the formula unit. Consequently, the eigenvalues of the Co-$3d$ occupation matrices, reported in Tables~\ref{tab:magncoo2_expt} and \ref{tab:magncoo2}, provide a more informative description of the electronic configuration. Nevertheless, as pointed out by Sit \textit{et al.}~\cite{Sit2011}, occupation-matrix eigenvalues should be interpreted with caution in metallic systems because of the delocalized character of the electronic states.

We first analyze in detail Fig.~\ref{fig:doscoo2_exp}, which shows the PDOS of CoO$_2$ computed using the experimental crystal structure. Figure~\ref{fig:doscoo2_exp}(a) presents the results obtained with \pbesol. The corresponding electronic structure is metallic, in agreement with experimental observations. As reported in Table~\ref{tab:magncoo2_expt}, the three spin-up eigenvalues ($\lambda_3^\uparrow$, $\lambda_4^\uparrow$, and $\lambda_5^\uparrow$) obtained with \pbesol\ are close to unity, indicating nearly fully occupied states. In the spin-down channel, however, two eigenvalues ($\lambda_4^\downarrow$ and $\lambda_5^\downarrow$) are reduced to approximately 0.8, while $\lambda_3^\downarrow$, which would nominally correspond to an unoccupied state, exhibits a significant occupation of 0.71. The remaining eigenvalues, $\lambda_1$ and $\lambda_2$, are close to 0.5 in both spin channels, indicating partial occupation of the $e_g$ manifold. Overall, the PBEsol+rVV10 description of CoO$_2$ provides a physically reasonable electronic structure compared with the other approaches discussed below. In particular, the presence of non-integer occupation-matrix eigenvalues $\lambda_3^\downarrow$, $\lambda_4^\downarrow$, and $\lambda_5^\downarrow$ is consistent with the metallic character of the system, where the Co-$3d$ states are partially delocalized, and therefore should not be interpreted as a failure of the method~\cite{Sit2011}. The PDOS obtained using the relaxed crystal structure (Fig.~\ref{fig:doscoo2}(a)) remains qualitatively unchanged, indicating that structural relaxation does not significantly modify the electronic structure within PBEsol+rVV10.

The inclusion of Hubbard corrections on top of \pbesol\ at the experimental geometry leads to a substantial modification of the electronic structure of CoO$_2$. Importantly, for all cases shown in Fig.~\ref{fig:doscoo2_exp}(b,c), which include $U_\mathrm{Co}$ (with or without $U_\mathrm{O}$) as well as inter-site $V$ interactions, and for the corresponding calculations without $V$ (Fig.~S7 in the SI), the magnetic energy landscape exploration consistently identifies the global minimum. This state corresponds to a low-spin Co$^{4+}$ configuration. The eigenvalues of the occupation matrices are consistent with this picture, with three eigenvalues close to unity in the spin-up channel and two eigenvalues close to unity in the spin-down channel, characteristic of a low-spin $d^5$ configuration. However, these Hubbard-corrected solutions are insulating and therefore contradict the experimentally observed metallic character of CoO$_2$. If the magnetic energy landscape exploration is not performed, the calculations converge instead to metastable high-spin-like solutions even at the experimental geometry (Fig.~S8 in the SI). These metastable states exhibit a metallic electronic structure when only $U_\mathrm{Co}$ is applied, with or without inter-site $V$, although the distribution of spectral weight appears physically questionable. When $U_\mathrm{O}$ is additionally included, the resulting solution becomes insulating. Since these states do not correspond to the lowest-energy solutions identified from the landscape exploration, they are not considered further. To further interpret these results, we also performed HSE06 calculations for CoO$_2$ using the experimental geometry. The resulting PDOS, shown in Fig.~\ref{fig:doscoo2_exp}(d), also corresponds to an insulating solution, in contrast with experiment. Although the PDOS obtained from HSE06 and from the various Hubbard-corrected calculations differ qualitatively, these differences may originate from the specific values of the Hubbard parameters, the choice of Hubbard projectors, and the fraction of exact exchange employed in HSE06 (0.25 in the present calculations). Nevertheless, the occupation-matrix eigenvalues obtained with HSE06 are consistent with those from the Hubbard-corrected calculations, as reported in Table~\ref{tab:magncoo2_expt}.

\begin{table*}[t]
    \centering
    \caption{Eigenvalues of the diagonal ($I=J$) occupation matrix for the Co-$3d$ for the experimental unit cell of CoO$_2$, shown separately for the spin-up ($\lambda_i^\uparrow$) and spin-down ($\lambda_i^\downarrow$) channels. Also reported are the total L\"owdin occupation, $n = \sum_{i=1}^5 (\lambda_i^\uparrow + \lambda_i^\downarrow)$~\cite{Sit2011}, and the local magnetic moment, $m = \sum_{i=1}^5 (\lambda_i^\uparrow - \lambda_i^\downarrow)$, of the Co ion, as well as the total magnetization of the simulation cell, $M$. Eigenvalues shown in bold correspond to fully occupied states.}
    \begin{tabular}{l|ccccc|cccccc|cccc} 
    \hline\hline
    Method &$\lambda_1^\uparrow$ & $\lambda_2^\uparrow$ & $\lambda_3^\uparrow$ & $\lambda_4^\uparrow$ & $\lambda_5^\uparrow$ & &  $\lambda_1^\downarrow$ & $\lambda_2^\downarrow$ & $\lambda_3^\downarrow$ & $\lambda_4^\downarrow$ & $\lambda_5^\downarrow$ & & $n$ & $m$ ($\mu_\mathrm{B}$)& $M$ ($\mu_\mathrm{B}$)\\
    \hline
    \pbesol\ & 0.51 & 0.51 & \textbf{0.97} & \textbf{0.99} & \textbf{0.99} & & 0.47 & 0.47 & 0.71 & 0.80 & 0.81 & & 7.21 & 0.70 & 0.95\\
    \pbesolu\ & 0.72 & 0.72 & \textbf{0.99} & \textbf{0.99} & \textbf{0.99} & & 0.06 & 0.30 & 0.30 & \textbf{1.00} & \textbf{1.00} & & 7.07&1.76&1.00\\
    \pbesoluv& 0.68 &  0.68 & \textbf{0.99} & \textbf{0.99} & \textbf{0.99}& & 0.08 & 0.34 & 0.34 & \textbf{1.00} & \textbf{1.00} & & 7.08& 1.59&1.00\\
    \pbesoluo\ & 0.74 & 0.74 & \textbf{0.99} & \textbf{0.99} & \textbf{1.00} & & 0.05 & 0.21 & 0.21 & \textbf{1.00} & \textbf{1.00}& & 6.94& 2.00&1.00\\
    \pbesoluov\ & 0.74 & 0.74 & \textbf{0.99} & \textbf{0.99} & \textbf{1.00} & & 0.05 & 0.22 & 0.22 & \textbf{1.00} & \textbf{1.00} & & 6.94&1.98&1.00\\
    HSE06 & 0.47 & 0.47 & \textbf{0.90} & \textbf{0.90} & \textbf{0.98} & & 0.43 & 0.43 & 0.53 & \textbf{0.99} & \textbf{0.99} & & 7.09&0.35&0.31\\  
    \hline
    Nominal (high-spin)& \textbf{1.00} & \textbf{1.00} & \textbf{1.00} & \textbf{1.00} & \textbf{1.00} & & 0.00 & 0.00 & 0.00 & 0.00 & 0.00 & & 5.00 & 5.00&5.00\\
    Nominal (low-spin)& 0.00 & 0.00 & \textbf{1.00} & \textbf{1.00} & \textbf{1.00} & & 0.00 & 0.00 & 0.00 & \textbf{1.00} & \textbf{1.00} & & 5.00 & 1.00&1.00\\  
    \hline\hline
    \end{tabular}
    \label{tab:magncoo2_expt}
\end{table*}

\begin{table*}[t]
    \centering
    \caption{Eigenvalues of the diagonal ($I=J$) occupation matrix (see Eq.~\eqref{eq:occ_matrix_0}) for the Co-$3d$ states for the relaxed unit cell of in CoO$_2$, shown separately for the spin-up ($\lambda_i^\uparrow$) and spin-down ($\lambda_i^\downarrow$) channels. Also reported are the total L\"owdin occupation, $n = \sum_{i=1}^5 (\lambda_i^\uparrow + \lambda_i^\downarrow)$, and the local magnetic moment, $m = \sum_{i=1}^5 (\lambda_i^\uparrow - \lambda_i^\downarrow)$, of the Co ion~\cite{Sit2011}, as well as the total magnetization of the simulation cell, $M$. Eigenvalues shown in bold correspond to fully occupied states.}
    \begin{tabular}{l|ccccc|cccccc|cccc} 
    \hline\hline
    Method&$\lambda_1^\uparrow$ & $\lambda_2^\uparrow$ & $\lambda_3^\uparrow$ & $\lambda_4^\uparrow$ & $\lambda_5^\uparrow$ & &  $\lambda_1^\downarrow$ & $\lambda_2^\downarrow$ & $\lambda_3^\downarrow$ & $\lambda_4^\downarrow$ & $\lambda_5^\downarrow$ & & $n$ & $m$ ($\mu_\mathrm{B}$)& $M$ ($\mu_\mathrm{B}$)\\
    \hline
    \pbesol\ & 0.50 & 0.50 & \textbf{0.92} & \textbf{0.98} & \textbf{0.98} & & 0.47 & 0.47 & 0.75 & 0.81 & 0.81 & & 7.22 & 0.57 & 0.75\\
    \pbesolu\ & \textbf{0.90} & \textbf{0.90} & \textbf{1.00} & \textbf{1.00} & \textbf{1.00} & & 0.03 & 0.13 & 0.13 & \textbf{1.00} & \textbf{1.00} & & 7.08&2.50&1.00\\
    \pbesoluv\ & \textbf{0.87} & \textbf{0.87} & \textbf{0.99} & \textbf{0.99} & \textbf{1.00}& & 0.03 & 0.17 & 0.17 & \textbf{1.00} & \textbf{1.00} & & 7.08& 2.35&1.00\\
    \pbesoluo& \textbf{0.96} & \textbf{0.96} & \textbf{1.00} & \textbf{1.00} & \textbf{1.00} & & 0.01 & 0.06 & 0.06 & \textbf{1.00} & \textbf{1.00}& & 7.04& 2.77&1.00\\
    \pbesoluov& \textbf{0.90} & \textbf{0.90} & \textbf{1.00} & \textbf{1.00} & \textbf{1.00} & & 0.02 & 0.09 & 0.09 & \textbf{1.00} & \textbf{1.00} & & 6.99&2.60&1.00\\ 
    \hline
    Nominal (high-spin)& \textbf{1.00} & \textbf{1.00} & \textbf{1.00} & \textbf{1.00} & \textbf{1.00} & & 0.00 & 0.00 & 0.00 & 0.00 & 0.00 & & 5.00 & 5.00&5.00\\
    Nominal (low-spin)& 0.00 & 0.00 & \textbf{1.00} & \textbf{1.00} & \textbf{1.00} & & 0.00 & 0.00 & 0.00 & \textbf{1.00} & \textbf{1.00} & & 5.00 & 1.00&1.00\\  
    \hline\hline
    \end{tabular}
    \label{tab:magncoo2}
\end{table*}

We now analyze the PDOS and occupation-matrix eigenvalues obtained for the relaxed crystal structure. The results are shown in Figs.~\ref{fig:doscoo2}(b,c), with the corresponding calculations without inter-site $V$ reported in Fig.~S9 in the SI~\cite{Supplementary_Information}. In contrast to the Hubbard-corrected calculations performed at the experimental geometry (Figs.~\ref{fig:doscoo2_exp}(b,c)), the relaxed structure leads to qualitatively different electronic structure. Figure~\ref{fig:doscoo2}(b) shows the PDOS obtained with $U_\mathrm{Co}$ and inter-site $V$ applied to the optimized structure. Unlike the insulating solution obtained for the experimental geometry, this state is metallic. Moreover, its electronic structure differs significantly from the \pbesol\ result shown in Fig.~\ref{fig:doscoo2}(a). In the latter case, the metallic character originates primarily from the Co-$3d$ states, with a smaller contribution from the O-$2p$ states crossing the Fermi level. In contrast, for the Hubbard-corrected calculation, the states at the Fermi level are dominated almost exclusively by O-$2p$ character, while the Co-$3d$ states are shifted away from the Fermi level. The inclusion of the $U_\mathrm{O}$ correction does not lead to substantial changes, as shown in Fig.~\ref{fig:doscoo2}(c).
Further insight is obtained from the occupation-matrix eigenvalues reported in Table~\ref{tab:magncoo2}. When Hubbard corrections are applied, the spin-up Co-$3d$ states become nearly fully occupied. This is also evident from the two pronounced Co-derived peaks around $-7$ and $-6$~eV in the PDOS, which can be attributed to the spin-up $t_{2g}$ and $e_g$ states. If the spin-down channel were completely empty, this configuration would correspond to a high-spin Co$^{4+}$ state. However, the occupation-matrix eigenvalues show a more complex picture: $\lambda_1^\downarrow$, $\lambda_2^\downarrow$, and $\lambda_3^\downarrow$ are close to zero, whereas $\lambda_4^\downarrow$ and $\lambda_5^\downarrow$ are close to unity. Therefore, the resulting electronic configuration corresponds neither to the high-spin nor to the low-spin Co$^{4+}$ state.
This occupation of the Co-$3d$ manifold results from a partial transfer of electronic charge from the O-$2p$ states, which become spin-polarized in the process. Consequently, the application of Hubbard corrections stabilizes a metastable electronic configuration that is distinct from the low-spin ground state identified through the magnetic energy landscape exploration at the experimental geometry. This behavior arises because, during the self-consistent evolution of the relaxed structure, the electronic solution moves away from the true ground state toward a metastable state. This also explains the significant deviation of the optimized structural parameters from experiment (Sec.~\ref{ssec:cryst_struc_CoO2}). A similar ``runaway'' behavior of the electronic configuration has also been reported in other systems, such as FeSb$_3$~\cite{DiLucente:2026}. Overall, these results demonstrate the strong interplay between structural and electronic degrees of freedom in CoO$_2$. While constrained DFT+$U$ allows the low-spin Co$^{4+}$ configuration to be stabilized for the experimental geometry, structural relaxation drives the system toward a distorted structure associated with a different electronic solution.

The results presented above provide several important insights into the description of CoO$_2$. Among the investigated xc functionals, \pbesol\ provides the most consistent description of the experimental electronic structure, yielding a metallic state and an electronic configuration that is qualitatively compatible with a low-spin Co$^{4+}$ picture. In contrast, the inclusion of first-principles Hubbard corrections at the experimental geometry stabilizes the insulating low-spin solution with a band gap that is inconsistent with the experimentally observed metallic character of CoO$_2$. This finding is consistent with earlier DFT+$U$ studies employing empirical Hubbard corrections, which showed that the insulating solution is obtained for $U$ values larger than $\sim 2.5$~eV~\cite{Lee2005}. Upon structural relaxation, the first-principles Hubbard-corrected calculations evolve toward a different metallic state; however, this solution is associated with a different redistribution of charge between Co-$3d$ and O-$2p$ states and does not correspond to either the low-spin or high-spin Co$^{4+}$ configurations. The origin of this behavior can be traced to the large self-consistent Hubbard parameter obtained for the Co-$3d$ states in CoO$_2$. The resulting $U_\mathrm{Co}$ of $\sim 7-8$ eV strongly localizes the Co-$3d$ states and opens a gap, in direct conflict with the experimentally observed metallic character of the material. This electronic structure consequently produces large forces and stresses during structural relaxation that drive a substantial expansion of the lattice. The resulting structural distortion further modifies the electronic configuration and favors a transition from the low-spin solution toward the high-spin-like state found for the relaxed structure. Thus, the limitation of the self-consistent Hubbard-corrected approach for CoO$_2$ appears to originate primarily from the excessively large self-consistent $U$, rather than from the magnetic configuration itself.

It is therefore natural to examine whether hybrid functionals provide a more reliable description of the electronic structure. However, it turns out that HSE06 calculations performed for the experimental geometry also result in an insulating electronic structure, again in disagreement with experiment. Furthermore, the HSE06 calculation at the experimental geometry yields in-plane stresses exceeding 10~GPa, implying that hybrid functionals, much like Hubbard-corrected approaches, tend to favor strong volume expansion. This behavior can be understood from the perspective of dielectric-dependent hybrid functionals. In the latter, as we pointed out in Sec.~\ref{sec:LCO_electronic_structure}, the fraction of exact exchange is related to the inverse of the high-frequency dielectric constant of the material. Since metallic systems exhibit strong electronic screening and consequently a very large dielectric response, the physically appropriate amount of exact exchange is expected to be substantially reduced compared with the standard HSE06 value of 0.25. In the limiting case of complete screening, this would correspond to vanishing exact exchange, recovering a semi-local xc functional. This reasoning is consistent with the observation that \pbesol\ already provides a satisfactory description of the electronic structure of CoO$_2$. A similar argument can be made for Hubbard corrections. Since DFT+$U$ and hybrid functionals are conceptually related approaches that both mitigate SIEs~\cite{Ivady:2014}, the strong screening present in a metallic system suggests that large on-site interaction corrections may not be appropriate. Thus, based on the present electronic-structure analysis and general considerations for metallic systems, neither Hubbard corrections nor standard hybrid functionals appear necessary for describing the ground-state electronic structure of CoO$_2$. Nevertheless, CoO$_2$ is not only metallic but also experimentally paramagnetic, meaning that its electronic structure involves both itinerant carriers and magnetic fluctuations that are not captured by conventional spin-polarized DFT. Therefore, more advanced approaches that explicitly account for dynamical electronic correlations, such DMFT~\cite{Georges1996}, may provide a more appropriate framework for a quantitative description of this material.

\subsection{Voltages}
\label{sec:Voltages}

Intercalation voltages are computed from total-energy differences using the corresponding self-consistent Hubbard parameters for the Hubbard-corrected calculations~\cite{Cococcioni2019,Timrov2022,Timrov2023}. We focus on the average voltage between the two end members of Li$_x$CoO$_2$, namely LiCoO$_2$ ($x=1$) and CoO$_2$ ($x=0$), whose structural and electronic properties were analyzed in detail in the previous sections. Because structural relaxation produces substantial distortions in CoO$_2$, we first evaluate the voltages using the experimental crystal structures of both compounds and then compare them with those obtained using the fully optimized geometries.

\begin{figure}[h!]
    \centering
    \includegraphics[width=\linewidth]{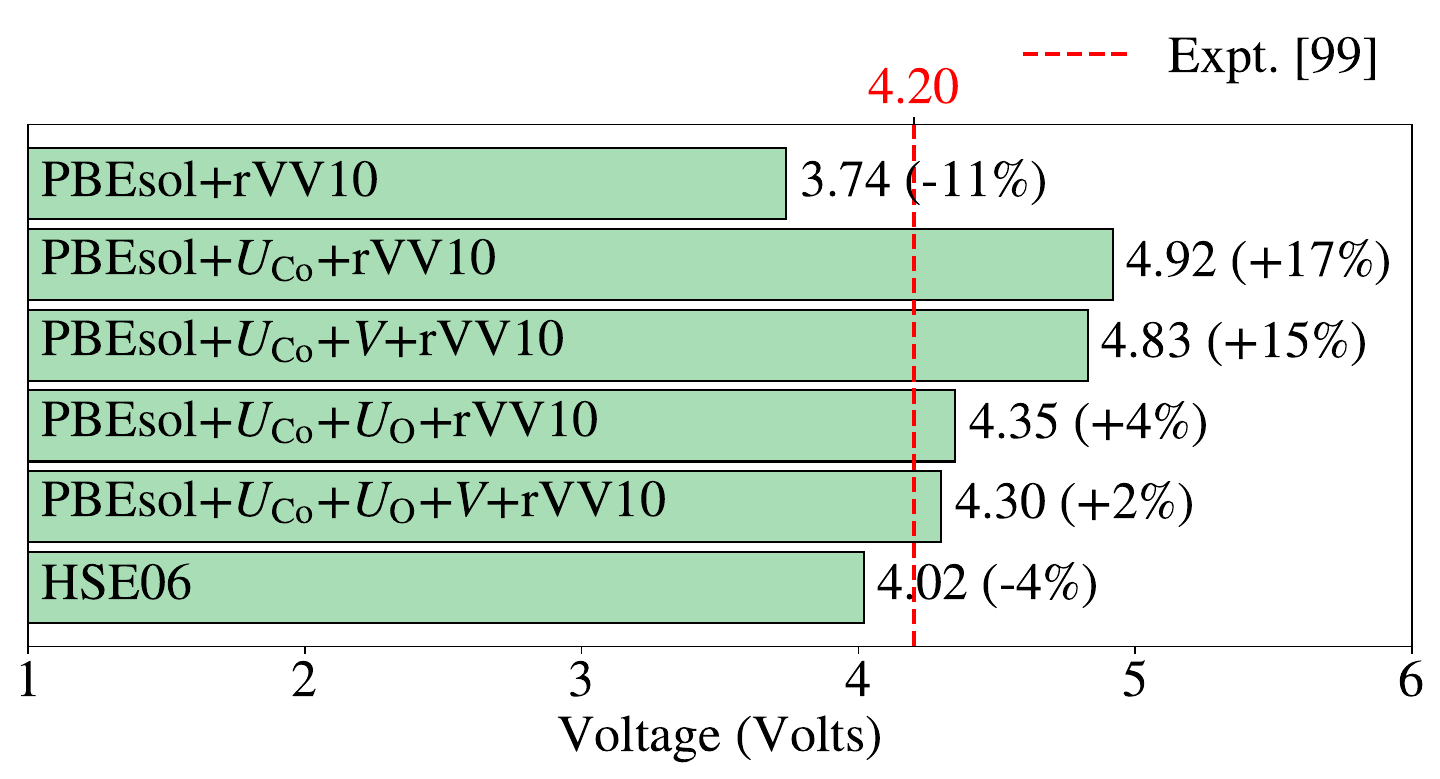}
    \caption{Average intercalation voltages versus Li/Li$^+$ (in V) for Li$_x$CoO$_2$ ($x=0$ and $x=1$) computed using different Hubbard corrections and xc functional and using the experimental geometry. Experimental value (shown as a vertical dashed line) is taken from Ref.~\cite{Amatucci1996}.}
    \label{fig:expt_voltages}
\end{figure}

Figure~\ref{fig:expt_voltages} summarizes the results obtained for the experimental crystal structures. We find that PBEsol+rVV10 underestimates the intercalation voltage by 11\%, consistent with previous studies of olivine and spinel cathode materials~\cite{Timrov2022,Timrov2023}. Introducing the Co on-site Hubbard correction increases this error, leading to a 17\% overestimation. The inclusion of the inter-site interaction $V$ reduces this error only slightly, lowering the overestimation to 15\%. We therefore investigate the effect of also applying a Hubbard correction to oxygen. Including both $U_\mathrm{Co}$ and $U_\mathrm{O}$ (without $V$) reduces the error substantially, yielding a voltage that is overestimated by only 4\%. When all three corrections ($U_\mathrm{Co}$, $U_\mathrm{O}$, and $V$) are included simultaneously, the agreement with experiment improves further, with the voltage overestimated by just 2\%. For comparison, the HSE06 hybrid functional underestimates the voltage by 4\%. Thus, among all xc functionals considered, \pbesoluov\ provides the best agreement with experiment, even outperforming HSE06. Nevertheless, as discussed in the previous sections, this apparent success should be interpreted with caution because the underlying electronic structure of LiCoO$_2$ and CoO$_2$ is inaccurately described.

\begin{figure}[h!]
    \centering
    \includegraphics[width=\linewidth]{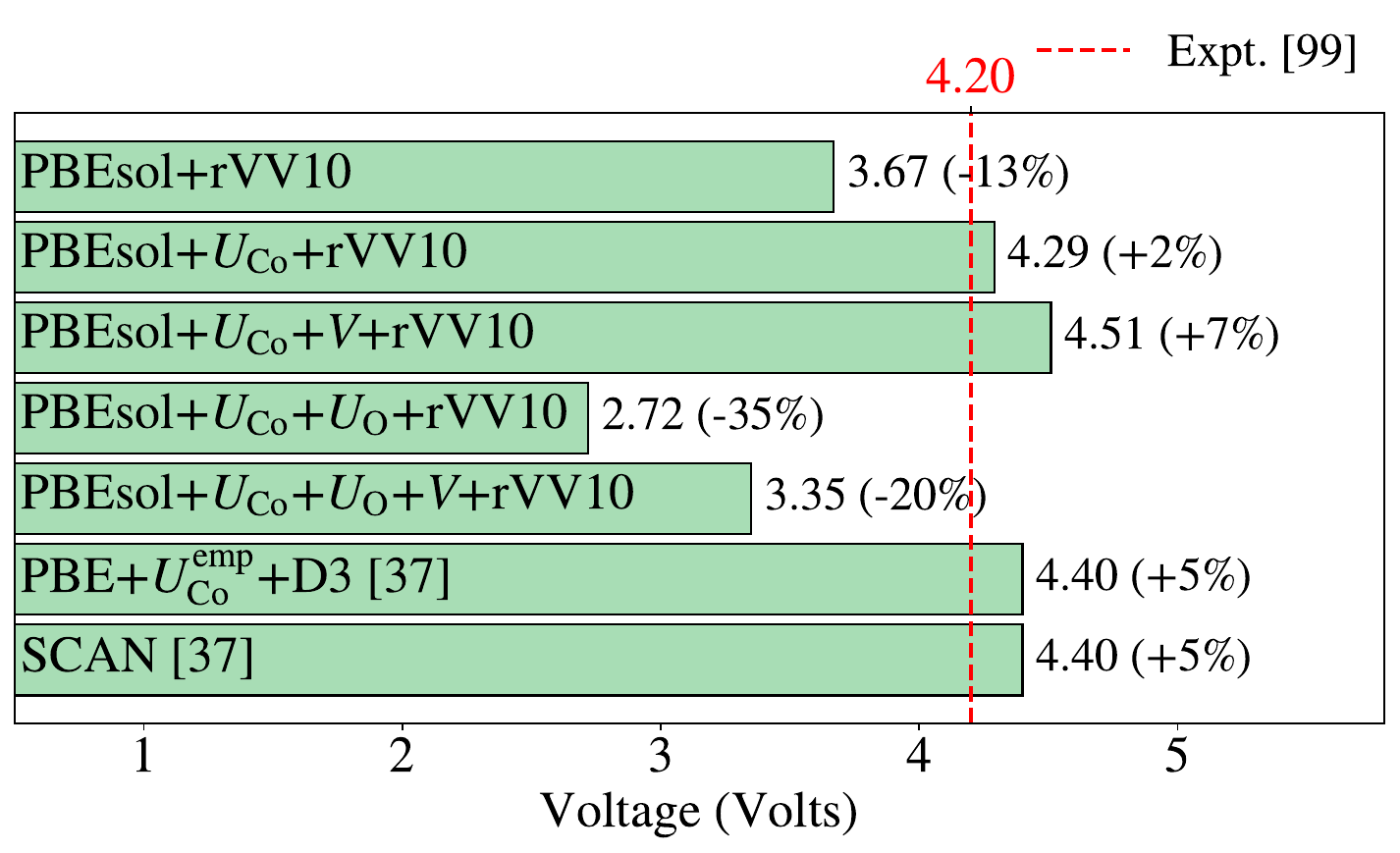}
    \caption{Average intercalation voltages versus Li/Li$^+$ (in V) for Li$_x$CoO$_2$ ($x=0$ and $x=1$) computed using different Hubbard corrections and respective optimized geometries. Experimental value (shown as a vertical dashed line) is taken from Ref.~\cite{Amatucci1996}. Computational results obtained with SCAN and PBE+$U_\mathrm{Co}^\mathrm{emp}$+D3 (using an empirical $U_\mathrm{Co}^\mathrm{emp} \approx 3$~eV for the Co-$3d$ states) are taken from Ref.~\cite{Chakraborty2018}.}
    \label{fig:voltages}
\end{figure}

We now turn to the voltages obtained using the fully relaxed crystal structures (Fig.~\ref{fig:voltages}). PBEsol+rVV10 underestimates the voltage by 13\%, very similar to the result obtained using the experimental geometries. In contrast, the inclusion of $U_\mathrm{Co}$ yields a voltage within 2\% of the experimental value. Unlike the results obtained for the experimental structures and previous studies on other cathode materials~\cite{Timrov2022,Timrov2023}, the addition of the inter-site interaction $V$ deteriorates the agreement, increasing the overestimation to 7\%. A markedly different trend is found when $U_\mathrm{O}$ is included. The \pbesoluo\ functional underestimates the voltage by 35\%, while the additional inclusion of $V$ partially compensates this error but still results in a voltage that is underestimated by 20\%. These trends can be directly related to the structural and electronic properties discussed above. In particular, the pronounced structural distortion of CoO$_2$ induced by structural relaxation in the Hubbard-corrected calculations, together with the accompanying changes in the electronic structure, has a strong impact on the computed voltages, especially when $U_\mathrm{O}$ is included. Surprisingly, calculations including only $U_\mathrm{Co}$ still yield voltages in very good agreement with experiment despite predicting an inaccurate crystal structure and electronic ground state for CoO$_2$.

Overall, the most accurate intercalation voltage is obtained with \pbesoluov\ when the experimental crystal structures are used. However, this agreement results from the electronic structures for LiCoO$_2$ and CoO$_2$ that are inconsistent with experiment. Conversely, for the optimized geometries, the best agreement with experiment is obtained using \pbesolu, despite the fact that the corresponding structural and electronic properties are likewise unsatisfactory. These observations demonstrate that agreement with the experimental intercalation voltage alone is not a sufficient criterion for assessing the quality of an xc functional, which should instead simultaneously reproduce the structural, electronic, and electrochemical properties of the material. 

\section{Conclusions}
\label{sec:conclusions}

We have presented a comprehensive first-principles investigation of the structural properties, electronic structure, and average intercalation voltage of Li$_x$CoO$_2$ ($x=0$ and $x=1$). The role of different Hubbard corrections was systematically examined by considering on-site interactions on Co-$3d$ and O-$2p$ states ($U_\mathrm{Co}$ and $U_\mathrm{O}$), together with the inter-site Co--O interaction ($V$). All Hubbard parameters were computed fully self-consistently within DFPT~\cite{Timrov2018,Timrov2021}, including structural optimization, using L\"owdin-orthogonalized atomic orbitals as Hubbard projectors. To account for dispersion interactions in these layered materials, all calculations employed the rVV10 nonlocal van der Waals functional. The performance of the different approaches was assessed through direct comparison with available experimental data and HSE06 hybrid-functional calculations.

For LiCoO$_2$, we find that the \pbesoluov\ functional provides an excellent description of the crystal structure, with lattice parameters and cell volumes differing by at most 2\% from experiment. In contrast, the electronic structure exhibits noticeable discrepancies with respect to experiment and HSE06, particularly in the distribution of the occupied states, although all approaches consistently predict the expected low-spin Co$^{3+}$ configuration. Motivated by previous work on LiCoO$_2$~\cite{Ting2023}, we further show that frontier Wannier functions substantially improve the electronic structure compared to L\"owdin-orthogonalized atomic orbitals, highlighting that the choice of Hubbard projectors is as important as the values of the first-principles Hubbard parameters.

CoO$_2$ presents a significantly more challenging case. Owing to the coexistence of multiple metastable electronic solutions, we demonstrate that a systematic exploration of the magnetic energy landscape is essential for identifying the true ground state. For the experimental crystal structure, this procedure consistently yields the low-spin Co$^{4+}$ ground state. However, the corresponding Hubbard-corrected electronic structure is insulating, in agreement with HSE06 but in contradiction with the experimentally observed metallic character of CoO$_2$. Structural relaxation further complicates the picture: the optimized geometry deviates substantially from experiment and drives a redistribution of charge between the Co-$3d$ and O-$2p$ states, yielding a metallic electronic structure in which the Co electronic configuration corresponds to neither the conventional low-spin nor high-spin Co$^{4+}$ state. 
In contrast, the PBEsol functional yields a metallic electronic structure in agreement with experiment.
These findings indicate that static mean-field approaches, including both DFT+$U$ and standard hybrid functionals, face fundamental limitations in describing metallic paramagnetic systems such as CoO$_2$. More advanced many-body approaches that explicitly account for dynamical electronic correlations, such as dynamical mean-field theory~\cite{Georges1996}, are therefore expected to provide a more appropriate description.

Despite these shortcomings in the structural and electronic properties, the calculated intercalation voltages are in surprisingly good agreement with experiment. Using the experimental crystal structures, \pbesoluov\ reproduces the average intercalation voltage within 2\% of the experimental value, outperforming all other approaches considered, including HSE06. For the fully relaxed structures, the best agreement is instead obtained with \pbesolu, which also predicts the voltage within 2\% of experiment. These results demonstrate that an accurate prediction of electrochemical properties does not necessarily imply an accurate description of the underlying crystal and electronic structures. The apparent agreement with experiment likely reflects a favorable cancellation of errors in the total-energy differences entering the voltage calculation.
Nevertheless, the good agreement obtained with \pbesoluov\ indicates that the inclusion of first-principles on-site and inter-site Hubbard interactions can provide a promising description of the energetics of layered battery materials, even when the corresponding electronic structure is not fully satisfactory.

Overall, our results demonstrate that the predictive performance of exchange-correlation functionals for battery materials must be evaluated across multiple physical properties. Agreement with experimental voltages alone is not a sufficient measure of predictive accuracy, particularly in high-throughput workflows and the generation of large computational datasets for machine learning and AI-guided materials discovery, where total-energy differences are often considered without a detailed assessment of the underlying ground state. Such inaccuracies can consequently be propagated into machine-learning models and affect predictions and the AI-guided design of novel cathode materials. Instead, a reliable assessment of first-principles frameworks should consider a broad range of properties, including structural, electronic, magnetic, and electrochemical observables. Finally, our work highlights the need for further investigations of layered transition-metal oxides, particularly for compositions exhibiting metallicity and paramagnetism, where methods beyond conventional DFT+$U$ and hybrid functionals are likely required~\cite{Georges1996, Macke2024, Chiarotti2024}.

\section{Methods}
\label{sec:methods}

\subsection{The DFT+$U$+$V$ approach}

In DFT+$U$+$V$~\cite{Campo2010, Himmetoglu2013}, a corrective Hubbard energy term $E_{U+V}$ is added to the standard DFT energy $E_{\mathrm{DFT}}$:
\begin{equation}
    E_{\mathrm{DFT}+U+V} = E_{\mathrm{DFT}} + E_{U+V} .
    \label{eq:Edft_plus_uv}
\end{equation}
While DFT+$U$ accounts only for onsite interactions parameterized by $U$, DFT+$U$+$V$ extends the correction to include intersite interactions between an atom and its neighbors through the parameter $V$. In the simplified rotationally invariant formalism~\cite{Dudarev1998}, the extended Hubbard energy takes the form:
\begin{eqnarray}
    E_{U+V} & = & \frac{1}{2} \sum_I \sum_{\sigma m m'}
    U^I \left( \delta_{m m'} - n^{II \sigma}_{m m'} \right) n^{II \sigma}_{m' m} \nonumber \\
    & & - \frac{1}{2} \sum_{I} \sum_{J (J \ne I)}^{*} \sum_{\sigma m m'} V^{I J}
    n^{I J \sigma}_{m m'} n^{J I \sigma}_{m' m} \,,
    \label{eq:Edftu}
\end{eqnarray}
where $I$ and $J$ label atomic sites, $m$ and $m'$ denote magnetic quantum numbers, and $U^I$ and $V^{IJ}$ are the onsite and intersite effective Hubbard parameters, respectively. The asterisk indicates that, for each atom $I$, the summation over $J$ is restricted to neighbors within a chosen cutoff distance. The generalized occupation matrices $n^{IJ \sigma}_{m m'}$ are defined by projecting the Kohn-Sham (KS) states $\psi^\sigma_{v,\mathbf{k}}$ onto localized orbitals $\phi^I_m(\mathbf{r})$:
\begin{equation}
    n^{I J \sigma}_{m m'} = \sum_{v,\mathbf{k}} f^\sigma_{v,\mathbf{k}}
    \braket{\psi^\sigma_{v,\mathbf{k}}}{\phi^{J}_{m'}} \braket{\phi^{I}_{m}}{\psi^\sigma_{v,\mathbf{k}}} \,,
    \label{eq:occ_matrix_0}
\end{equation}
where $v$ and $\sigma$ denote band and spin indices, $\mathbf{k}$ runs over the first Brillouin zone (BZ), and $f^\sigma_{v,\mathbf{k}}$ are the KS occupations. In Eq.~\eqref{eq:occ_matrix_0}, $m \in I$ and $m' \in J$; that is, $m$ runs over the magnetic quantum numbers corresponding to the chosen principal quantum number $n$ and orbital angular momentum $l$ of atom $I$, and $m'$ analogously for atom $J$. The onsite and intersite terms in Eq.~\eqref{eq:Edftu} play opposing roles: $U^I$ promotes electron localization and suppresses hybridization, whereas $V^{IJ}$ enhances hybridization with neighboring sites. Accurately determining $U^I$ and $V^{IJ}$ from first principles is therefore essential to capture the correct balance between localization and hybridization. Finally, the numerical values of $U^I$ and $V^{IJ}$ depend on the choice of Hubbard projector functions $\phi^I_{m}(\mathbf{r})$ used to construct the occupation matrices in Eq.~\eqref{eq:occ_matrix_0}. In this work, we employ atomic orbitals orthogonalized via the L\"owdin method~\cite{Lowdin1950, Mayer2002}.

\subsection{Hubbard $U$ and $V$ parameters from linear-response theory}
Within the LR-cDFT framework, the Hubbard parameters are obtained from the requirement that the total energy in Eq.~\eqref{eq:Edft_plus_uv} remains piecewise linear with respect to variations in the occupation of the Hubbard manifold~\cite{Cococcioni2005, Campo2010}. They are computed as
\begin{equation}
U^I = \left(\chi_0^{-1} - \chi^{-1}\right)_{II} \,,
\label{eq:Ucalc}
\end{equation}
and
\begin{equation}
V^{IJ} = \left(\chi_0^{-1} - \chi^{-1}\right)_{IJ} \,,
\label{eq:Vcalc}
\end{equation}
where $\chi_0$ and $\chi$ denote the bare and self-consistent response matrices, respectively. These matrices quantify how the occupation matrices respond to localized perturbations. The self-consistent response is defined as
\begin{equation}
\chi_{IJ} = \sum_{m\sigma} \frac{dn^{I \sigma}_{mm}}{d\alpha^J} \,,
\end{equation}
where $n^{I\sigma}_{mm'} \equiv n^{II\sigma}_{mm'}$ is the onsite occupation matrix and $\alpha^J$ is the strength of the perturbation on site $J$. The bare response $\chi_0$ is defined analogously, but evaluated prior to the self-consistent adjustment of the Hartree and xc potentials.

DFPT reformulates LR-cDFT in terms of monochromatic reciprocal-space perturbations applied within the primitive unit cell, thus eliminating the need for supercells and greatly reducing computational cost~\cite{Timrov2018, Timrov2021}. In this formalism, the variation of the occupation matrix reads:
\begin{equation}
\frac{dn^{I \sigma}_{mm'}}{d\alpha^J} = \frac{1}{N_{\mathbf{q}}}\sum_{\mathbf{q}}^{N_{\mathbf{q}}} e^{i\mathbf{q}\cdot(\mathbf{R}_{l} - \mathbf{R}_{l'})}\Delta_{\mathbf{q}}^{s'} n^{s \sigma}_{mm'} \,,
\label{eq:dnq}
\end{equation}
where $\mathbf{q}$ is the wavevector of the perturbation, $N_\mathbf{q}$ is the total number of $\mathbf{q}$ points, and $\Delta_{\mathbf{q}}^{s'} n^{s \sigma}_{mm'}$ is the lattice-periodic response of the occupation matrix to the perturbation at wavevector $\mathbf{q}$. Indices $I\equiv(l,s)$ and $J\equiv(l',s')$ label atoms within cell $l$ or $l'$, with $s$ and $s'$ identifying the atoms in the unit cell, while $\mathbf{R}_l$ and $\mathbf{R}_{l'}$ are the corresponding lattice vectors. The quantities $\Delta_\mathbf{q}^{s'} n^{s \sigma}_{mm'}$ are obtained from the response KS wavefunctions, which are computed by solving the Sternheimer equations for each $\mathbf{q}$-specific perturbation. Further details on the DFPT formulation are given in Refs.~\citenum{Timrov2018, Timrov2021}. The efficiency and robustness of DFPT for evaluating Hubbard parameters have been demonstrated in numerous applications, including recent works~\cite{Mahajan2021, Mahajan2022, Gebreyesus2023, Binci2023, Gelin2024, Macke2024, Haddadi2024, Bonfa2024, Uhrin2025, Bastonero2025, Chang2025, Binci2025, dosSantos2025, Warda2025}.

\subsection{Computational details}
\label{comp_details}

\begin{table*}[t]
    \renewcommand{\arraystretch}{1.3}
    \centering
    \caption{Self-consistent Hubbard parameters (in eV) computed within DFPT using L\"owdin-orthogonalized atomic orbitals as Hubbard projectors, including rVV10: on-site $U_\mathrm{Co}$ for Co-$3d$ states, on-site $U_\mathrm{O}$ for O-$2p$ states, and inter-site $V$ between Co-$3d$ and O-$2p$ states, for LiCoO$_2$ and CoO$_2$. Structural optimizations are included in the self-consistent protocol (``relaxed" geometry) or excluded (``experimental" geometry).}
    \begin{tabular}{lccccccccccc} 
    \hline\hline
    &\multicolumn{3}{c}{LiCoO$_2$} & \phantom{a} & \multicolumn{7}{c}{CoO$_2$} \\ 
    \hline
    Crystal structure &\multicolumn{3}{c}{relaxed} & \phantom{a} & \multicolumn{3}{c}{relaxed} & & \multicolumn{3}{c}{experimental} \\
    \hline
    Functional & $U_\mathrm{Co}$ & $U_\mathrm{O}$ & $V$ & & $U_\mathrm{Co}$ & $U_\mathrm{O}$ & $V$ & & $U_\mathrm{Co}$ & $U_\mathrm{O}$ & $V$ \\
    \hline
    \pbesoluno   & 7.38 &      &      & & 8.21 &      &      & & 7.99  &       &      \\
    \pbesoluvno  & 7.51 &      & 0.65 & & 8.01 &      & 0.43 & & 7.25  &       & 0.15 \\
    \pbesoluono  & 7.08 & 9.36 &      & & 8.13 & 7.81 &      & & 7.16  & 7.71  &      \\
    \pbesoluovno & 7.15 & 9.42 & 0.63 & & 7.62 & 7.58 & 0.02 & & 7.14  & 7.73  & 0.10 \\
    \hline\hline
    \end{tabular}
    \label{tab:UOV}
\end{table*}

All first-principles calculations are performed using the \textsc{Quantum ESPRESSO} (QE) package~\cite{Giannozzi2009, Giannozzi2017, Giannozzi2020}. Pseudopotentials for Li, Co, and O are taken from the SSSP Efficiency library (v1.3.0)~\cite{Prandini2018}. The base xc functional is GGA with the PBEsol parametrization~\cite{Perdew2008}. A non-spin-polarized setup is used for LiCoO$_2$, while CoO$_2$ is treated in a spin-polarized framework (using the ferromagnetic ordering). Crystal structures are fully optimized using each xc functional considered in this work utilizing the Broyden-Fletcher-Goldfarb-Shanno (BFGS) algorithm~\cite{Fletcher1987}. Convergence thresholds are set to $10^{-6}$~Ry for the total energy, $10^{-5}$~Ry/Bohr for forces, and 0.01~kbar for the pressure. For structural optimization, kinetic-energy cutoffs for the KS wavefunctions (and charge density) are 90~Ry (1080~Ry) for LiCoO$_2$ and 110~Ry (1320~Ry) for CoO$_2$. The corresponding $\Gamma$-centered $\mathbf{k}$-point meshes are $20 \times 20 \times 4$ for LiCoO$_2$ and $20 \times 20 \times 10$ for CoO$_2$. All results presented in the main text employ the rVV10 functional to account for vdW interactions~\cite{Sabatini2013}. We chose rVV10 because it is formulated as a non-local functional that can be combined with different xc functionals, allowing its consistent application together with the spin-polarized PBEsol functional used throughout this work. Additional calculations using vdW-DF2-C09$_x$~\cite{Lee2010}, optB88-vdW~\cite{Klimes2009}, optB86b-vdW~\cite{Klimes2011}, and DFT-D3~\cite{Grimme2010} are provided in the SI~\cite{Supplementary_Information}. PDOS calculations are performed using Gaussian smearing with a broadening parameter of 0.005~Ry.

Hubbard parameters are computed using the \textsc{HP} code of QE~\cite{Timrov2022b} via DFPT in a basis of L\"owdin-orthogonalized atomic orbitals. The kinetic-energy cutoff for KS wavefunctions (and charge density) is set to 65~Ry (780~Ry). The $\Gamma$-centered \textbf{k}-point meshes are $10\times10\times2$ for LiCoO$_2$ and $12\times12\times6$ for CoO$_2$, while a $4\times4\times2$ \textbf{q}-point mesh is used for the DFPT calculations. Hubbard parameters are determined self-consistently (which includes structural relaxations) until convergence within 0.1~eV is achieved~\cite{Timrov2021}, and reported in Table~\ref{tab:UOV}. Additionally, Table~\ref{tab:UOV} reports the self-consistent Hubbard parameters for CoO$_2$ computed using the experimental geometry (i.e. without structural optimizations). Figure~S10 in the SI reports also the Hubbard parameters for LiCoO$_2$ computed using various other xc functionals.

In addition, we computed the Hubbard parameters for LiCoO$_2$ using the development version of the \textsc{HP} code (it will be made publicly available in the future releases of \textsc{Quantum ESPRESSO}), which enables DFPT calculations with Wannier functions as Hubbard projectors~\cite{Andolfatto2026}. For this purpose, we employed the interface with the \textsc{Wannier90} code~\cite{Pizzi2020} introduced in Ref.~\cite{Carta2025}. The kinetic-energy cutoff for KS wavefunctions (and charge density) is set to 65~Ry (780~Ry), The $\Gamma$-centered \textbf{k}- and \textbf{q}-point meshes were chosen as $10\times10\times2$ and $4\times4\times2$, respectively.  The wannierization was performed using the Kohn-Sham states inside the blue rectangle in Fig.~S5(a) of the SI, resulting in 15 frontier Wannier functions (5 functions per Co atom). To obtain Wannier functions aligned with the local CoO$_6$ octahedral environment, the initial local coordinate system for the projections was defined from the local Co–O bonding geometry, with the local $z$ axis chosen along a Co–O bond and the remaining axes obtained by constructing an orthonormal local reference frame.  We carried out a ``one-shot'' calculation of $U$ (i.e., without the cyclic self-consistency procedure used in Ref.~\cite{Timrov2021}) using the experimental crystal structure, yielding a converged value of $U=3.29$~eV with an accuracy of $\sim 0.2$~eV. The singular-value decomposition (SVD) was employed to invert the response matrices. 

For the HSE06 calculations~\cite{Heyd2003,Heyd2006}, we use kinetic-energy cutoffs of 100~Ry for the KS wavefunctions, 400~Ry for the charge density and potentials, and 100~Ry for the exact-exchange part. The $\Gamma$-centered \textbf{k}-point meshes are $10\times10\times2$ for LiCoO$_2$ and $12\times12\times6$ for CoO$_2$, and a $4\times4\times2$ \textbf{q}-point mesh is employed for evaluating the exact exchange. The $\mathbf{q}\!\rightarrow\!0$ limit in HSE06 is treated using the Gygi–Baldereschi scheme~\cite{Gygi1986}. We use PBE~\cite{Perdew1996} norm-conserving pseudopotentials from the \textsc{PseudoDojo} v.0.5 library~\cite{vanSetten2018}.

The magnetic energy landscape scan for CoO$_2$ is performed following the approach introduced in Ref.~\cite{Ponet2024}, employing random initializations of the local ($I=J$) occupation matrices defined in Eq.~\eqref{eq:occ_matrix_0}. We automated this workflow using the AiiDA framework~\cite{Pizzi2016-aiida, Wang2026}. 

Bulk Li is modeled at the PBEsol level in its \textit{bcc} structure, using a conventional unit cell with one Li atom at the origin. The optimized lattice parameter is 3.493~\AA. The BZ is sampled using a uniform $\Gamma$-centered $10\times10\times10$ \textbf{k}-point mesh, together with Marzari–Vanderbilt smearing~\cite{Marzari1999} (broadening of 0.02~Ry). KS wavefunctions and potentials are expanded in plane waves up to kinetic-energy cutoffs of 65~Ry and 780~Ry, respectively.

\section*{ACKNOWLEDGEMENTS}
We thank Matteo Cococcioni and Stefano de Gironcoli for fruitful discussions. V.S. acknowledges support from the MARVEL INSPIRE Potentials Master’s Fellowship. C.M. acknowledges support by the European Commission through the MaX Centre of Excellence for supercomputing applications (Grant No.~101093374). This research was supported by the NCCR MARVEL, a National Centre of Competence in Research, funded by the Swiss National Science Foundation (grant number 205602), and by the Swiss National Science Foundation Grant No.~200021-227641 and No.~200021-236507. Computer time was provided by the Swiss National Supercomputing Centre (CSCS) under project No.~lp18, s1326, s1335, and mr33.





\begin{thebibliography}{149}%
\makeatletter
\providecommand \@ifxundefined [1]{%
 \@ifx{#1\undefined}
}%
\providecommand \@ifnum [1]{%
 \ifnum #1\expandafter \@firstoftwo
 \else \expandafter \@secondoftwo
 \fi
}%
\providecommand \@ifx [1]{%
 \ifx #1\expandafter \@firstoftwo
 \else \expandafter \@secondoftwo
 \fi
}%
\providecommand \natexlab [1]{#1}%
\providecommand \enquote  [1]{``#1''}%
\providecommand \bibnamefont  [1]{#1}%
\providecommand \bibfnamefont [1]{#1}%
\providecommand \citenamefont [1]{#1}%
\providecommand \href@noop [0]{\@secondoftwo}%
\providecommand \href [0]{\begingroup \@sanitize@url \@href}%
\providecommand \@href[1]{\@@startlink{#1}\@@href}%
\providecommand \@@href[1]{\endgroup#1\@@endlink}%
\providecommand \@sanitize@url [0]{\catcode `\\12\catcode `\$12\catcode
  `\&12\catcode `\#12\catcode `\^12\catcode `\_12\catcode `\%12\relax}%
\providecommand \@@startlink[1]{}%
\providecommand \@@endlink[0]{}%
\providecommand \url  [0]{\begingroup\@sanitize@url \@url }%
\providecommand \@url [1]{\endgroup\@href {#1}{\urlprefix }}%
\providecommand \urlprefix  [0]{URL }%
\providecommand \Eprint [0]{\href }%
\providecommand \doibase [0]{https://doi.org/}%
\providecommand \selectlanguage [0]{\@gobble}%
\providecommand \bibinfo  [0]{\@secondoftwo}%
\providecommand \bibfield  [0]{\@secondoftwo}%
\providecommand \translation [1]{[#1]}%
\providecommand \BibitemOpen [0]{}%
\providecommand \bibitemStop [0]{}%
\providecommand \bibitemNoStop [0]{.\EOS\space}%
\providecommand \EOS [0]{\spacefactor3000\relax}%
\providecommand \BibitemShut  [1]{\csname bibitem#1\endcsname}%
\let\auto@bib@innerbib\@empty
\bibitem [{\citenamefont {Scrosati}(2005)}]{Scrosati2005}%
  \BibitemOpen
  \bibfield  {author} {\bibinfo {author} {\bibfnamefont {B.}~\bibnamefont
  {Scrosati}},\ }\bibfield  {title} {\bibinfo {title} {Power sources for
  portable electronics and hybrid cars: lithium batteries and fuel cells},\
  }\href {https://doi.org/10.1002/tcr.20054} {\bibfield  {journal} {\bibinfo
  {journal} {Chem. Rec.}\ }\textbf {\bibinfo {volume} {5}},\ \bibinfo {pages}
  {286} (\bibinfo {year} {2005})}\BibitemShut {NoStop}%
\bibitem [{\citenamefont {Chen}\ \emph {et~al.}(2012)\citenamefont {Chen},
  \citenamefont {Shen}, \citenamefont {Vo}, \citenamefont {Cao},\ and\
  \citenamefont {Kapoor}}]{Chen2012}%
  \BibitemOpen
  \bibfield  {author} {\bibinfo {author} {\bibfnamefont {X.}~\bibnamefont
  {Chen}}, \bibinfo {author} {\bibfnamefont {W.}~\bibnamefont {Shen}}, \bibinfo
  {author} {\bibfnamefont {T.~T.}\ \bibnamefont {Vo}}, \bibinfo {author}
  {\bibfnamefont {Z.}~\bibnamefont {Cao}},\ and\ \bibinfo {author}
  {\bibfnamefont {A.}~\bibnamefont {Kapoor}},\ }\bibfield  {title} {\bibinfo
  {title} {An overview of lithium-ion batteries for electric vehicles},\ }in\
  \href {https://doi.org/10.1109/ASSCC.2012.6523269} {\emph {\bibinfo
  {booktitle} {Proc. Int. Power Energy Conf. (IPEC)}}}\ (\bibinfo {year}
  {2012})\ p.\ \bibinfo {pages} {230}\BibitemShut {NoStop}%
\bibitem [{\citenamefont {Thackeray}\ \emph {et~al.}(2012)\citenamefont
  {Thackeray}, \citenamefont {Wolverton},\ and\ \citenamefont
  {Isaacs}}]{Thackeray2012}%
  \BibitemOpen
  \bibfield  {author} {\bibinfo {author} {\bibfnamefont {M.~M.}\ \bibnamefont
  {Thackeray}}, \bibinfo {author} {\bibfnamefont {C.}~\bibnamefont
  {Wolverton}},\ and\ \bibinfo {author} {\bibfnamefont {E.~D.}\ \bibnamefont
  {Isaacs}},\ }\bibfield  {title} {\bibinfo {title} {Electrical energy storage
  for transportation—approaching the limits of{,} and going beyond{,}
  lithium-ion batteries},\ }\href {https://doi.org/10.1039/C2EE21892E}
  {\bibfield  {journal} {\bibinfo  {journal} {Energy Environ. Sci.}\ }\textbf
  {\bibinfo {volume} {5}},\ \bibinfo {pages} {7854} (\bibinfo {year}
  {2012})}\BibitemShut {NoStop}%
\bibitem [{\citenamefont {Mizushima}\ \emph {et~al.}(1980)\citenamefont
  {Mizushima}, \citenamefont {Jones}, \citenamefont {Wiseman},\ and\
  \citenamefont {Goodenough}}]{Goodenough1980}%
  \BibitemOpen
  \bibfield  {author} {\bibinfo {author} {\bibfnamefont {K.}~\bibnamefont
  {Mizushima}}, \bibinfo {author} {\bibfnamefont {P.~C.}\ \bibnamefont
  {Jones}}, \bibinfo {author} {\bibfnamefont {P.~J.}\ \bibnamefont {Wiseman}},\
  and\ \bibinfo {author} {\bibfnamefont {J.~B.}\ \bibnamefont {Goodenough}},\
  }\bibfield  {title} {\bibinfo {title} {{Li$_x$CoO$_2$ ($0<x<1$)}: A new
  cathode material for batteries of high energy density},\ }\href
  {https://doi.org/10.1016/0025-5408(80)90012-4} {\bibfield  {journal}
  {\bibinfo  {journal} {Mater. Res. Bull.}\ }\textbf {\bibinfo {volume} {15}},\
  \bibinfo {pages} {783} (\bibinfo {year} {1980})}\BibitemShut {NoStop}%
\bibitem [{\citenamefont {Nayak}\ \emph {et~al.}(2018)\citenamefont {Nayak},
  \citenamefont {Yang}, \citenamefont {Brehm},\ and\ \citenamefont
  {Adelhelm}}]{Nayak2018}%
  \BibitemOpen
  \bibfield  {author} {\bibinfo {author} {\bibfnamefont {P.~K.}\ \bibnamefont
  {Nayak}}, \bibinfo {author} {\bibfnamefont {L.}~\bibnamefont {Yang}},
  \bibinfo {author} {\bibfnamefont {W.}~\bibnamefont {Brehm}},\ and\ \bibinfo
  {author} {\bibfnamefont {P.}~\bibnamefont {Adelhelm}},\ }\bibfield  {title}
  {\bibinfo {title} {From lithium-ion to sodium-ion batteries: advantages,
  challenges, and surprises},\ }\href {https://doi.org/10.1002/anie.201703114}
  {\bibfield  {journal} {\bibinfo  {journal} {Angew. Chem. Int. Ed.}\ }\textbf
  {\bibinfo {volume} {57}},\ \bibinfo {pages} {102} (\bibinfo {year}
  {2018})}\BibitemShut {NoStop}%
\bibitem [{\citenamefont {Yabuuchi}\ and\ \citenamefont
  {Ohzuku}(2003)}]{Yabuuchi2003}%
  \BibitemOpen
  \bibfield  {author} {\bibinfo {author} {\bibfnamefont {N.}~\bibnamefont
  {Yabuuchi}}\ and\ \bibinfo {author} {\bibfnamefont {T.}~\bibnamefont
  {Ohzuku}},\ }\bibfield  {title} {\bibinfo {title} {Novel lithium insertion
  material of {LiCo$_{1/3}$Ni$_{1/3}$Mn$_{1/3}$O$_2$} for advanced lithium-ion
  batteries},\ }\href {https://doi.org/10.1016/S0378-7753(03)00173-3}
  {\bibfield  {journal} {\bibinfo  {journal} {J. Power Sources}\ }\textbf
  {\bibinfo {volume} {119}},\ \bibinfo {pages} {171} (\bibinfo {year}
  {2003})}\BibitemShut {NoStop}%
\bibitem [{\citenamefont {Noh}\ \emph {et~al.}(2013)\citenamefont {Noh},
  \citenamefont {Youn}, \citenamefont {Yoon},\ and\ \citenamefont
  {Sun}}]{Noh2013}%
  \BibitemOpen
  \bibfield  {author} {\bibinfo {author} {\bibfnamefont {H.-J.}\ \bibnamefont
  {Noh}}, \bibinfo {author} {\bibfnamefont {S.}~\bibnamefont {Youn}}, \bibinfo
  {author} {\bibfnamefont {C.~S.}\ \bibnamefont {Yoon}},\ and\ \bibinfo
  {author} {\bibfnamefont {Y.-K.}\ \bibnamefont {Sun}},\ }\bibfield  {title}
  {\bibinfo {title} {Comparison of the structural and electrochemical
  properties of layered {Li[Ni$_x$Co$_y$Mn$_z$]O2} (x = 1/3, 0.5, 0.6, 0.7, 0.8
  and 0.85) cathode material for lithium-ion batteries},\ }\href
  {https://doi.org/10.1016/j.jpowsour.2013.01.063} {\bibfield  {journal}
  {\bibinfo  {journal} {J. Power Sources}\ }\textbf {\bibinfo {volume} {233}},\
  \bibinfo {pages} {121} (\bibinfo {year} {2013})}\BibitemShut {NoStop}%
\bibitem [{\citenamefont {Xuan}\ \emph {et~al.}(2019)\citenamefont {Xuan},
  \citenamefont {Otsuki},\ and\ \citenamefont {Chagnes}}]{Xuan2019}%
  \BibitemOpen
  \bibfield  {author} {\bibinfo {author} {\bibfnamefont {W.}~\bibnamefont
  {Xuan}}, \bibinfo {author} {\bibfnamefont {A.}~\bibnamefont {Otsuki}},\ and\
  \bibinfo {author} {\bibfnamefont {A.}~\bibnamefont {Chagnes}},\ }\bibfield
  {title} {\bibinfo {title} {Investigation of the leaching mechanism of {NMC}
  811 ({LiNi$_{0.8}$Mn$_{0.1}$Co$_{0.1}$O$_2$}) by hydrochloric acid for
  recycling lithium ion battery cathodes},\ }\href
  {https://doi.org/10.1039/C9RA06686A} {\bibfield  {journal} {\bibinfo
  {journal} {RSC Adv.}\ }\textbf {\bibinfo {volume} {9}},\ \bibinfo {pages}
  {38612} (\bibinfo {year} {2019})}\BibitemShut {NoStop}%
\bibitem [{\citenamefont {Chen}\ \emph {et~al.}(2004)\citenamefont {Chen},
  \citenamefont {Liu}, \citenamefont {Stoll}, \citenamefont {Henriksen},
  \citenamefont {Vissers},\ and\ \citenamefont {Amine}}]{Chen2004}%
  \BibitemOpen
  \bibfield  {author} {\bibinfo {author} {\bibfnamefont {C.~H.}\ \bibnamefont
  {Chen}}, \bibinfo {author} {\bibfnamefont {J.}~\bibnamefont {Liu}}, \bibinfo
  {author} {\bibfnamefont {M.~E.}\ \bibnamefont {Stoll}}, \bibinfo {author}
  {\bibfnamefont {G.}~\bibnamefont {Henriksen}}, \bibinfo {author}
  {\bibfnamefont {D.~R.}\ \bibnamefont {Vissers}},\ and\ \bibinfo {author}
  {\bibfnamefont {K.}~\bibnamefont {Amine}},\ }\bibfield  {title} {\bibinfo
  {title} {Aluminum-doped lithium nickel cobalt oxide electrodes for high-power
  lithium-ion batteries},\ }\href
  {https://doi.org/10.1016/j.jpowsour.2003.10.009} {\bibfield  {journal}
  {\bibinfo  {journal} {J. Power Sources}\ }\textbf {\bibinfo {volume} {128}},\
  \bibinfo {pages} {278} (\bibinfo {year} {2004})}\BibitemShut {NoStop}%
\bibitem [{\citenamefont {Hohenberg}\ and\ \citenamefont
  {Kohn}(1964)}]{Hohenberg1964}%
  \BibitemOpen
  \bibfield  {author} {\bibinfo {author} {\bibfnamefont {P.}~\bibnamefont
  {Hohenberg}}\ and\ \bibinfo {author} {\bibfnamefont {W.}~\bibnamefont
  {Kohn}},\ }\bibfield  {title} {\bibinfo {title} {Inhomogeneous electron
  gas},\ }\href {https://doi.org/10.1103/PhysRev.136.B864} {\bibfield
  {journal} {\bibinfo  {journal} {Phys. Rev.}\ }\textbf {\bibinfo {volume}
  {136}},\ \bibinfo {pages} {B864} (\bibinfo {year} {1964})}\BibitemShut
  {NoStop}%
\bibitem [{\citenamefont {Kohn}\ and\ \citenamefont {Sham}(1965)}]{Kohn1965}%
  \BibitemOpen
  \bibfield  {author} {\bibinfo {author} {\bibfnamefont {W.}~\bibnamefont
  {Kohn}}\ and\ \bibinfo {author} {\bibfnamefont {L.~J.}\ \bibnamefont
  {Sham}},\ }\bibfield  {title} {\bibinfo {title} {Self-consistent equations
  including exchange and correlation effects},\ }\href
  {https://doi.org/10.1103/PhysRev.140.A1133} {\bibfield  {journal} {\bibinfo
  {journal} {Phys. Rev.}\ }\textbf {\bibinfo {volume} {140}},\ \bibinfo {pages}
  {A1133} (\bibinfo {year} {1965})}\BibitemShut {NoStop}%
\bibitem [{\citenamefont {Perdew}\ and\ \citenamefont
  {Wang}(1992)}]{Perdew1992}%
  \BibitemOpen
  \bibfield  {author} {\bibinfo {author} {\bibfnamefont {J.~P.}\ \bibnamefont
  {Perdew}}\ and\ \bibinfo {author} {\bibfnamefont {Y.}~\bibnamefont {Wang}},\
  }\bibfield  {title} {\bibinfo {title} {Accurate and simple analytic
  representation of the electron-gas correlation energy},\ }\href
  {https://doi.org/10.1103/PhysRevB.45.13244} {\bibfield  {journal} {\bibinfo
  {journal} {Phys. Rev. B}\ }\textbf {\bibinfo {volume} {45}},\ \bibinfo
  {pages} {13244} (\bibinfo {year} {1992})}\BibitemShut {NoStop}%
\bibitem [{\citenamefont {Perdew}\ and\ \citenamefont
  {Zunger}(1981)}]{Perdew1981}%
  \BibitemOpen
  \bibfield  {author} {\bibinfo {author} {\bibfnamefont {J.~P.}\ \bibnamefont
  {Perdew}}\ and\ \bibinfo {author} {\bibfnamefont {A.}~\bibnamefont
  {Zunger}},\ }\bibfield  {title} {\bibinfo {title} {Self-interaction
  correction to density-functional approximations for many-electron systems},\
  }\href {https://doi.org/10.1103/PhysRevB.23.5048} {\bibfield  {journal}
  {\bibinfo  {journal} {Phys. Rev. B}\ }\textbf {\bibinfo {volume} {23}},\
  \bibinfo {pages} {5048} (\bibinfo {year} {1981})}\BibitemShut {NoStop}%
\bibitem [{\citenamefont {Mori-Sánchez}\ \emph {et~al.}(2006)\citenamefont
  {Mori-Sánchez}, \citenamefont {Cohen},\ and\ \citenamefont
  {Yang}}]{MoriSanchez2006}%
  \BibitemOpen
  \bibfield  {author} {\bibinfo {author} {\bibfnamefont {P.}~\bibnamefont
  {Mori-Sánchez}}, \bibinfo {author} {\bibfnamefont {A.~J.}\ \bibnamefont
  {Cohen}},\ and\ \bibinfo {author} {\bibfnamefont {W.}~\bibnamefont {Yang}},\
  }\bibfield  {title} {\bibinfo {title} {Many-electron self-interaction error
  in approximate density functionals},\ }\href
  {https://doi.org/10.1063/1.2403848} {\bibfield  {journal} {\bibinfo
  {journal} {J. Chem. Phys.}\ }\textbf {\bibinfo {volume} {125}},\ \bibinfo
  {pages} {201102} (\bibinfo {year} {2006})}\BibitemShut {NoStop}%
\bibitem [{\citenamefont {Anisimov}\ \emph {et~al.}(1991)\citenamefont
  {Anisimov}, \citenamefont {Zaanen},\ and\ \citenamefont
  {Andersen}}]{Anisimov1991}%
  \BibitemOpen
  \bibfield  {author} {\bibinfo {author} {\bibfnamefont {V.~I.}\ \bibnamefont
  {Anisimov}}, \bibinfo {author} {\bibfnamefont {J.}~\bibnamefont {Zaanen}},\
  and\ \bibinfo {author} {\bibfnamefont {O.~K.}\ \bibnamefont {Andersen}},\
  }\bibfield  {title} {\bibinfo {title} {Band theory and mott insulators:
  Hubbard {$U$} instead of stoner {I}},\ }\href
  {https://doi.org/10.1103/PhysRevB.44.943} {\bibfield  {journal} {\bibinfo
  {journal} {Phys. Rev. B}\ }\textbf {\bibinfo {volume} {44}},\ \bibinfo
  {pages} {943} (\bibinfo {year} {1991})}\BibitemShut {NoStop}%
\bibitem [{\citenamefont {Liechtenstein}\ \emph {et~al.}(1995)\citenamefont
  {Liechtenstein}, \citenamefont {Anisimov},\ and\ \citenamefont
  {Zaanen}}]{Liechtenstein1995}%
  \BibitemOpen
  \bibfield  {author} {\bibinfo {author} {\bibfnamefont {A.~I.}\ \bibnamefont
  {Liechtenstein}}, \bibinfo {author} {\bibfnamefont {V.~I.}\ \bibnamefont
  {Anisimov}},\ and\ \bibinfo {author} {\bibfnamefont {J.}~\bibnamefont
  {Zaanen}},\ }\bibfield  {title} {\bibinfo {title} {Density-functional theory
  and strong interactions: Orbital ordering in mott-hubbard insulators},\
  }\href {https://doi.org/10.1103/PhysRevB.52.R5467} {\bibfield  {journal}
  {\bibinfo  {journal} {Phys. Rev. B}\ }\textbf {\bibinfo {volume} {52}},\
  \bibinfo {pages} {R5467} (\bibinfo {year} {1995})}\BibitemShut {NoStop}%
\bibitem [{\citenamefont {Dudarev}\ \emph {et~al.}(1998)\citenamefont
  {Dudarev}, \citenamefont {Botton}, \citenamefont {Savrasov}, \citenamefont
  {Humphreys},\ and\ \citenamefont {Sutton}}]{Dudarev1998}%
  \BibitemOpen
  \bibfield  {author} {\bibinfo {author} {\bibfnamefont {S.~L.}\ \bibnamefont
  {Dudarev}}, \bibinfo {author} {\bibfnamefont {G.~A.}\ \bibnamefont {Botton}},
  \bibinfo {author} {\bibfnamefont {S.~Y.}\ \bibnamefont {Savrasov}}, \bibinfo
  {author} {\bibfnamefont {C.~J.}\ \bibnamefont {Humphreys}},\ and\ \bibinfo
  {author} {\bibfnamefont {A.~P.}\ \bibnamefont {Sutton}},\ }\bibfield  {title}
  {\bibinfo {title} {Electron-energy-loss spectra and the structural stability
  of nickel oxide: An {LSDA+$U$} study},\ }\href
  {https://doi.org/10.1103/PhysRevB.57.1505} {\bibfield  {journal} {\bibinfo
  {journal} {Phys. Rev. B}\ }\textbf {\bibinfo {volume} {57}},\ \bibinfo
  {pages} {1505} (\bibinfo {year} {1998})}\BibitemShut {NoStop}%
\bibitem [{\citenamefont {Campo}\ and\ \citenamefont
  {Cococcioni}(2010)}]{Campo2010}%
  \BibitemOpen
  \bibfield  {author} {\bibinfo {author} {\bibfnamefont {V.}~\bibnamefont
  {Campo}}\ and\ \bibinfo {author} {\bibfnamefont {M.}~\bibnamefont
  {Cococcioni}},\ }\bibfield  {title} {\bibinfo {title} {Extended {DFT+$U$+$V$}
  method with on-site and inter-site electronic interactions},\ }\href
  {https://doi.org/10.1088/0953-8984/22/5/055602} {\bibfield  {journal}
  {\bibinfo  {journal} {J. Phys. Condens. Matter}\ }\textbf {\bibinfo {volume}
  {22}},\ \bibinfo {pages} {055602} (\bibinfo {year} {2010})}\BibitemShut
  {NoStop}%
\bibitem [{\citenamefont {Lee}\ and\ \citenamefont {Son}(2020)}]{Lee2020}%
  \BibitemOpen
  \bibfield  {author} {\bibinfo {author} {\bibfnamefont {S.-H.}\ \bibnamefont
  {Lee}}\ and\ \bibinfo {author} {\bibfnamefont {Y.-W.}\ \bibnamefont {Son}},\
  }\bibfield  {title} {\bibinfo {title} {First-principles approach with a
  pseudohybrid density functional for extended hubbard interactions},\ }\href
  {https://doi.org/10.1103/PhysRevResearch.2.043410} {\bibfield  {journal}
  {\bibinfo  {journal} {Phys. Rev. Res.}\ }\textbf {\bibinfo {volume} {2}},\
  \bibinfo {pages} {043410} (\bibinfo {year} {2020})}\BibitemShut {NoStop}%
\bibitem [{\citenamefont {Tancogne-Dejean}\ and\ \citenamefont
  {Rubio}(2020)}]{TancogneDejean2020}%
  \BibitemOpen
  \bibfield  {author} {\bibinfo {author} {\bibfnamefont {N.}~\bibnamefont
  {Tancogne-Dejean}}\ and\ \bibinfo {author} {\bibfnamefont {A.}~\bibnamefont
  {Rubio}},\ }\bibfield  {title} {\bibinfo {title} {Parameter-free hybridlike
  functional based on an extended hubbard model: {DFT+$U+V$}},\ }\href
  {https://doi.org/10.1103/PhysRevB.102.155117} {\bibfield  {journal} {\bibinfo
   {journal} {Phys. Rev. B}\ }\textbf {\bibinfo {volume} {102}},\ \bibinfo
  {pages} {155117} (\bibinfo {year} {2020})}\BibitemShut {NoStop}%
\bibitem [{\citenamefont {Sun}\ \emph {et~al.}(2015)\citenamefont {Sun},
  \citenamefont {Ruzsinszky},\ and\ \citenamefont {Perdew}}]{Sun2015}%
  \BibitemOpen
  \bibfield  {author} {\bibinfo {author} {\bibfnamefont {J.}~\bibnamefont
  {Sun}}, \bibinfo {author} {\bibfnamefont {A.}~\bibnamefont {Ruzsinszky}},\
  and\ \bibinfo {author} {\bibfnamefont {J.~P.}\ \bibnamefont {Perdew}},\
  }\bibfield  {title} {\bibinfo {title} {Strongly constrained and appropriately
  normed semilocal density functional},\ }\href
  {https://doi.org/10.1103/PhysRevLett.115.036402} {\bibfield  {journal}
  {\bibinfo  {journal} {Phys. Rev. Lett.}\ }\textbf {\bibinfo {volume} {115}},\
  \bibinfo {pages} {036402} (\bibinfo {year} {2015})}\BibitemShut {NoStop}%
\bibitem [{\citenamefont {Bartók}\ and\ \citenamefont
  {Yates}(2019)}]{Bartok2019}%
  \BibitemOpen
  \bibfield  {author} {\bibinfo {author} {\bibfnamefont {A.~P.}\ \bibnamefont
  {Bartók}}\ and\ \bibinfo {author} {\bibfnamefont {J.~R.}\ \bibnamefont
  {Yates}},\ }\bibfield  {title} {\bibinfo {title} {Regularized {SCAN}
  functional},\ }\href {https://doi.org/10.1063/1.5098388} {\bibfield
  {journal} {\bibinfo  {journal} {J. Chem. Phys.}\ }\textbf {\bibinfo {volume}
  {150}},\ \bibinfo {pages} {161101} (\bibinfo {year} {2019})}\BibitemShut
  {NoStop}%
\bibitem [{\citenamefont {Furness}\ \emph {et~al.}(2020)\citenamefont
  {Furness}, \citenamefont {Kaplan}, \citenamefont {Ning}, \citenamefont
  {Perdew},\ and\ \citenamefont {Sun}}]{Furness2020}%
  \BibitemOpen
  \bibfield  {author} {\bibinfo {author} {\bibfnamefont {J.~W.}\ \bibnamefont
  {Furness}}, \bibinfo {author} {\bibfnamefont {A.~D.}\ \bibnamefont {Kaplan}},
  \bibinfo {author} {\bibfnamefont {J.}~\bibnamefont {Ning}}, \bibinfo {author}
  {\bibfnamefont {J.~P.}\ \bibnamefont {Perdew}},\ and\ \bibinfo {author}
  {\bibfnamefont {J.}~\bibnamefont {Sun}},\ }\bibfield  {title} {\bibinfo
  {title} {Accurate and numerically efficient {r$^2$SCAN} meta-generalized
  gradient approximation},\ }\href
  {https://doi.org/10.1021/acs.jpclett.0c02405} {\bibfield  {journal} {\bibinfo
   {journal} {J. Phys. Chem. Lett.}\ }\textbf {\bibinfo {volume} {11}},\
  \bibinfo {pages} {8208} (\bibinfo {year} {2020})}\BibitemShut {NoStop}%
\bibitem [{\citenamefont {Adamo}\ and\ \citenamefont
  {Barone}(1999)}]{Adamo1999}%
  \BibitemOpen
  \bibfield  {author} {\bibinfo {author} {\bibfnamefont {C.}~\bibnamefont
  {Adamo}}\ and\ \bibinfo {author} {\bibfnamefont {V.}~\bibnamefont {Barone}},\
  }\bibfield  {title} {\bibinfo {title} {Toward reliable density functional
  methods without adjustable parameters: The {PBE0} model},\ }\href
  {https://doi.org/10.1063/1.478522} {\bibfield  {journal} {\bibinfo  {journal}
  {J. Chem. Phys.}\ }\textbf {\bibinfo {volume} {110}},\ \bibinfo {pages}
  {6158} (\bibinfo {year} {1999})}\BibitemShut {NoStop}%
\bibitem [{\citenamefont {Heyd}\ \emph {et~al.}(2003)\citenamefont {Heyd},
  \citenamefont {Scuseria},\ and\ \citenamefont {Ernzerhof}}]{Heyd2003}%
  \BibitemOpen
  \bibfield  {author} {\bibinfo {author} {\bibfnamefont {J.}~\bibnamefont
  {Heyd}}, \bibinfo {author} {\bibfnamefont {G.~E.}\ \bibnamefont {Scuseria}},\
  and\ \bibinfo {author} {\bibfnamefont {M.}~\bibnamefont {Ernzerhof}},\
  }\bibfield  {title} {\bibinfo {title} {Hybrid functionals based on a screened
  coulomb potential},\ }\href {https://doi.org/10.1063/1.1564060} {\bibfield
  {journal} {\bibinfo  {journal} {J. Chem. Phys.}\ }\textbf {\bibinfo {volume}
  {118}},\ \bibinfo {pages} {8207} (\bibinfo {year} {2003})}\BibitemShut
  {NoStop}%
\bibitem [{\citenamefont {Heyd}\ \emph {et~al.}(2006)\citenamefont {Heyd},
  \citenamefont {Scuseria},\ and\ \citenamefont {Ernzerhof}}]{Heyd2006}%
  \BibitemOpen
  \bibfield  {author} {\bibinfo {author} {\bibfnamefont {J.}~\bibnamefont
  {Heyd}}, \bibinfo {author} {\bibfnamefont {G.~E.}\ \bibnamefont {Scuseria}},\
  and\ \bibinfo {author} {\bibfnamefont {M.}~\bibnamefont {Ernzerhof}},\
  }\bibfield  {title} {\bibinfo {title} {Erratum: Hybrid functionals based on a
  screened coulomb potential},\ }\href {https://doi.org/10.1063/1.2204597}
  {\bibfield  {journal} {\bibinfo  {journal} {J. Chem. Phys.}\ }\textbf
  {\bibinfo {volume} {124}},\ \bibinfo {pages} {219906} (\bibinfo {year}
  {2006})}\BibitemShut {NoStop}%
\bibitem [{\citenamefont {Timrov}\ \emph
  {et~al.}(2022{\natexlab{a}})\citenamefont {Timrov}, \citenamefont
  {Aquilante}, \citenamefont {Cococcioni},\ and\ \citenamefont
  {Marzari}}]{Timrov2022}%
  \BibitemOpen
  \bibfield  {author} {\bibinfo {author} {\bibfnamefont {I.}~\bibnamefont
  {Timrov}}, \bibinfo {author} {\bibfnamefont {F.}~\bibnamefont {Aquilante}},
  \bibinfo {author} {\bibfnamefont {M.}~\bibnamefont {Cococcioni}},\ and\
  \bibinfo {author} {\bibfnamefont {N.}~\bibnamefont {Marzari}},\ }\bibfield
  {title} {\bibinfo {title} {Accurate electronic properties and intercalation
  voltages of olivine-type {Li}-ion cathode materials from extended hubbard
  functionals},\ }\href {https://doi.org/10.1103/PRXEnergy.1.033003} {\bibfield
   {journal} {\bibinfo  {journal} {PRX Energy}\ }\textbf {\bibinfo {volume}
  {1}},\ \bibinfo {pages} {033003} (\bibinfo {year}
  {2022}{\natexlab{a}})}\BibitemShut {NoStop}%
\bibitem [{\citenamefont {Long}\ \emph {et~al.}(2020)\citenamefont {Long},
  \citenamefont {Gautam},\ and\ \citenamefont {Carter}}]{Long2020}%
  \BibitemOpen
  \bibfield  {author} {\bibinfo {author} {\bibfnamefont {O.~Y.}\ \bibnamefont
  {Long}}, \bibinfo {author} {\bibfnamefont {G.~S.}\ \bibnamefont {Gautam}},\
  and\ \bibinfo {author} {\bibfnamefont {E.~A.}\ \bibnamefont {Carter}},\
  }\bibfield  {title} {\bibinfo {title} {Evaluating optimal {$U$} for {$3d$}
  transition-metal oxides within the {SCAN+$U$} framework},\ }\href
  {https://doi.org/10.1103/PhysRevMaterials.4.045401} {\bibfield  {journal}
  {\bibinfo  {journal} {Phys. Rev. Mater.}\ }\textbf {\bibinfo {volume} {4}},\
  \bibinfo {pages} {045401} (\bibinfo {year} {2020})}\BibitemShut {NoStop}%
\bibitem [{\citenamefont {Swathilakshmi}\ \emph {et~al.}(2023)\citenamefont
  {Swathilakshmi}, \citenamefont {Devi},\ and\ \citenamefont
  {Gautam}}]{Swathilakshmi2023}%
  \BibitemOpen
  \bibfield  {author} {\bibinfo {author} {\bibfnamefont {S.}~\bibnamefont
  {Swathilakshmi}}, \bibinfo {author} {\bibfnamefont {R.}~\bibnamefont
  {Devi}},\ and\ \bibinfo {author} {\bibfnamefont {G.~S.}\ \bibnamefont
  {Gautam}},\ }\bibfield  {title} {\bibinfo {title} {Performance of the
  {r$^2$SCAN} functional in transition metal oxides},\ }\href
  {https://doi.org/10.1021/acs.jctc.3c00030} {\bibfield  {journal} {\bibinfo
  {journal} {J. Chem. Theory Comput.}\ }\textbf {\bibinfo {volume} {19}},\
  \bibinfo {pages} {4202} (\bibinfo {year} {2023})}\BibitemShut {NoStop}%
\bibitem [{\citenamefont {Georges}\ \emph {et~al.}(1996)\citenamefont
  {Georges}, \citenamefont {Kotliar}, \citenamefont {Krauth},\ and\
  \citenamefont {Rozenberg}}]{Georges1996}%
  \BibitemOpen
  \bibfield  {author} {\bibinfo {author} {\bibfnamefont {A.}~\bibnamefont
  {Georges}}, \bibinfo {author} {\bibfnamefont {G.}~\bibnamefont {Kotliar}},
  \bibinfo {author} {\bibfnamefont {W.}~\bibnamefont {Krauth}},\ and\ \bibinfo
  {author} {\bibfnamefont {M.~J.}\ \bibnamefont {Rozenberg}},\ }\bibfield
  {title} {\bibinfo {title} {Dynamical mean-field theory of strongly correlated
  fermion systems and the limit of infinite dimensions},\ }\href
  {https://doi.org/10.1103/RevModPhys.68.13} {\bibfield  {journal} {\bibinfo
  {journal} {Rev. Mod. Phys.}\ }\textbf {\bibinfo {volume} {68}},\ \bibinfo
  {pages} {13} (\bibinfo {year} {1996})}\BibitemShut {NoStop}%
\bibitem [{\citenamefont {Himmetoglu}\ \emph {et~al.}(2013)\citenamefont
  {Himmetoglu}, \citenamefont {Floris}, \citenamefont {de~Gironcoli},\ and\
  \citenamefont {Cococcioni}}]{Himmetoglu2013}%
  \BibitemOpen
  \bibfield  {author} {\bibinfo {author} {\bibfnamefont {B.}~\bibnamefont
  {Himmetoglu}}, \bibinfo {author} {\bibfnamefont {A.}~\bibnamefont {Floris}},
  \bibinfo {author} {\bibfnamefont {S.}~\bibnamefont {de~Gironcoli}},\ and\
  \bibinfo {author} {\bibfnamefont {M.}~\bibnamefont {Cococcioni}},\ }\bibfield
   {title} {\bibinfo {title} {Hubbard-corrected {DFT} energy functionals: The
  {LDA+$U$} description of correlated systems},\ }\href
  {https://doi.org/10.1002/qua.24521} {\bibfield  {journal} {\bibinfo
  {journal} {Int. J. Quantum Chem.}\ }\textbf {\bibinfo {volume} {114}},\
  \bibinfo {pages} {14} (\bibinfo {year} {2013})}\BibitemShut {NoStop}%
\bibitem [{\citenamefont {Kulik}\ and\ \citenamefont
  {Marzari}(2008)}]{Kulik2008}%
  \BibitemOpen
  \bibfield  {author} {\bibinfo {author} {\bibfnamefont {H.~J.}\ \bibnamefont
  {Kulik}}\ and\ \bibinfo {author} {\bibfnamefont {N.}~\bibnamefont
  {Marzari}},\ }\bibfield  {title} {\bibinfo {title} {A self-consistent hubbard
  {$U$} density-functional theory approach to the addition-elimination
  reactions of hydrocarbons on bare {FeO$^+$}},\ }\href
  {https://doi.org/10.1063/1.2974249} {\bibfield  {journal} {\bibinfo
  {journal} {J. Chem. Phys.}\ }\textbf {\bibinfo {volume} {129}},\ \bibinfo
  {pages} {134314} (\bibinfo {year} {2008})}\BibitemShut {NoStop}%
\bibitem [{\citenamefont {Kulik}\ and\ \citenamefont
  {Marzari}(2011)}]{Kulik2011}%
  \BibitemOpen
  \bibfield  {author} {\bibinfo {author} {\bibfnamefont {H.~J.}\ \bibnamefont
  {Kulik}}\ and\ \bibinfo {author} {\bibfnamefont {N.}~\bibnamefont
  {Marzari}},\ }\bibfield  {title} {\bibinfo {title} {Transition-metal
  dioxides: A case for the intersite term in hubbard-model functionals},\
  }\href {https://doi.org/10.1063/1.3544211} {\bibfield  {journal} {\bibinfo
  {journal} {J. Chem. Phys.}\ }\textbf {\bibinfo {volume} {134}},\ \bibinfo
  {pages} {094103} (\bibinfo {year} {2011})}\BibitemShut {NoStop}%
\bibitem [{\citenamefont {Hautier}\ \emph {et~al.}(2011)\citenamefont
  {Hautier}, \citenamefont {Jain}, \citenamefont {Ong}, \citenamefont {Kang},
  \citenamefont {Moore}, \citenamefont {Doe},\ and\ \citenamefont
  {Ceder}}]{Hautier2011}%
  \BibitemOpen
  \bibfield  {author} {\bibinfo {author} {\bibfnamefont {G.}~\bibnamefont
  {Hautier}}, \bibinfo {author} {\bibfnamefont {A.}~\bibnamefont {Jain}},
  \bibinfo {author} {\bibfnamefont {S.~P.}\ \bibnamefont {Ong}}, \bibinfo
  {author} {\bibfnamefont {B.}~\bibnamefont {Kang}}, \bibinfo {author}
  {\bibfnamefont {C.}~\bibnamefont {Moore}}, \bibinfo {author} {\bibfnamefont
  {R.}~\bibnamefont {Doe}},\ and\ \bibinfo {author} {\bibfnamefont
  {G.}~\bibnamefont {Ceder}},\ }\bibfield  {title} {\bibinfo {title}
  {Phosphates as lithium-ion battery cathodes: An evaluation based on
  high-throughput ab initio calculations},\ }\href
  {https://doi.org/10.1021/cm200949v} {\bibfield  {journal} {\bibinfo
  {journal} {Chem. Mater.}\ }\textbf {\bibinfo {volume} {23}},\ \bibinfo
  {pages} {3495} (\bibinfo {year} {2011})}\BibitemShut {NoStop}%
\bibitem [{\citenamefont {Aykol}\ and\ \citenamefont
  {Wolverton}(2014)}]{Aykol2014}%
  \BibitemOpen
  \bibfield  {author} {\bibinfo {author} {\bibfnamefont {M.}~\bibnamefont
  {Aykol}}\ and\ \bibinfo {author} {\bibfnamefont {C.}~\bibnamefont
  {Wolverton}},\ }\bibfield  {title} {\bibinfo {title} {Local environment
  dependent {GGA+$U$} method for accurate thermochemistry of transition metal
  compounds},\ }\href {https://doi.org/10.1103/PhysRevB.90.115105} {\bibfield
  {journal} {\bibinfo  {journal} {Phys. Rev. B}\ }\textbf {\bibinfo {volume}
  {90}},\ \bibinfo {pages} {115105} (\bibinfo {year} {2014})}\BibitemShut
  {NoStop}%
\bibitem [{\citenamefont {Urban}\ \emph {et~al.}(2016)\citenamefont {Urban},
  \citenamefont {Seo},\ and\ \citenamefont {Ceder}}]{Urban2016}%
  \BibitemOpen
  \bibfield  {author} {\bibinfo {author} {\bibfnamefont {A.}~\bibnamefont
  {Urban}}, \bibinfo {author} {\bibfnamefont {D.-H.}\ \bibnamefont {Seo}},\
  and\ \bibinfo {author} {\bibfnamefont {G.}~\bibnamefont {Ceder}},\ }\bibfield
   {title} {\bibinfo {title} {Computational understanding of {Li}-ion
  batteries},\ }\href {https://doi.org/10.1038/npjcompumats.2016.2} {\bibfield
  {journal} {\bibinfo  {journal} {npj Comput. Mater.}\ }\textbf {\bibinfo
  {volume} {2}},\ \bibinfo {pages} {16002} (\bibinfo {year}
  {2016})}\BibitemShut {NoStop}%
\bibitem [{\citenamefont {Chakraborty}\ \emph {et~al.}(2018)\citenamefont
  {Chakraborty}, \citenamefont {Dixit}, \citenamefont {Aurbach},\ and\
  \citenamefont {Major}}]{Chakraborty2018}%
  \BibitemOpen
  \bibfield  {author} {\bibinfo {author} {\bibfnamefont {A.}~\bibnamefont
  {Chakraborty}}, \bibinfo {author} {\bibfnamefont {M.}~\bibnamefont {Dixit}},
  \bibinfo {author} {\bibfnamefont {D.}~\bibnamefont {Aurbach}},\ and\ \bibinfo
  {author} {\bibfnamefont {D.~T.}\ \bibnamefont {Major}},\ }\bibfield  {title}
  {\bibinfo {title} {Predicting accurate cathode properties of layered oxide
  materials using the {SCAN} meta-{GGA} density functional},\ }\href
  {https://doi.org/10.1038/s41524-018-0117-4} {\bibfield  {journal} {\bibinfo
  {journal} {npj Comput. Mater.}\ }\textbf {\bibinfo {volume} {4}},\ \bibinfo
  {pages} {117} (\bibinfo {year} {2018})}\BibitemShut {NoStop}%
\bibitem [{\citenamefont {Isaacs}\ \emph {et~al.}(2020)\citenamefont {Isaacs},
  \citenamefont {Patel},\ and\ \citenamefont {Wolverton}}]{Isaacs2020}%
  \BibitemOpen
  \bibfield  {author} {\bibinfo {author} {\bibfnamefont {E.~B.}\ \bibnamefont
  {Isaacs}}, \bibinfo {author} {\bibfnamefont {S.}~\bibnamefont {Patel}},\ and\
  \bibinfo {author} {\bibfnamefont {C.}~\bibnamefont {Wolverton}},\ }\bibfield
  {title} {\bibinfo {title} {Prediction of {Li} intercalation voltages in
  rechargeable battery cathode materials: Effects of exchange-correlation
  functional, van der waals interactions, and hubbard {$U$}},\ }\href
  {https://doi.org/10.1103/PhysRevMaterials.4.065405} {\bibfield  {journal}
  {\bibinfo  {journal} {Phys. Rev. Mater.}\ }\textbf {\bibinfo {volume} {4}},\
  \bibinfo {pages} {065405} (\bibinfo {year} {2020})}\BibitemShut {NoStop}%
\bibitem [{\citenamefont {Dederichs}\ \emph {et~al.}(1984)\citenamefont
  {Dederichs}, \citenamefont {Blügel}, \citenamefont {Zeller},\ and\
  \citenamefont {Akai}}]{Dederichs1984}%
  \BibitemOpen
  \bibfield  {author} {\bibinfo {author} {\bibfnamefont {P.~H.}\ \bibnamefont
  {Dederichs}}, \bibinfo {author} {\bibfnamefont {S.}~\bibnamefont {Blügel}},
  \bibinfo {author} {\bibfnamefont {R.}~\bibnamefont {Zeller}},\ and\ \bibinfo
  {author} {\bibfnamefont {H.}~\bibnamefont {Akai}},\ }\bibfield  {title}
  {\bibinfo {title} {Ground states of constrained systems: Application to
  cerium impurities},\ }\href {https://doi.org/10.1103/PhysRevLett.53.2512}
  {\bibfield  {journal} {\bibinfo  {journal} {Phys. Rev. Lett.}\ }\textbf
  {\bibinfo {volume} {53}},\ \bibinfo {pages} {2512} (\bibinfo {year}
  {1984})}\BibitemShut {NoStop}%
\bibitem [{\citenamefont {McMahan}\ \emph {et~al.}(1988)\citenamefont
  {McMahan}, \citenamefont {Martin},\ and\ \citenamefont
  {Satpathy}}]{Mcmahan1988}%
  \BibitemOpen
  \bibfield  {author} {\bibinfo {author} {\bibfnamefont {A.~K.}\ \bibnamefont
  {McMahan}}, \bibinfo {author} {\bibfnamefont {R.~M.}\ \bibnamefont
  {Martin}},\ and\ \bibinfo {author} {\bibfnamefont {S.}~\bibnamefont
  {Satpathy}},\ }\bibfield  {title} {\bibinfo {title} {Calculated effective
  hamiltonian for {La$_2$CuO$_4$} and solution in the impurity anderson
  approximation},\ }\href {https://doi.org/10.1103/PhysRevB.38.6650} {\bibfield
   {journal} {\bibinfo  {journal} {Phys. Rev. B}\ }\textbf {\bibinfo {volume}
  {38}},\ \bibinfo {pages} {6650} (\bibinfo {year} {1988})}\BibitemShut
  {NoStop}%
\bibitem [{\citenamefont {Gunnarsson}\ \emph {et~al.}(1989)\citenamefont
  {Gunnarsson}, \citenamefont {Andersen}, \citenamefont {Jepsen},\ and\
  \citenamefont {Zaanen}}]{Gunnarsson1989}%
  \BibitemOpen
  \bibfield  {author} {\bibinfo {author} {\bibfnamefont {O.}~\bibnamefont
  {Gunnarsson}}, \bibinfo {author} {\bibfnamefont {O.~K.}\ \bibnamefont
  {Andersen}}, \bibinfo {author} {\bibfnamefont {O.}~\bibnamefont {Jepsen}},\
  and\ \bibinfo {author} {\bibfnamefont {J.}~\bibnamefont {Zaanen}},\
  }\bibfield  {title} {\bibinfo {title} {Density-functional calculation of the
  parameters in the anderson model: Application to {Mn} in {CdTe}},\ }\href
  {https://doi.org/10.1103/PhysRevB.39.1708} {\bibfield  {journal} {\bibinfo
  {journal} {Phys. Rev. B}\ }\textbf {\bibinfo {volume} {39}},\ \bibinfo
  {pages} {1708} (\bibinfo {year} {1989})}\BibitemShut {NoStop}%
\bibitem [{\citenamefont {Hybertsen}\ \emph {et~al.}(1989)\citenamefont
  {Hybertsen}, \citenamefont {Schl\"uter},\ and\ \citenamefont
  {Christensen}}]{Hybertsen1989}%
  \BibitemOpen
  \bibfield  {author} {\bibinfo {author} {\bibfnamefont {M.~S.}\ \bibnamefont
  {Hybertsen}}, \bibinfo {author} {\bibfnamefont {M.}~\bibnamefont
  {Schl\"uter}},\ and\ \bibinfo {author} {\bibfnamefont {N.~E.}\ \bibnamefont
  {Christensen}},\ }\bibfield  {title} {\bibinfo {title} {Calculation of
  coulomb-interaction parameters for {La$_2$CuO$_4$} using a
  constrained-density-functional approach},\ }\href
  {https://doi.org/10.1103/PhysRevB.39.9028} {\bibfield  {journal} {\bibinfo
  {journal} {Phys. Rev. B}\ }\textbf {\bibinfo {volume} {39}},\ \bibinfo
  {pages} {9028} (\bibinfo {year} {1989})}\BibitemShut {NoStop}%
\bibitem [{\citenamefont {Gunnarsson}(1990)}]{Gunnarsson1990}%
  \BibitemOpen
  \bibfield  {author} {\bibinfo {author} {\bibfnamefont {O.}~\bibnamefont
  {Gunnarsson}},\ }\bibfield  {title} {\bibinfo {title} {Calculation of
  parameters in model hamiltonians},\ }\href
  {https://doi.org/10.1103/PhysRevB.41.514} {\bibfield  {journal} {\bibinfo
  {journal} {Phys. Rev. B}\ }\textbf {\bibinfo {volume} {41}},\ \bibinfo
  {pages} {514} (\bibinfo {year} {1990})}\BibitemShut {NoStop}%
\bibitem [{\citenamefont {Pickett}\ \emph {et~al.}(1998)\citenamefont
  {Pickett}, \citenamefont {Erwin},\ and\ \citenamefont
  {Ethridge}}]{Pickett1998}%
  \BibitemOpen
  \bibfield  {author} {\bibinfo {author} {\bibfnamefont {W.~E.}\ \bibnamefont
  {Pickett}}, \bibinfo {author} {\bibfnamefont {S.~C.}\ \bibnamefont {Erwin}},\
  and\ \bibinfo {author} {\bibfnamefont {E.~C.}\ \bibnamefont {Ethridge}},\
  }\bibfield  {title} {\bibinfo {title} {Reformulation of the {LDA+$U$} method
  for a local-orbital basis},\ }\href
  {https://doi.org/10.1103/PhysRevB.58.1201} {\bibfield  {journal} {\bibinfo
  {journal} {Phys. Rev. B}\ }\textbf {\bibinfo {volume} {58}},\ \bibinfo
  {pages} {1201} (\bibinfo {year} {1998})}\BibitemShut {NoStop}%
\bibitem [{\citenamefont {Solovyev}\ and\ \citenamefont
  {Imada}(2005)}]{Solovyev2005}%
  \BibitemOpen
  \bibfield  {author} {\bibinfo {author} {\bibfnamefont {I.~V.}\ \bibnamefont
  {Solovyev}}\ and\ \bibinfo {author} {\bibfnamefont {M.}~\bibnamefont
  {Imada}},\ }\bibfield  {title} {\bibinfo {title} {Screening of coulomb
  interactions in transition metals},\ }\href
  {https://doi.org/10.1103/PhysRevB.71.045103} {\bibfield  {journal} {\bibinfo
  {journal} {Phys. Rev. B}\ }\textbf {\bibinfo {volume} {71}},\ \bibinfo
  {pages} {045103} (\bibinfo {year} {2005})}\BibitemShut {NoStop}%
\bibitem [{\citenamefont {Nakamura}\ \emph {et~al.}(2006)\citenamefont
  {Nakamura}, \citenamefont {Arita}, \citenamefont {Yoshimoto},\ and\
  \citenamefont {Tsuneyuki}}]{Nakamura2006}%
  \BibitemOpen
  \bibfield  {author} {\bibinfo {author} {\bibfnamefont {K.}~\bibnamefont
  {Nakamura}}, \bibinfo {author} {\bibfnamefont {R.}~\bibnamefont {Arita}},
  \bibinfo {author} {\bibfnamefont {Y.}~\bibnamefont {Yoshimoto}},\ and\
  \bibinfo {author} {\bibfnamefont {S.}~\bibnamefont {Tsuneyuki}},\ }\bibfield
  {title} {\bibinfo {title} {First-principles calculation of effective onsite
  coulomb interactions of $3d$ transition metals: Constrained local density
  functional approach with maximally localized wannier functions},\ }\href
  {https://doi.org/10.1103/PhysRevB.74.235113} {\bibfield  {journal} {\bibinfo
  {journal} {Phys. Rev. B}\ }\textbf {\bibinfo {volume} {74}},\ \bibinfo
  {pages} {235113} (\bibinfo {year} {2006})}\BibitemShut {NoStop}%
\bibitem [{\citenamefont {Shishkin}\ and\ \citenamefont
  {Sato}(2016)}]{Shishkin2016}%
  \BibitemOpen
  \bibfield  {author} {\bibinfo {author} {\bibfnamefont {M.}~\bibnamefont
  {Shishkin}}\ and\ \bibinfo {author} {\bibfnamefont {H.}~\bibnamefont
  {Sato}},\ }\bibfield  {title} {\bibinfo {title} {Self-consistent
  parametrization of {DFT+$U$} framework using linear response approach:
  Application to evaluation of redox potentials of battery cathodes},\ }\href
  {https://doi.org/10.1103/PhysRevB.93.085135} {\bibfield  {journal} {\bibinfo
  {journal} {Phys. Rev. B}\ }\textbf {\bibinfo {volume} {93}},\ \bibinfo
  {pages} {085135} (\bibinfo {year} {2016})}\BibitemShut {NoStop}%
\bibitem [{\citenamefont {Springer}\ and\ \citenamefont
  {Aryasetiawan}(1998)}]{Springer1998}%
  \BibitemOpen
  \bibfield  {author} {\bibinfo {author} {\bibfnamefont {M.}~\bibnamefont
  {Springer}}\ and\ \bibinfo {author} {\bibfnamefont {F.}~\bibnamefont
  {Aryasetiawan}},\ }\bibfield  {title} {\bibinfo {title} {Frequency-dependent
  screened interaction in ni within the random-phase approximation},\ }\href
  {https://doi.org/10.1103/PhysRevB.57.4364} {\bibfield  {journal} {\bibinfo
  {journal} {Phys. Rev. B}\ }\textbf {\bibinfo {volume} {57}},\ \bibinfo
  {pages} {4364} (\bibinfo {year} {1998})}\BibitemShut {NoStop}%
\bibitem [{\citenamefont {Kotani}(2000)}]{Kotani2000}%
  \BibitemOpen
  \bibfield  {author} {\bibinfo {author} {\bibfnamefont {T.}~\bibnamefont
  {Kotani}},\ }\bibfield  {title} {\bibinfo {title} {{\it Ab initio}
  random-phase-approximation calculation of the frequency-dependent effective
  interaction between {$3d$} electrons: {Ni, Fe, and MnO}},\ }\href
  {https://doi.org/10.1088/0953-8984/12/10/301} {\bibfield  {journal} {\bibinfo
   {journal} {J. Phys. Condens. Matter}\ }\textbf {\bibinfo {volume} {12}},\
  \bibinfo {pages} {2413} (\bibinfo {year} {2000})}\BibitemShut {NoStop}%
\bibitem [{\citenamefont {Aryasetiawan}\ \emph {et~al.}(2004)\citenamefont
  {Aryasetiawan}, \citenamefont {Imada}, \citenamefont {Georges}, \citenamefont
  {Kotliar}, \citenamefont {Biermann},\ and\ \citenamefont
  {Lichtenstein}}]{Aryasetiawan2004}%
  \BibitemOpen
  \bibfield  {author} {\bibinfo {author} {\bibfnamefont {F.}~\bibnamefont
  {Aryasetiawan}}, \bibinfo {author} {\bibfnamefont {M.}~\bibnamefont {Imada}},
  \bibinfo {author} {\bibfnamefont {A.}~\bibnamefont {Georges}}, \bibinfo
  {author} {\bibfnamefont {G.}~\bibnamefont {Kotliar}}, \bibinfo {author}
  {\bibfnamefont {S.}~\bibnamefont {Biermann}},\ and\ \bibinfo {author}
  {\bibfnamefont {A.~I.}\ \bibnamefont {Lichtenstein}},\ }\bibfield  {title}
  {\bibinfo {title} {Frequency-dependent local interactions and low-energy
  effective models from electronic structure calculations},\ }\href
  {https://doi.org/10.1103/PhysRevB.70.195104} {\bibfield  {journal} {\bibinfo
  {journal} {Phys. Rev. B}\ }\textbf {\bibinfo {volume} {70}},\ \bibinfo
  {pages} {195104} (\bibinfo {year} {2004})}\BibitemShut {NoStop}%
\bibitem [{\citenamefont {Aryasetiawan}\ \emph {et~al.}(2006)\citenamefont
  {Aryasetiawan}, \citenamefont {Karlsson}, \citenamefont {Jepsen},\ and\
  \citenamefont {Schönberger}}]{Aryasetiawan2006}%
  \BibitemOpen
  \bibfield  {author} {\bibinfo {author} {\bibfnamefont {F.}~\bibnamefont
  {Aryasetiawan}}, \bibinfo {author} {\bibfnamefont {K.}~\bibnamefont
  {Karlsson}}, \bibinfo {author} {\bibfnamefont {O.}~\bibnamefont {Jepsen}},\
  and\ \bibinfo {author} {\bibfnamefont {U.}~\bibnamefont {Schönberger}},\
  }\bibfield  {title} {\bibinfo {title} {Calculations of hubbard {$U$} from
  first principles},\ }\href {https://doi.org/10.1103/PhysRevB.74.125106}
  {\bibfield  {journal} {\bibinfo  {journal} {Phys. Rev. B}\ }\textbf {\bibinfo
  {volume} {74}},\ \bibinfo {pages} {125106} (\bibinfo {year}
  {2006})}\BibitemShut {NoStop}%
\bibitem [{\citenamefont {Mosey}\ and\ \citenamefont
  {Carter}(2007)}]{Mosey2007}%
  \BibitemOpen
  \bibfield  {author} {\bibinfo {author} {\bibfnamefont {N.~J.}\ \bibnamefont
  {Mosey}}\ and\ \bibinfo {author} {\bibfnamefont {E.~A.}\ \bibnamefont
  {Carter}},\ }\bibfield  {title} {\bibinfo {title} {Ab initio evaluation of
  coulomb and exchange parameters for {DFT+$U$} calculations},\ }\href
  {https://doi.org/10.1103/PhysRevB.76.155123} {\bibfield  {journal} {\bibinfo
  {journal} {Phys. Rev. B}\ }\textbf {\bibinfo {volume} {76}},\ \bibinfo
  {pages} {155123} (\bibinfo {year} {2007})}\BibitemShut {NoStop}%
\bibitem [{\citenamefont {Mosey}\ \emph {et~al.}(2008)\citenamefont {Mosey},
  \citenamefont {Liao},\ and\ \citenamefont {Carter}}]{Mosey2008}%
  \BibitemOpen
  \bibfield  {author} {\bibinfo {author} {\bibfnamefont {N.~J.}\ \bibnamefont
  {Mosey}}, \bibinfo {author} {\bibfnamefont {P.}~\bibnamefont {Liao}},\ and\
  \bibinfo {author} {\bibfnamefont {E.~A.}\ \bibnamefont {Carter}},\ }\bibfield
   {title} {\bibinfo {title} {Rotationally invariant ab initio evaluation of
  coulomb and exchange parameters for {DFT+$U$} calculations},\ }\href
  {https://doi.org/10.1063/1.2943138} {\bibfield  {journal} {\bibinfo
  {journal} {J. Chem. Phys.}\ }\textbf {\bibinfo {volume} {129}},\ \bibinfo
  {pages} {014103} (\bibinfo {year} {2008})}\BibitemShut {NoStop}%
\bibitem [{\citenamefont {Andriotis}\ \emph {et~al.}(2010)\citenamefont
  {Andriotis}, \citenamefont {Sheetz},\ and\ \citenamefont
  {Menon}}]{Andriotis2010}%
  \BibitemOpen
  \bibfield  {author} {\bibinfo {author} {\bibfnamefont {A.~N.}\ \bibnamefont
  {Andriotis}}, \bibinfo {author} {\bibfnamefont {R.~M.}\ \bibnamefont
  {Sheetz}},\ and\ \bibinfo {author} {\bibfnamefont {M.}~\bibnamefont
  {Menon}},\ }\bibfield  {title} {\bibinfo {title} {{LSDA+$U$} method: A
  calculation of the {$U$} values at the hartree-fock level of approximation},\
  }\href {https://doi.org/10.1103/PhysRevB.81.245103} {\bibfield  {journal}
  {\bibinfo  {journal} {Phys. Rev. B}\ }\textbf {\bibinfo {volume} {81}},\
  \bibinfo {pages} {245103} (\bibinfo {year} {2010})}\BibitemShut {NoStop}%
\bibitem [{\citenamefont {Agapito}\ \emph {et~al.}(2015)\citenamefont
  {Agapito}, \citenamefont {Curtarolo},\ and\ \citenamefont
  {Nardelli}}]{Agapito2015}%
  \BibitemOpen
  \bibfield  {author} {\bibinfo {author} {\bibfnamefont {L.~A.}\ \bibnamefont
  {Agapito}}, \bibinfo {author} {\bibfnamefont {S.}~\bibnamefont {Curtarolo}},\
  and\ \bibinfo {author} {\bibfnamefont {M.~B.}\ \bibnamefont {Nardelli}},\
  }\bibfield  {title} {\bibinfo {title} {Reformulation of {DFT+$U$} as a
  pseudohybrid hubbard density functional for accelerated materials
  discovery},\ }\href {https://doi.org/10.1103/PhysRevX.5.011006} {\bibfield
  {journal} {\bibinfo  {journal} {Phys. Rev. X}\ }\textbf {\bibinfo {volume}
  {5}},\ \bibinfo {pages} {011006} (\bibinfo {year} {2015})}\BibitemShut
  {NoStop}%
\bibitem [{\citenamefont {Cococcioni}\ and\ \citenamefont
  {de~Gironcoli}(2005)}]{Cococcioni2005}%
  \BibitemOpen
  \bibfield  {author} {\bibinfo {author} {\bibfnamefont {M.}~\bibnamefont
  {Cococcioni}}\ and\ \bibinfo {author} {\bibfnamefont {S.}~\bibnamefont
  {de~Gironcoli}},\ }\bibfield  {title} {\bibinfo {title} {Linear response
  approach to the calculation of the effective interaction parameters in the
  {LDA+$U$} method},\ }\href {https://doi.org/10.1103/PhysRevB.71.035105}
  {\bibfield  {journal} {\bibinfo  {journal} {Phys. Rev. B}\ }\textbf {\bibinfo
  {volume} {71}},\ \bibinfo {pages} {035105} (\bibinfo {year}
  {2005})}\BibitemShut {NoStop}%
\bibitem [{\citenamefont {Timrov}\ \emph {et~al.}(2018)\citenamefont {Timrov},
  \citenamefont {Marzari},\ and\ \citenamefont {Cococcioni}}]{Timrov2018}%
  \BibitemOpen
  \bibfield  {author} {\bibinfo {author} {\bibfnamefont {I.}~\bibnamefont
  {Timrov}}, \bibinfo {author} {\bibfnamefont {N.}~\bibnamefont {Marzari}},\
  and\ \bibinfo {author} {\bibfnamefont {M.}~\bibnamefont {Cococcioni}},\
  }\bibfield  {title} {\bibinfo {title} {Hubbard parameters from
  density-functional perturbation theory},\ }\href
  {https://doi.org/10.1103/PhysRevB.98.085127} {\bibfield  {journal} {\bibinfo
  {journal} {Phys. Rev. B}\ }\textbf {\bibinfo {volume} {98}},\ \bibinfo
  {pages} {085127} (\bibinfo {year} {2018})}\BibitemShut {NoStop}%
\bibitem [{\citenamefont {Timrov}\ \emph {et~al.}(2021)\citenamefont {Timrov},
  \citenamefont {Marzari},\ and\ \citenamefont {Cococcioni}}]{Timrov2021}%
  \BibitemOpen
  \bibfield  {author} {\bibinfo {author} {\bibfnamefont {I.}~\bibnamefont
  {Timrov}}, \bibinfo {author} {\bibfnamefont {N.}~\bibnamefont {Marzari}},\
  and\ \bibinfo {author} {\bibfnamefont {M.}~\bibnamefont {Cococcioni}},\
  }\bibfield  {title} {\bibinfo {title} {Self-consistent hubbard parameters
  from density-functional perturbation theory in the ultrasoft and
  projector-augmented wave formulations},\ }\href
  {https://doi.org/10.1103/PhysRevB.103.045141} {\bibfield  {journal} {\bibinfo
   {journal} {Phys. Rev. B}\ }\textbf {\bibinfo {volume} {103}},\ \bibinfo
  {pages} {045141} (\bibinfo {year} {2021})}\BibitemShut {NoStop}%
\bibitem [{\citenamefont {Cococcioni}\ and\ \citenamefont
  {Marzari}(2019)}]{Cococcioni2019}%
  \BibitemOpen
  \bibfield  {author} {\bibinfo {author} {\bibfnamefont {M.}~\bibnamefont
  {Cococcioni}}\ and\ \bibinfo {author} {\bibfnamefont {N.}~\bibnamefont
  {Marzari}},\ }\bibfield  {title} {\bibinfo {title} {Energetics and cathode
  voltages of {Li$M$PO$_4$} olivines ({$M$=Fe, Mn}) from extended hubbard
  functionals},\ }\href
  {https://link.aps.org/doi/10.1103/PhysRevMaterials.3.033801} {\bibfield
  {journal} {\bibinfo  {journal} {Phys. Rev. Mater.}\ }\textbf {\bibinfo
  {volume} {3}},\ \bibinfo {pages} {033801} (\bibinfo {year}
  {2019})}\BibitemShut {NoStop}%
\bibitem [{\citenamefont {Timrov}\ \emph {et~al.}(2023)\citenamefont {Timrov},
  \citenamefont {Kotiuga},\ and\ \citenamefont {Marzari}}]{Timrov2023}%
  \BibitemOpen
  \bibfield  {author} {\bibinfo {author} {\bibfnamefont {I.}~\bibnamefont
  {Timrov}}, \bibinfo {author} {\bibfnamefont {M.}~\bibnamefont {Kotiuga}},\
  and\ \bibinfo {author} {\bibfnamefont {N.}~\bibnamefont {Marzari}},\
  }\bibfield  {title} {\bibinfo {title} {Unraveling the effects of inter-site
  hubbard interactions in spinel {Li}-ion cathode materials},\ }\href
  {https://doi.org/10.1039/D3CP00419H} {\bibfield  {journal} {\bibinfo
  {journal} {Phys. Chem. Chem. Phys.}\ }\textbf {\bibinfo {volume} {25}},\
  \bibinfo {pages} {9061} (\bibinfo {year} {2023})}\BibitemShut {NoStop}%
\bibitem [{\citenamefont {Malica}\ and\ \citenamefont
  {Marzari}(2025)}]{Malica2024}%
  \BibitemOpen
  \bibfield  {author} {\bibinfo {author} {\bibfnamefont {C.}~\bibnamefont
  {Malica}}\ and\ \bibinfo {author} {\bibfnamefont {N.}~\bibnamefont
  {Marzari}},\ }\bibfield  {title} {\bibinfo {title} {Teaching oxidation states
  to neural networks},\ }\href
  {https://www.nature.com/articles/s41524-025-01709-z} {\bibfield  {journal}
  {\bibinfo  {journal} {npj Comput Mater 11, 212 (2025)}\ } (\bibinfo {year}
  {2025})}\BibitemShut {NoStop}%
\bibitem [{\citenamefont {Grimme}\ \emph {et~al.}(2016)\citenamefont {Grimme},
  \citenamefont {Hansen}, \citenamefont {Brandenburg},\ and\ \citenamefont
  {Bannwarth}}]{Grimme2016}%
  \BibitemOpen
  \bibfield  {author} {\bibinfo {author} {\bibfnamefont {S.}~\bibnamefont
  {Grimme}}, \bibinfo {author} {\bibfnamefont {A.}~\bibnamefont {Hansen}},
  \bibinfo {author} {\bibfnamefont {J.~G.}\ \bibnamefont {Brandenburg}},\ and\
  \bibinfo {author} {\bibfnamefont {C.}~\bibnamefont {Bannwarth}},\ }\bibfield
  {title} {\bibinfo {title} {Dispersion-corrected mean-field electronic
  structure methods},\ }\href {https://doi.org/10.1021/acs.chemrev.5b00533}
  {\bibfield  {journal} {\bibinfo  {journal} {Chem. Rev.}\ }\textbf {\bibinfo
  {volume} {116}},\ \bibinfo {pages} {5105} (\bibinfo {year}
  {2016})}\BibitemShut {NoStop}%
\bibitem [{\citenamefont {Grimme}\ \emph {et~al.}(2010)\citenamefont {Grimme},
  \citenamefont {Antony}, \citenamefont {Ehrlich},\ and\ \citenamefont
  {Krieg}}]{Grimme2010}%
  \BibitemOpen
  \bibfield  {author} {\bibinfo {author} {\bibfnamefont {S.}~\bibnamefont
  {Grimme}}, \bibinfo {author} {\bibfnamefont {J.}~\bibnamefont {Antony}},
  \bibinfo {author} {\bibfnamefont {S.}~\bibnamefont {Ehrlich}},\ and\ \bibinfo
  {author} {\bibfnamefont {H.}~\bibnamefont {Krieg}},\ }\bibfield  {title}
  {\bibinfo {title} {A consistent and accurate ab initio parametrization of
  density functional dispersion correction ({DFT-D}) for the 94 elements
  {H-Pu}},\ }\href {https://doi.org/10.1063/1.3382344} {\bibfield  {journal}
  {\bibinfo  {journal} {J. Chem. Phys.}\ }\textbf {\bibinfo {volume} {132}},\
  \bibinfo {pages} {154104} (\bibinfo {year} {2010})}\BibitemShut {NoStop}%
\bibitem [{\citenamefont {Dion}\ \emph {et~al.}(2004)\citenamefont {Dion},
  \citenamefont {Rydberg}, \citenamefont {Schröder}, \citenamefont
  {Langreth},\ and\ \citenamefont {Lundqvist}}]{Dion2004}%
  \BibitemOpen
  \bibfield  {author} {\bibinfo {author} {\bibfnamefont {M.}~\bibnamefont
  {Dion}}, \bibinfo {author} {\bibfnamefont {H.}~\bibnamefont {Rydberg}},
  \bibinfo {author} {\bibfnamefont {E.}~\bibnamefont {Schröder}}, \bibinfo
  {author} {\bibfnamefont {D.~C.}\ \bibnamefont {Langreth}},\ and\ \bibinfo
  {author} {\bibfnamefont {B.~I.}\ \bibnamefont {Lundqvist}},\ }\bibfield
  {title} {\bibinfo {title} {Van der waals density functional for general
  geometries},\ }\href {https://doi.org/10.1103/PhysRevLett.92.246401}
  {\bibfield  {journal} {\bibinfo  {journal} {Phys. Rev. Lett.}\ }\textbf
  {\bibinfo {volume} {92}},\ \bibinfo {pages} {246401} (\bibinfo {year}
  {2004})}\BibitemShut {NoStop}%
\bibitem [{\citenamefont {Thonhauser}\ \emph {et~al.}(2015)\citenamefont
  {Thonhauser}, \citenamefont {Zuluaga}, \citenamefont {Arter}, \citenamefont
  {Berland}, \citenamefont {Schr\"oder},\ and\ \citenamefont
  {Hyldgaard}}]{Thonhauser2015}%
  \BibitemOpen
  \bibfield  {author} {\bibinfo {author} {\bibfnamefont {T.}~\bibnamefont
  {Thonhauser}}, \bibinfo {author} {\bibfnamefont {S.}~\bibnamefont {Zuluaga}},
  \bibinfo {author} {\bibfnamefont {C.~A.}\ \bibnamefont {Arter}}, \bibinfo
  {author} {\bibfnamefont {K.}~\bibnamefont {Berland}}, \bibinfo {author}
  {\bibfnamefont {E.}~\bibnamefont {Schr\"oder}},\ and\ \bibinfo {author}
  {\bibfnamefont {P.}~\bibnamefont {Hyldgaard}},\ }\bibfield  {title} {\bibinfo
  {title} {Spin signature of nonlocal correlation binding in metal-organic
  frameworks},\ }\href {https://doi.org/10.1103/PhysRevLett.115.136402}
  {\bibfield  {journal} {\bibinfo  {journal} {Phys. Rev. Lett.}\ }\textbf
  {\bibinfo {volume} {115}},\ \bibinfo {pages} {136402} (\bibinfo {year}
  {2015})}\BibitemShut {NoStop}%
\bibitem [{\citenamefont {Lee}\ \emph {et~al.}(2010)\citenamefont {Lee},
  \citenamefont {Murray}, \citenamefont {Kong}, \citenamefont {Lundqvist},\
  and\ \citenamefont {Langreth}}]{Lee2010}%
  \BibitemOpen
  \bibfield  {author} {\bibinfo {author} {\bibfnamefont {K.}~\bibnamefont
  {Lee}}, \bibinfo {author} {\bibfnamefont {E.~D.}\ \bibnamefont {Murray}},
  \bibinfo {author} {\bibfnamefont {L.}~\bibnamefont {Kong}}, \bibinfo {author}
  {\bibfnamefont {B.~I.}\ \bibnamefont {Lundqvist}},\ and\ \bibinfo {author}
  {\bibfnamefont {D.~C.}\ \bibnamefont {Langreth}},\ }\bibfield  {title}
  {\bibinfo {title} {Higher-accuracy van der waals density functional},\ }\href
  {https://doi.org/10.1103/PhysRevB.82.081101} {\bibfield  {journal} {\bibinfo
  {journal} {Phys. Rev. B}\ }\textbf {\bibinfo {volume} {82}},\ \bibinfo
  {pages} {081101} (\bibinfo {year} {2010})}\BibitemShut {NoStop}%
\bibitem [{\citenamefont {Chakraborty}\ \emph {et~al.}(2020)\citenamefont
  {Chakraborty}, \citenamefont {Berland},\ and\ \citenamefont
  {Thonhauser}}]{Chakraborty2020}%
  \BibitemOpen
  \bibfield  {author} {\bibinfo {author} {\bibfnamefont {D.}~\bibnamefont
  {Chakraborty}}, \bibinfo {author} {\bibfnamefont {K.}~\bibnamefont
  {Berland}},\ and\ \bibinfo {author} {\bibfnamefont {T.}~\bibnamefont
  {Thonhauser}},\ }\bibfield  {title} {\bibinfo {title} {Next-generation
  nonlocal van der waals density functional},\ }\href
  {https://doi.org/10.1021/acs.jctc.0c00471} {\bibfield  {journal} {\bibinfo
  {journal} {J. Chem. Theory Comput.}\ }\textbf {\bibinfo {volume} {16}},\
  \bibinfo {pages} {5893} (\bibinfo {year} {2020})}\BibitemShut {NoStop}%
\bibitem [{\citenamefont {Sabatini}\ \emph {et~al.}(2013)\citenamefont
  {Sabatini}, \citenamefont {Gorni},\ and\ \citenamefont
  {de~Gironcoli}}]{Sabatini2013}%
  \BibitemOpen
  \bibfield  {author} {\bibinfo {author} {\bibfnamefont {R.}~\bibnamefont
  {Sabatini}}, \bibinfo {author} {\bibfnamefont {T.}~\bibnamefont {Gorni}},\
  and\ \bibinfo {author} {\bibfnamefont {S.}~\bibnamefont {de~Gironcoli}},\
  }\bibfield  {title} {\bibinfo {title} {Nonlocal van der waals density
  functional made simple and efficient},\ }\href
  {https://doi.org/10.1103/PhysRevB.87.041108} {\bibfield  {journal} {\bibinfo
  {journal} {Phys. Rev. B}\ }\textbf {\bibinfo {volume} {87}},\ \bibinfo
  {pages} {041108} (\bibinfo {year} {2013})}\BibitemShut {NoStop}%
\bibitem [{\citenamefont {Klimeš}\ \emph {et~al.}(2009)\citenamefont
  {Klimeš}, \citenamefont {Bowler},\ and\ \citenamefont
  {Michaelides}}]{Klimes2009}%
  \BibitemOpen
  \bibfield  {author} {\bibinfo {author} {\bibfnamefont {J.}~\bibnamefont
  {Klimeš}}, \bibinfo {author} {\bibfnamefont {D.~R.}\ \bibnamefont
  {Bowler}},\ and\ \bibinfo {author} {\bibfnamefont {A.}~\bibnamefont
  {Michaelides}},\ }\bibfield  {title} {\bibinfo {title} {Chemical accuracy for
  the van der waals density functional},\ }\href
  {https://doi.org/10.1088/0953-8984/22/2/022201} {\bibfield  {journal}
  {\bibinfo  {journal} {J. Phys. Condens. Matter}\ }\textbf {\bibinfo {volume}
  {22}},\ \bibinfo {pages} {022201} (\bibinfo {year} {2009})}\BibitemShut
  {NoStop}%
\bibitem [{\citenamefont {Klimeš}\ \emph {et~al.}(2011)\citenamefont
  {Klimeš}, \citenamefont {Bowler},\ and\ \citenamefont
  {Michaelides}}]{Klimes2011}%
  \BibitemOpen
  \bibfield  {author} {\bibinfo {author} {\bibfnamefont {J.}~\bibnamefont
  {Klimeš}}, \bibinfo {author} {\bibfnamefont {D.~R.}\ \bibnamefont
  {Bowler}},\ and\ \bibinfo {author} {\bibfnamefont {A.}~\bibnamefont
  {Michaelides}},\ }\bibfield  {title} {\bibinfo {title} {Van der waals density
  functionals applied to solids},\ }\href
  {https://doi.org/10.1103/PhysRevB.83.195131} {\bibfield  {journal} {\bibinfo
  {journal} {Phys. Rev. B}\ }\textbf {\bibinfo {volume} {83}},\ \bibinfo
  {pages} {195131} (\bibinfo {year} {2011})}\BibitemShut {NoStop}%
\bibitem [{\citenamefont {Wolverton}\ and\ \citenamefont
  {Zunger}(1998)}]{Wolverton1998}%
  \BibitemOpen
  \bibfield  {author} {\bibinfo {author} {\bibfnamefont {C.}~\bibnamefont
  {Wolverton}}\ and\ \bibinfo {author} {\bibfnamefont {A.}~\bibnamefont
  {Zunger}},\ }\bibfield  {title} {\bibinfo {title} {Prediction of {Li}
  intercalation and battery voltages in layered vs. cubic {Li$_x$CoO$_2$}},\
  }\href {https://doi.org/10.1149/1.1838653} {\bibfield  {journal} {\bibinfo
  {journal} {J. Electrochem. Soc.}\ }\textbf {\bibinfo {volume} {145}},\
  \bibinfo {pages} {2424} (\bibinfo {year} {1998})}\BibitemShut {NoStop}%
\bibitem [{\citenamefont {Carlier}\ \emph {et~al.}(2002)\citenamefont
  {Carlier}, \citenamefont {Ven}, \citenamefont {Ceder}, \citenamefont
  {Croguennec}, \citenamefont {Ménétrier},\ and\ \citenamefont
  {Delmas}}]{Carlier2002}%
  \BibitemOpen
  \bibfield  {author} {\bibinfo {author} {\bibfnamefont {D.}~\bibnamefont
  {Carlier}}, \bibinfo {author} {\bibfnamefont {A.}~\bibnamefont {Ven}},
  \bibinfo {author} {\bibfnamefont {G.}~\bibnamefont {Ceder}}, \bibinfo
  {author} {\bibfnamefont {L.}~\bibnamefont {Croguennec}}, \bibinfo {author}
  {\bibfnamefont {M.}~\bibnamefont {Ménétrier}},\ and\ \bibinfo {author}
  {\bibfnamefont {C.}~\bibnamefont {Delmas}},\ }\bibfield  {title} {\bibinfo
  {title} {Lithium electrochemical deintercalation from {O2-LiCoO$_2$}:
  Structural study and first principles calculations},\ }\href
  {https://doi.org/10.1557/PROC-756-EE5.9} {\bibfield  {journal} {\bibinfo
  {journal} {MRS Proc.}\ }\textbf {\bibinfo {volume} {756}},\ \bibinfo {pages}
  {EE5.9} (\bibinfo {year} {2002})}\BibitemShut {NoStop}%
\bibitem [{\citenamefont {Kim}\ \emph {et~al.}(2021)\citenamefont {Kim},
  \citenamefont {Kim},\ and\ \citenamefont {Kim}}]{Kim2021}%
  \BibitemOpen
  \bibfield  {author} {\bibinfo {author} {\bibfnamefont {B.}~\bibnamefont
  {Kim}}, \bibinfo {author} {\bibfnamefont {K.}~\bibnamefont {Kim}},\ and\
  \bibinfo {author} {\bibfnamefont {S.}~\bibnamefont {Kim}},\ }\bibfield
  {title} {\bibinfo {title} {Quantification of coulomb interactions in layered
  lithium and sodium battery cathode materials},\ }\href
  {https://doi.org/10.1103/PhysRevMaterials.5.035404} {\bibfield  {journal}
  {\bibinfo  {journal} {Phys. Rev. Mater.}\ }\textbf {\bibinfo {volume} {5}},\
  \bibinfo {pages} {035404} (\bibinfo {year} {2021})}\BibitemShut {NoStop}%
\bibitem [{\citenamefont {Perdew}\ \emph {et~al.}(1996)\citenamefont {Perdew},
  \citenamefont {Burke},\ and\ \citenamefont {Ernzerhof}}]{Perdew1996}%
  \BibitemOpen
  \bibfield  {author} {\bibinfo {author} {\bibfnamefont {J.~P.}\ \bibnamefont
  {Perdew}}, \bibinfo {author} {\bibfnamefont {K.}~\bibnamefont {Burke}},\ and\
  \bibinfo {author} {\bibfnamefont {M.}~\bibnamefont {Ernzerhof}},\ }\bibfield
  {title} {\bibinfo {title} {Generalized gradient approximation made simple},\
  }\href {https://doi.org/10.1103/PhysRevLett.77.3865} {\bibfield  {journal}
  {\bibinfo  {journal} {Phys. Rev. Lett.}\ }\textbf {\bibinfo {volume} {77}},\
  \bibinfo {pages} {3865} (\bibinfo {year} {1996})}\BibitemShut {NoStop}%
\bibitem [{\citenamefont {Peng}\ \emph {et~al.}(2016)\citenamefont {Peng},
  \citenamefont {Yang}, \citenamefont {Perdew},\ and\ \citenamefont
  {Sun}}]{Peng2016}%
  \BibitemOpen
  \bibfield  {author} {\bibinfo {author} {\bibfnamefont {H.}~\bibnamefont
  {Peng}}, \bibinfo {author} {\bibfnamefont {Z.-H.}\ \bibnamefont {Yang}},
  \bibinfo {author} {\bibfnamefont {J.~P.}\ \bibnamefont {Perdew}},\ and\
  \bibinfo {author} {\bibfnamefont {J.}~\bibnamefont {Sun}},\ }\bibfield
  {title} {\bibinfo {title} {Versatile van der waals density functional based
  on a meta-generalized gradient approximation},\ }\href
  {https://doi.org/10.1103/PhysRevX.6.041005} {\bibfield  {journal} {\bibinfo
  {journal} {Phys. Rev. X}\ }\textbf {\bibinfo {volume} {6}},\ \bibinfo {pages}
  {041005} (\bibinfo {year} {2016})}\BibitemShut {NoStop}%
\bibitem [{\citenamefont {Aykol}\ \emph {et~al.}(2015)\citenamefont {Aykol},
  \citenamefont {Kim},\ and\ \citenamefont {Wolverton}}]{Aykol2015}%
  \BibitemOpen
  \bibfield  {author} {\bibinfo {author} {\bibfnamefont {M.}~\bibnamefont
  {Aykol}}, \bibinfo {author} {\bibfnamefont {S.}~\bibnamefont {Kim}},\ and\
  \bibinfo {author} {\bibfnamefont {C.}~\bibnamefont {Wolverton}},\ }\bibfield
  {title} {\bibinfo {title} {Van der waals interactions in layered lithium
  cobalt oxides},\ }\href {https://doi.org/10.1021/acs.jpcc.5b06240} {\bibfield
   {journal} {\bibinfo  {journal} {J. Phys. Chem. C}\ }\textbf {\bibinfo
  {volume} {119}},\ \bibinfo {pages} {19053} (\bibinfo {year}
  {2015})}\BibitemShut {NoStop}%
\bibitem [{\citenamefont {Shishkin}\ and\ \citenamefont
  {Sato}(2021)}]{Shishkin2021}%
  \BibitemOpen
  \bibfield  {author} {\bibinfo {author} {\bibfnamefont {M.}~\bibnamefont
  {Shishkin}}\ and\ \bibinfo {author} {\bibfnamefont {H.}~\bibnamefont
  {Sato}},\ }\bibfield  {title} {\bibinfo {title} {Evaluation of redox
  potentials of cathode materials of alkali-ion batteries using extended
  {DFT+$U$} +{$U\uparrow\downarrow$} method: The role of interactions between
  the electrons with opposite spins},\ }\href
  {https://doi.org/10.1063/5.0039594} {\bibfield  {journal} {\bibinfo
  {journal} {J. Chem. Phys.}\ }\textbf {\bibinfo {volume} {154}},\ \bibinfo
  {pages} {114709} (\bibinfo {year} {2021})}\BibitemShut {NoStop}%
\bibitem [{\citenamefont {Zhou}\ \emph {et~al.}(2004)\citenamefont {Zhou},
  \citenamefont {Cococcioni}, \citenamefont {Marianetti}, \citenamefont
  {Morgan},\ and\ \citenamefont {Ceder}}]{Zhou2004}%
  \BibitemOpen
  \bibfield  {author} {\bibinfo {author} {\bibfnamefont {F.}~\bibnamefont
  {Zhou}}, \bibinfo {author} {\bibfnamefont {M.}~\bibnamefont {Cococcioni}},
  \bibinfo {author} {\bibfnamefont {C.~A.}\ \bibnamefont {Marianetti}},
  \bibinfo {author} {\bibfnamefont {D.}~\bibnamefont {Morgan}},\ and\ \bibinfo
  {author} {\bibfnamefont {G.}~\bibnamefont {Ceder}},\ }\bibfield  {title}
  {\bibinfo {title} {First-principles prediction of redox potentials in
  transition-metal compounds with {LDA+$U$}},\ }\href
  {https://doi.org/10.1103/PhysRevB.70.235121} {\bibfield  {journal} {\bibinfo
  {journal} {Phys. Rev. B}\ }\textbf {\bibinfo {volume} {70}},\ \bibinfo
  {pages} {235121} (\bibinfo {year} {2004})}\BibitemShut {NoStop}%
\bibitem [{\citenamefont {Kresse}\ and\ \citenamefont
  {Hafner}(1993)}]{Kresse1993}%
  \BibitemOpen
  \bibfield  {author} {\bibinfo {author} {\bibfnamefont {G.}~\bibnamefont
  {Kresse}}\ and\ \bibinfo {author} {\bibfnamefont {J.}~\bibnamefont
  {Hafner}},\ }\bibfield  {title} {\bibinfo {title} {Ab initio molecular
  dynamics for liquid metals},\ }\href
  {https://doi.org/10.1103/PhysRevB.47.558} {\bibfield  {journal} {\bibinfo
  {journal} {Phys. Rev. B}\ }\textbf {\bibinfo {volume} {47}},\ \bibinfo
  {pages} {558} (\bibinfo {year} {1993})}\BibitemShut {NoStop}%
\bibitem [{\citenamefont {Blöchl}(1994)}]{Blochl1994}%
  \BibitemOpen
  \bibfield  {author} {\bibinfo {author} {\bibfnamefont {P.~E.}\ \bibnamefont
  {Blöchl}},\ }\bibfield  {title} {\bibinfo {title} {Projector augmented-wave
  method},\ }\href {https://doi.org/10.1103/PhysRevB.50.17953} {\bibfield
  {journal} {\bibinfo  {journal} {Phys. Rev. B}\ }\textbf {\bibinfo {volume}
  {50}},\ \bibinfo {pages} {17953} (\bibinfo {year} {1994})}\BibitemShut
  {NoStop}%
\bibitem [{\citenamefont {Isaacs}\ and\ \citenamefont
  {Marianetti}(2020)}]{Isaacs2020b}%
  \BibitemOpen
  \bibfield  {author} {\bibinfo {author} {\bibfnamefont {E.~B.}\ \bibnamefont
  {Isaacs}}\ and\ \bibinfo {author} {\bibfnamefont {C.~A.}\ \bibnamefont
  {Marianetti}},\ }\bibfield  {title} {\bibinfo {title} {Compositional phase
  stability of correlated electron materials within
  $\mathrm{DFT}\text{+}\mathrm{DMFT}$},\ }\href
  {https://doi.org/10.1103/PhysRevB.102.045146} {\bibfield  {journal} {\bibinfo
   {journal} {Phys. Rev. B}\ }\textbf {\bibinfo {volume} {102}},\ \bibinfo
  {pages} {045146} (\bibinfo {year} {2020})}\BibitemShut {NoStop}%
\bibitem [{\citenamefont {Nekrasov}\ \emph {et~al.}(2000)\citenamefont
  {Nekrasov}, \citenamefont {Korotin},\ and\ \citenamefont
  {Anisimov}}]{Nekrasov2000}%
  \BibitemOpen
  \bibfield  {author} {\bibinfo {author} {\bibfnamefont {I.~A.}\ \bibnamefont
  {Nekrasov}}, \bibinfo {author} {\bibfnamefont {M.~A.}\ \bibnamefont
  {Korotin}},\ and\ \bibinfo {author} {\bibfnamefont {V.~I.}\ \bibnamefont
  {Anisimov}},\ }\bibfield  {title} {\bibinfo {title} {Coulomb interaction in
  oxygen $p$-shell in {LDA+$U$} method and its influence on calculated spectral
  and magnetic properties of transition metal oxides},\ }\href
  {https://arxiv.org/abs/cond-mat/0009107} {\bibfield  {journal} {\bibinfo
  {journal} {arXiv preprint}\ } (\bibinfo {year} {2000})}\BibitemShut {NoStop}%
\bibitem [{\citenamefont {Kirchner-Hall}\ \emph {et~al.}(2021)\citenamefont
  {Kirchner-Hall}, \citenamefont {Zhao}, \citenamefont {Xiong}, \citenamefont
  {Timrov},\ and\ \citenamefont {Dabo}}]{KirchnerHall2021}%
  \BibitemOpen
  \bibfield  {author} {\bibinfo {author} {\bibfnamefont {N.~E.}\ \bibnamefont
  {Kirchner-Hall}}, \bibinfo {author} {\bibfnamefont {W.}~\bibnamefont {Zhao}},
  \bibinfo {author} {\bibfnamefont {Y.}~\bibnamefont {Xiong}}, \bibinfo
  {author} {\bibfnamefont {I.}~\bibnamefont {Timrov}},\ and\ \bibinfo {author}
  {\bibfnamefont {I.}~\bibnamefont {Dabo}},\ }\bibfield  {title} {\bibinfo
  {title} {Extensive benchmarking of {DFT+$U$} calculations for predicting band
  gaps},\ }\href {https://doi.org/10.3390/app11052395} {\bibfield  {journal}
  {\bibinfo  {journal} {Appl. Sci.}\ }\textbf {\bibinfo {volume} {11}},\
  \bibinfo {pages} {2395} (\bibinfo {year} {2021})}\BibitemShut {NoStop}%
\bibitem [{\citenamefont {Xiong}\ \emph {et~al.}(2021)\citenamefont {Xiong},
  \citenamefont {Campbell}, \citenamefont {Fanghanel}, \citenamefont {Badding},
  \citenamefont {Wang}, \citenamefont {Kirchner-Hall}, \citenamefont
  {Theibault}, \citenamefont {Khan}, \citenamefont {Rivera}, \citenamefont
  {Smith}, \citenamefont {Timrov}, \citenamefont {Montes-Santi}, \citenamefont
  {Abruña}, \citenamefont {Toksoz},\ and\ \citenamefont
  {Velivelli}}]{Xiong2021}%
  \BibitemOpen
  \bibfield  {author} {\bibinfo {author} {\bibfnamefont {Y.}~\bibnamefont
  {Xiong}}, \bibinfo {author} {\bibfnamefont {Q.~T.}\ \bibnamefont {Campbell}},
  \bibinfo {author} {\bibfnamefont {J.}~\bibnamefont {Fanghanel}}, \bibinfo
  {author} {\bibfnamefont {C.~K.}\ \bibnamefont {Badding}}, \bibinfo {author}
  {\bibfnamefont {H.}~\bibnamefont {Wang}}, \bibinfo {author} {\bibfnamefont
  {N.~E.}\ \bibnamefont {Kirchner-Hall}}, \bibinfo {author} {\bibfnamefont
  {M.~J.}\ \bibnamefont {Theibault}}, \bibinfo {author} {\bibfnamefont {M.~M.}\
  \bibnamefont {Khan}}, \bibinfo {author} {\bibfnamefont {T.}~\bibnamefont
  {Rivera}}, \bibinfo {author} {\bibfnamefont {S.}~\bibnamefont {Smith}},
  \bibinfo {author} {\bibfnamefont {I.}~\bibnamefont {Timrov}}, \bibinfo
  {author} {\bibfnamefont {D.}~\bibnamefont {Montes-Santi}}, \bibinfo {author}
  {\bibfnamefont {H.}~\bibnamefont {Abruña}}, \bibinfo {author} {\bibfnamefont
  {D.~I.}\ \bibnamefont {Toksoz}},\ and\ \bibinfo {author} {\bibfnamefont
  {N.}~\bibnamefont {Velivelli}},\ }\bibfield  {title} {\bibinfo {title}
  {Optimizing accuracy and efficacy in data-driven materials discovery for the
  solar production of hydrogen},\ }\href {https://doi.org/10.1039/D0EE02389F}
  {\bibfield  {journal} {\bibinfo  {journal} {Energy Environ. Sci.}\ }\textbf
  {\bibinfo {volume} {14}},\ \bibinfo {pages} {2335} (\bibinfo {year}
  {2021})}\BibitemShut {NoStop}%
\bibitem [{\citenamefont {Lechermann}(2024)}]{Lechermann2024}%
  \BibitemOpen
  \bibfield  {author} {\bibinfo {author} {\bibfnamefont {F.}~\bibnamefont
  {Lechermann}},\ }\bibfield  {title} {\bibinfo {title} {A theoretical
  perspective on transition-metal-based magnets},\ }\href
  {https://arxiv.org/abs/2410.06891} {\bibfield  {journal} {\bibinfo  {journal}
  {arXiv preprint}\ } (\bibinfo {year} {2024})}\BibitemShut {NoStop}%
\bibitem [{\citenamefont {May}\ and\ \citenamefont {Kolpak}(2020)}]{May2020}%
  \BibitemOpen
  \bibfield  {author} {\bibinfo {author} {\bibfnamefont {K.~J.}\ \bibnamefont
  {May}}\ and\ \bibinfo {author} {\bibfnamefont {A.~M.}\ \bibnamefont
  {Kolpak}},\ }\bibfield  {title} {\bibinfo {title} {Improved description of
  perovskite oxide crystal structure and electronic properties using
  self-consistent hubbard {$U$} corrections from {ACBN0}},\ }\href
  {https://doi.org/10.1103/PhysRevB.101.165117} {\bibfield  {journal} {\bibinfo
   {journal} {Phys. Rev. B}\ }\textbf {\bibinfo {volume} {101}},\ \bibinfo
  {pages} {165117} (\bibinfo {year} {2020})}\BibitemShut {NoStop}%
\bibitem [{\citenamefont {Berman}\ \emph {et~al.}(2023)\citenamefont {Berman},
  \citenamefont {Zhussupbekova}, \citenamefont {Boschker}, \citenamefont
  {Schwarzkopf}, \citenamefont {O'Regan}, \citenamefont {Shvets},\ and\
  \citenamefont {Zhussupbekov}}]{Berman2023}%
  \BibitemOpen
  \bibfield  {author} {\bibinfo {author} {\bibfnamefont {S.}~\bibnamefont
  {Berman}}, \bibinfo {author} {\bibfnamefont {A.}~\bibnamefont
  {Zhussupbekova}}, \bibinfo {author} {\bibfnamefont {J.~E.}\ \bibnamefont
  {Boschker}}, \bibinfo {author} {\bibfnamefont {J.}~\bibnamefont
  {Schwarzkopf}}, \bibinfo {author} {\bibfnamefont {D.~D.}\ \bibnamefont
  {O'Regan}}, \bibinfo {author} {\bibfnamefont {I.~V.}\ \bibnamefont
  {Shvets}},\ and\ \bibinfo {author} {\bibfnamefont {K.}~\bibnamefont
  {Zhussupbekov}},\ }\bibfield  {title} {\bibinfo {title} {Reconciling the
  theoretical and experimental electronic structure of {NbO$_2$}},\ }\href
  {https://doi.org/10.1103/PhysRevB.108.155141} {\bibfield  {journal} {\bibinfo
   {journal} {Phys. Rev. B}\ }\textbf {\bibinfo {volume} {108}},\ \bibinfo
  {pages} {155141} (\bibinfo {year} {2023})}\BibitemShut {NoStop}%
\bibitem [{\citenamefont {Orhan}\ and\ \citenamefont
  {O'Regan}(2020)}]{Orhan2020}%
  \BibitemOpen
  \bibfield  {author} {\bibinfo {author} {\bibfnamefont {O.~K.}\ \bibnamefont
  {Orhan}}\ and\ \bibinfo {author} {\bibfnamefont {D.~D.}\ \bibnamefont
  {O'Regan}},\ }\bibfield  {title} {\bibinfo {title} {First-principles hubbard
  {$U$} and hund's {$J$} corrected approximate density functional theory
  predicts an accurate fundamental gap in rutile and anatase {TiO$_2$}},\
  }\href {https://doi.org/10.1103/PhysRevB.101.245137} {\bibfield  {journal}
  {\bibinfo  {journal} {Phys. Rev. B}\ }\textbf {\bibinfo {volume} {101}},\
  \bibinfo {pages} {245137} (\bibinfo {year} {2020})}\BibitemShut {NoStop}%
\bibitem [{\citenamefont {Lambert}\ and\ \citenamefont
  {O'Regan}(2023)}]{Lambert2023}%
  \BibitemOpen
  \bibfield  {author} {\bibinfo {author} {\bibfnamefont {D.~S.}\ \bibnamefont
  {Lambert}}\ and\ \bibinfo {author} {\bibfnamefont {D.~D.}\ \bibnamefont
  {O'Regan}},\ }\bibfield  {title} {\bibinfo {title} {Use of {DFT+$U$+$J$} with
  linear response parameters to predict non-magnetic oxide band gaps with
  hybrid-functional accuracy},\ }\href
  {https://doi.org/10.1103/PhysRevResearch.5.013160} {\bibfield  {journal}
  {\bibinfo  {journal} {Phys. Rev. Res.}\ }\textbf {\bibinfo {volume} {5}},\
  \bibinfo {pages} {013160} (\bibinfo {year} {2023})}\BibitemShut {NoStop}%
\bibitem [{\citenamefont {Kam}\ \emph {et~al.}(2025)\citenamefont {Kam},
  \citenamefont {Binci}, \citenamefont {Kaplan}, \citenamefont {Persson},
  \citenamefont {Marzari},\ and\ \citenamefont {Ceder}}]{Kam2024}%
  \BibitemOpen
  \bibfield  {author} {\bibinfo {author} {\bibfnamefont {R.~L.}\ \bibnamefont
  {Kam}}, \bibinfo {author} {\bibfnamefont {L.}~\bibnamefont {Binci}}, \bibinfo
  {author} {\bibfnamefont {A.~D.}\ \bibnamefont {Kaplan}}, \bibinfo {author}
  {\bibfnamefont {K.~A.}\ \bibnamefont {Persson}}, \bibinfo {author}
  {\bibfnamefont {N.}~\bibnamefont {Marzari}},\ and\ \bibinfo {author}
  {\bibfnamefont {G.}~\bibnamefont {Ceder}},\ }\bibfield  {title} {\bibinfo
  {title} {Interplay between electron localization, magnetic order, and
  jahn-teller distortion dictates {LiMnO$_2$} phase stability},\ }\href
  {https://doi.org/10.1103/99jn-17v6} {\bibfield  {journal} {\bibinfo
  {journal} {Phys. Rev. B}\ }\textbf {\bibinfo {volume} {111}},\ \bibinfo
  {pages} {245132} (\bibinfo {year} {2025})}\BibitemShut {NoStop}%
\bibitem [{\citenamefont {Carta}\ \emph {et~al.}(2026)\citenamefont {Carta},
  \citenamefont {Panda},\ and\ \citenamefont {Ederer}}]{Carta2026}%
  \BibitemOpen
  \bibfield  {author} {\bibinfo {author} {\bibfnamefont {A.}~\bibnamefont
  {Carta}}, \bibinfo {author} {\bibfnamefont {A.}~\bibnamefont {Panda}},\ and\
  \bibinfo {author} {\bibfnamefont {C.}~\bibnamefont {Ederer}},\ }\bibfield
  {title} {\bibinfo {title} {Importance of ligand on-site interactions for the
  description of mott-insulators in {DFT+DMFT}},\ }\href
  {https://doi.org/10.1038/s41524-025-01928-4} {\bibfield  {journal} {\bibinfo
  {journal} {npj Comput. Mater.}\ }\textbf {\bibinfo {volume} {12}},\ \bibinfo
  {pages} {57} (\bibinfo {year} {2026})}\BibitemShut {NoStop}%
\bibitem [{\citenamefont {Löwdin}(1950)}]{Lowdin1950}%
  \BibitemOpen
  \bibfield  {author} {\bibinfo {author} {\bibfnamefont {P.-O.}\ \bibnamefont
  {Löwdin}},\ }\bibfield  {title} {\bibinfo {title} {On the
  non‐orthogonality problem connected with the use of atomic wave functions
  in the theory of molecules and crystals},\ }\href
  {https://doi.org/10.1063/1.1747632} {\bibfield  {journal} {\bibinfo
  {journal} {J. Chem. Phys.}\ }\textbf {\bibinfo {volume} {18}},\ \bibinfo
  {pages} {365} (\bibinfo {year} {1950})}\BibitemShut {NoStop}%
\bibitem [{\citenamefont {Mayer}(2002)}]{Mayer2002}%
  \BibitemOpen
  \bibfield  {author} {\bibinfo {author} {\bibfnamefont {I.}~\bibnamefont
  {Mayer}},\ }\bibfield  {title} {\bibinfo {title} {On löwdin's method of
  symmetric orthogonalization},\ }\href {https://doi.org/10.1002/qua.981}
  {\bibfield  {journal} {\bibinfo  {journal} {Int. J. Quantum Chem.}\ }\textbf
  {\bibinfo {volume} {90}},\ \bibinfo {pages} {63} (\bibinfo {year}
  {2002})}\BibitemShut {NoStop}%
\bibitem [{\citenamefont {Momma}\ and\ \citenamefont
  {Izumi}(2011)}]{Momma2011}%
  \BibitemOpen
  \bibfield  {author} {\bibinfo {author} {\bibfnamefont {K.}~\bibnamefont
  {Momma}}\ and\ \bibinfo {author} {\bibfnamefont {F.}~\bibnamefont {Izumi}},\
  }\bibfield  {title} {\bibinfo {title} {{VESTA3} for three-dimensional
  visualization of crystal, volumetric and morphology data},\ }\href
  {https://doi.org/10.1107/S0021889811038970} {\bibfield  {journal} {\bibinfo
  {journal} {J. Appl. Cryst.}\ }\textbf {\bibinfo {volume} {44}},\ \bibinfo
  {pages} {1272} (\bibinfo {year} {2011})}\BibitemShut {NoStop}%
\bibitem [{\citenamefont {Garcia}\ \emph {et~al.}(1995)\citenamefont {Garcia},
  \citenamefont {Barboux}, \citenamefont {Ribot}, \citenamefont {Kahn-Harari},
  \citenamefont {Mazerolles},\ and\ \citenamefont {Baffier}}]{Garcia1995}%
  \BibitemOpen
  \bibfield  {author} {\bibinfo {author} {\bibfnamefont {B.}~\bibnamefont
  {Garcia}}, \bibinfo {author} {\bibfnamefont {P.}~\bibnamefont {Barboux}},
  \bibinfo {author} {\bibfnamefont {F.}~\bibnamefont {Ribot}}, \bibinfo
  {author} {\bibfnamefont {A.}~\bibnamefont {Kahn-Harari}}, \bibinfo {author}
  {\bibfnamefont {L.}~\bibnamefont {Mazerolles}},\ and\ \bibinfo {author}
  {\bibfnamefont {N.}~\bibnamefont {Baffier}},\ }\bibfield  {title} {\bibinfo
  {title} {The structure of low temperature crystallized {LiCoO$_2$}},\ }\href
  {https://doi.org/10.1016/0167-2738(95)00117-O} {\bibfield  {journal}
  {\bibinfo  {journal} {Solid State Ion.}\ }\textbf {\bibinfo {volume} {80}},\
  \bibinfo {pages} {111} (\bibinfo {year} {1995})}\BibitemShut {NoStop}%
\bibitem [{\citenamefont {Antolini}(2004)}]{Antolini2004}%
  \BibitemOpen
  \bibfield  {author} {\bibinfo {author} {\bibfnamefont {E.}~\bibnamefont
  {Antolini}},\ }\bibfield  {title} {\bibinfo {title} {{LiCoO$_2$}: formation,
  structure, lithium and oxygen nonstoichiometry, electrochemical behavior, and
  transport properties},\ }\href {https://doi.org/10.1016/j.ssi.2004.04.003}
  {\bibfield  {journal} {\bibinfo  {journal} {Solid State Ion.}\ }\textbf
  {\bibinfo {volume} {170}},\ \bibinfo {pages} {159} (\bibinfo {year}
  {2004})}\BibitemShut {NoStop}%
\bibitem [{\citenamefont {van Elp}\ \emph {et~al.}(1991)\citenamefont {van
  Elp}, \citenamefont {Wieland}, \citenamefont {Eskes}, \citenamefont {Kuiper},
  \citenamefont {Sawatzky}, \citenamefont {de~Groot},\ and\ \citenamefont
  {Turner}}]{vanElp1991}%
  \BibitemOpen
  \bibfield  {author} {\bibinfo {author} {\bibfnamefont {J.}~\bibnamefont {van
  Elp}}, \bibinfo {author} {\bibfnamefont {J.~L.}\ \bibnamefont {Wieland}},
  \bibinfo {author} {\bibfnamefont {H.}~\bibnamefont {Eskes}}, \bibinfo
  {author} {\bibfnamefont {P.}~\bibnamefont {Kuiper}}, \bibinfo {author}
  {\bibfnamefont {G.~A.}\ \bibnamefont {Sawatzky}}, \bibinfo {author}
  {\bibfnamefont {F.~M.~F.}\ \bibnamefont {de~Groot}},\ and\ \bibinfo {author}
  {\bibfnamefont {T.~S.}\ \bibnamefont {Turner}},\ }\bibfield  {title}
  {\bibinfo {title} {Electronic structure of {CoO}, li-doped {CoO}, and
  {LiCoO$_2$}},\ }\href {https://doi.org/10.1103/PhysRevB.44.6090} {\bibfield
  {journal} {\bibinfo  {journal} {Phys. Rev. B}\ }\textbf {\bibinfo {volume}
  {44}},\ \bibinfo {pages} {6090} (\bibinfo {year} {1991})}\BibitemShut
  {NoStop}%
\bibitem [{\citenamefont {Ohzuku}\ \emph {et~al.}(1993)\citenamefont {Ohzuku},
  \citenamefont {Ueda}, \citenamefont {Nagayama}, \citenamefont {Iwakoshi},\
  and\ \citenamefont {Komori}}]{Ohzuku1993}%
  \BibitemOpen
  \bibfield  {author} {\bibinfo {author} {\bibfnamefont {T.}~\bibnamefont
  {Ohzuku}}, \bibinfo {author} {\bibfnamefont {A.}~\bibnamefont {Ueda}},
  \bibinfo {author} {\bibfnamefont {M.}~\bibnamefont {Nagayama}}, \bibinfo
  {author} {\bibfnamefont {Y.}~\bibnamefont {Iwakoshi}},\ and\ \bibinfo
  {author} {\bibfnamefont {H.}~\bibnamefont {Komori}},\ }\bibfield  {title}
  {\bibinfo {title} {Comparative study of {LiCoO$_2$},
  {LiNi$_{1/2}$Co$_{1/2}$O$_2$} and {LiNiO$_2$} for 4 volt secondary lithium
  cells},\ }\href {https://doi.org/10.1016/0013-4686(93)80046-3} {\bibfield
  {journal} {\bibinfo  {journal} {Electrochim. Acta}\ }\textbf {\bibinfo
  {volume} {38}},\ \bibinfo {pages} {1159} (\bibinfo {year}
  {1993})}\BibitemShut {NoStop}%
\bibitem [{\citenamefont {Amatucci}\ \emph {et~al.}(1996)\citenamefont
  {Amatucci}, \citenamefont {Tarascon},\ and\ \citenamefont
  {Klein}}]{Amatucci1996}%
  \BibitemOpen
  \bibfield  {author} {\bibinfo {author} {\bibfnamefont {G.~G.}\ \bibnamefont
  {Amatucci}}, \bibinfo {author} {\bibfnamefont {J.~M.}\ \bibnamefont
  {Tarascon}},\ and\ \bibinfo {author} {\bibfnamefont {L.~C.}\ \bibnamefont
  {Klein}},\ }\bibfield  {title} {\bibinfo {title} {Coo2, the end member of the
  {Li$_x$CoO$_2$} solid solution},\ }\href {https://doi.org/10.1149/1.1836594}
  {\bibfield  {journal} {\bibinfo  {journal} {J. Electrochem. Soc.}\ }\textbf
  {\bibinfo {volume} {143}},\ \bibinfo {pages} {1114} (\bibinfo {year}
  {1996})}\BibitemShut {NoStop}%
\bibitem [{\citenamefont {Motohashi}\ \emph {et~al.}(2007)\citenamefont
  {Motohashi}, \citenamefont {Katsumata}, \citenamefont {Ono}, \citenamefont
  {Kanno}, \citenamefont {Karppinen},\ and\ \citenamefont
  {Yamauchi}}]{Motohashi2007}%
  \BibitemOpen
  \bibfield  {author} {\bibinfo {author} {\bibfnamefont {T.}~\bibnamefont
  {Motohashi}}, \bibinfo {author} {\bibfnamefont {Y.}~\bibnamefont
  {Katsumata}}, \bibinfo {author} {\bibfnamefont {T.}~\bibnamefont {Ono}},
  \bibinfo {author} {\bibfnamefont {R.}~\bibnamefont {Kanno}}, \bibinfo
  {author} {\bibfnamefont {M.}~\bibnamefont {Karppinen}},\ and\ \bibinfo
  {author} {\bibfnamefont {H.}~\bibnamefont {Yamauchi}},\ }\bibfield  {title}
  {\bibinfo {title} {Synthesis and properties of {CoO$_2$}, the x = 0 end
  member of the {Li$_x$CoO$_2$} and {Na$_x$CoO$_2$} systems},\ }\href
  {https://doi.org/10.1021/cm0702464} {\bibfield  {journal} {\bibinfo
  {journal} {Chem. Mater.}\ }\textbf {\bibinfo {volume} {19}},\ \bibinfo
  {pages} {5063} (\bibinfo {year} {2007})}\BibitemShut {NoStop}%
\bibitem [{\citenamefont {Gerken}\ \emph {et~al.}(2011)\citenamefont {Gerken},
  \citenamefont {McAlpin}, \citenamefont {Chen}, \citenamefont {Rigsby},
  \citenamefont {Casey}, \citenamefont {Britt},\ and\ \citenamefont
  {Stahl}}]{Gerken2011}%
  \BibitemOpen
  \bibfield  {author} {\bibinfo {author} {\bibfnamefont {J.~B.}\ \bibnamefont
  {Gerken}}, \bibinfo {author} {\bibfnamefont {J.~G.}\ \bibnamefont {McAlpin}},
  \bibinfo {author} {\bibfnamefont {J.~Y.~C.}\ \bibnamefont {Chen}}, \bibinfo
  {author} {\bibfnamefont {M.~L.}\ \bibnamefont {Rigsby}}, \bibinfo {author}
  {\bibfnamefont {W.~H.}\ \bibnamefont {Casey}}, \bibinfo {author}
  {\bibfnamefont {R.~D.}\ \bibnamefont {Britt}},\ and\ \bibinfo {author}
  {\bibfnamefont {S.~S.}\ \bibnamefont {Stahl}},\ }\bibfield  {title} {\bibinfo
  {title} {Electrochemical water oxidation with cobalt-based electrocatalysts
  from {pH} 0--14: The thermodynamic basis for catalyst structure, stability,
  and activity},\ }\href {https://doi.org/10.1021/ja205647m} {\bibfield
  {journal} {\bibinfo  {journal} {J. Am. Chem. Soc.}\ }\textbf {\bibinfo
  {volume} {133}},\ \bibinfo {pages} {14431} (\bibinfo {year}
  {2011})}\BibitemShut {NoStop}%
\bibitem [{\citenamefont {de~Vaulx}\ \emph {et~al.}(2007)\citenamefont
  {de~Vaulx}, \citenamefont {Julien}, \citenamefont {Berthier}, \citenamefont
  {Hébert}, \citenamefont {Pralong},\ and\ \citenamefont
  {Maignan}}]{deVaulx2007}%
  \BibitemOpen
  \bibfield  {author} {\bibinfo {author} {\bibfnamefont {C.}~\bibnamefont
  {de~Vaulx}}, \bibinfo {author} {\bibfnamefont {M.-H.}\ \bibnamefont
  {Julien}}, \bibinfo {author} {\bibfnamefont {C.}~\bibnamefont {Berthier}},
  \bibinfo {author} {\bibfnamefont {S.}~\bibnamefont {Hébert}}, \bibinfo
  {author} {\bibfnamefont {V.}~\bibnamefont {Pralong}},\ and\ \bibinfo {author}
  {\bibfnamefont {A.}~\bibnamefont {Maignan}},\ }\bibfield  {title} {\bibinfo
  {title} {Electronic correlations in {CoO$_2$}, the parent compound of
  triangular cobaltates},\ }\href
  {https://doi.org/10.1103/PhysRevLett.98.246402} {\bibfield  {journal}
  {\bibinfo  {journal} {Phys. Rev. Lett.}\ }\textbf {\bibinfo {volume} {98}},\
  \bibinfo {pages} {246402} (\bibinfo {year} {2007})}\BibitemShut {NoStop}%
\bibitem [{\citenamefont {Mattila}\ and\ \citenamefont
  {Karttunen}(2022)}]{Mattila2022}%
  \BibitemOpen
  \bibfield  {author} {\bibinfo {author} {\bibfnamefont {N.}~\bibnamefont
  {Mattila}}\ and\ \bibinfo {author} {\bibfnamefont {A.}~\bibnamefont
  {Karttunen}},\ }\bibfield  {title} {\bibinfo {title} {Electronic properties
  and lattice dynamics of {Li$_x$CoO$_2$} and {Na$_x$CoO$_2$} (x = 0, 0.5, 1)
  studied by hybrid density functional theory},\ }\bibfield  {journal}
  {\bibinfo  {journal} {Phys. Status Solidi B}\ }\textbf {\bibinfo {volume}
  {259}},\ \href {https://doi.org/10.1002/pssb.202100665}
  {10.1002/pssb.202100665} (\bibinfo {year} {2022})\BibitemShut {NoStop}%
\bibitem [{supplementaryinformation()}]{Supplementary_Information}%
  \BibitemOpen
  supplementaryinformation,\ \href@noop {} {}\BibitemShut {NoStop}%
\bibitem [{\citenamefont {Sit}\ \emph {et~al.}(2011)\citenamefont {Sit},
  \citenamefont {Car}, \citenamefont {Cohen},\ and\ \citenamefont
  {Selloni}}]{Sit2011}%
  \BibitemOpen
  \bibfield  {author} {\bibinfo {author} {\bibfnamefont {P.~H.-L.}\
  \bibnamefont {Sit}}, \bibinfo {author} {\bibfnamefont {R.}~\bibnamefont
  {Car}}, \bibinfo {author} {\bibfnamefont {M.~H.}\ \bibnamefont {Cohen}},\
  and\ \bibinfo {author} {\bibfnamefont {A.}~\bibnamefont {Selloni}},\
  }\bibfield  {title} {\bibinfo {title} {Simple, unambiguous theoretical
  approach to oxidation state determination via first-principles
  calculations},\ }\href {https://doi.org/10.1021/ic2013107} {\bibfield
  {journal} {\bibinfo  {journal} {Inorg. Chem.}\ }\textbf {\bibinfo {volume}
  {50}},\ \bibinfo {pages} {10259} (\bibinfo {year} {2011})}\BibitemShut
  {NoStop}%
\bibitem [{\citenamefont {Ensling}\ \emph {et~al.}(2010)\citenamefont
  {Ensling}, \citenamefont {Thissen}, \citenamefont {Laubach}, \citenamefont
  {Schmidt},\ and\ \citenamefont {Jaegermann}}]{Ensling2010}%
  \BibitemOpen
  \bibfield  {author} {\bibinfo {author} {\bibfnamefont {D.}~\bibnamefont
  {Ensling}}, \bibinfo {author} {\bibfnamefont {A.}~\bibnamefont {Thissen}},
  \bibinfo {author} {\bibfnamefont {S.}~\bibnamefont {Laubach}}, \bibinfo
  {author} {\bibfnamefont {P.~C.}\ \bibnamefont {Schmidt}},\ and\ \bibinfo
  {author} {\bibfnamefont {W.}~\bibnamefont {Jaegermann}},\ }\bibfield  {title}
  {\bibinfo {title} {Electronic structure of {LiCoO}$_2$ thin films: A combined
  photoemission spectroscopy and density functional theory study},\ }\href
  {https://doi.org/10.1103/PhysRevB.82.195431} {\bibfield  {journal} {\bibinfo
  {journal} {Phys. Rev. B}\ }\textbf {\bibinfo {volume} {82}},\ \bibinfo
  {pages} {195431} (\bibinfo {year} {2010})}\BibitemShut {NoStop}%
\bibitem [{\citenamefont {Skone}\ \emph {et~al.}(2014)\citenamefont {Skone},
  \citenamefont {Govoni},\ and\ \citenamefont {Galli}}]{Skone2014}%
  \BibitemOpen
  \bibfield  {author} {\bibinfo {author} {\bibfnamefont {J.~H.}\ \bibnamefont
  {Skone}}, \bibinfo {author} {\bibfnamefont {M.}~\bibnamefont {Govoni}},\ and\
  \bibinfo {author} {\bibfnamefont {G.}~\bibnamefont {Galli}},\ }\bibfield
  {title} {\bibinfo {title} {{Self-consistent hybrid functional for condensed
  systems}},\ }\href {https://doi.org/10.1103/PhysRevB.89.195112} {\bibfield
  {journal} {\bibinfo  {journal} {Phys. Rev. B}\ }\textbf {\bibinfo {volume}
  {89}},\ \bibinfo {pages} {195112} (\bibinfo {year} {2014})}\BibitemShut
  {NoStop}%
\bibitem [{\citenamefont {Takahashi}\ \emph {et~al.}(2020)\citenamefont
  {Takahashi}, \citenamefont {Kumagai}, \citenamefont {Miyamoto}, \citenamefont
  {Mochizuki},\ and\ \citenamefont {Oba}}]{Takahashi:2020}%
  \BibitemOpen
  \bibfield  {author} {\bibinfo {author} {\bibfnamefont {A.}~\bibnamefont
  {Takahashi}}, \bibinfo {author} {\bibfnamefont {Y.}~\bibnamefont {Kumagai}},
  \bibinfo {author} {\bibfnamefont {J.}~\bibnamefont {Miyamoto}}, \bibinfo
  {author} {\bibfnamefont {Y.}~\bibnamefont {Mochizuki}},\ and\ \bibinfo
  {author} {\bibfnamefont {F.}~\bibnamefont {Oba}},\ }\bibfield  {title}
  {\bibinfo {title} {{Machine learning models for predicting the dielectric
  constants of oxides based on high-throughput first-principles
  calculations}},\ }\href {https://doi.org/10.1103/PhysRevMaterials.4.103801}
  {\bibfield  {journal} {\bibinfo  {journal} {Phys. Rev. Mater.}\ }\textbf
  {\bibinfo {volume} {4}},\ \bibinfo {pages} {103801} (\bibinfo {year}
  {2020})}\BibitemShut {NoStop}%
\bibitem [{\citenamefont {Bajaj}\ and\ \citenamefont
  {Kulik}(2021)}]{Bajaj2021}%
  \BibitemOpen
  \bibfield  {author} {\bibinfo {author} {\bibfnamefont {A.}~\bibnamefont
  {Bajaj}}\ and\ \bibinfo {author} {\bibfnamefont {H.~J.}\ \bibnamefont
  {Kulik}},\ }\bibfield  {title} {\bibinfo {title} {Molecular {DFT+$U$}: A
  transferable, low-cost approach to eliminate delocalization error},\ }\href
  {https://doi.org/10.1021/acs.jpclett.1c00796} {\bibfield  {journal} {\bibinfo
   {journal} {J. Phys. Chem. Lett.}\ }\textbf {\bibinfo {volume} {12}},\
  \bibinfo {pages} {3633} (\bibinfo {year} {2021})}\BibitemShut {NoStop}%
\bibitem [{\citenamefont {Dabo}\ \emph {et~al.}(2007)\citenamefont {Dabo},
  \citenamefont {Wieckowski},\ and\ \citenamefont {Marzari}}]{Dabo2007}%
  \BibitemOpen
  \bibfield  {author} {\bibinfo {author} {\bibfnamefont {I.}~\bibnamefont
  {Dabo}}, \bibinfo {author} {\bibfnamefont {A.}~\bibnamefont {Wieckowski}},\
  and\ \bibinfo {author} {\bibfnamefont {N.}~\bibnamefont {Marzari}},\
  }\bibfield  {title} {\bibinfo {title} {Vibrational recognition of adsorption
  sites for co on platinum and platinum-ruthenium surfaces},\ }\href
  {https://doi.org/10.1021/ja067944u} {\bibfield  {journal} {\bibinfo
  {journal} {J. Am. Chem. Soc.}\ }\textbf {\bibinfo {volume} {129}},\ \bibinfo
  {pages} {11045} (\bibinfo {year} {2007})}\BibitemShut {NoStop}%
\bibitem [{\citenamefont {Ting}\ and\ \citenamefont
  {Kowalski}(2023)}]{Ting2023}%
  \BibitemOpen
  \bibfield  {author} {\bibinfo {author} {\bibfnamefont {Y.-Y.}\ \bibnamefont
  {Ting}}\ and\ \bibinfo {author} {\bibfnamefont {P.~M.}\ \bibnamefont
  {Kowalski}},\ }\bibfield  {title} {\bibinfo {title} {Refined {DFT+$U$} method
  for computation of layered oxide cathode materials},\ }\href
  {https://doi.org/10.1016/j.electacta.2023.141912} {\bibfield  {journal}
  {\bibinfo  {journal} {Electrochim. Acta}\ }\textbf {\bibinfo {volume}
  {443}},\ \bibinfo {pages} {141912} (\bibinfo {year} {2023})}\BibitemShut
  {NoStop}%
\bibitem [{\citenamefont {Andolfatto}\ \emph {et~al.}()\citenamefont
  {Andolfatto}, \citenamefont {Botti}, \citenamefont {Sanella}, \citenamefont
  {Mocatti}, \citenamefont {Volpato}, \citenamefont {Marini}, \citenamefont
  {Calandra}, \citenamefont {Ederer},\ and\ \citenamefont
  {Timrov}}]{Andolfatto2026}%
  \BibitemOpen
  \bibfield  {author} {\bibinfo {author} {\bibfnamefont {M.}~\bibnamefont
  {Andolfatto}}, \bibinfo {author} {\bibfnamefont {E.}~\bibnamefont {Botti}},
  \bibinfo {author} {\bibfnamefont {V.}~\bibnamefont {Sanella}}, \bibinfo
  {author} {\bibfnamefont {S.}~\bibnamefont {Mocatti}}, \bibinfo {author}
  {\bibfnamefont {G.}~\bibnamefont {Volpato}}, \bibinfo {author} {\bibfnamefont
  {G.}~\bibnamefont {Marini}}, \bibinfo {author} {\bibfnamefont
  {M.}~\bibnamefont {Calandra}}, \bibinfo {author} {\bibfnamefont
  {C.}~\bibnamefont {Ederer}},\ and\ \bibinfo {author} {\bibfnamefont
  {I.}~\bibnamefont {Timrov}},\ }\href@noop {} {\bibinfo  {journal} {in
  preparation (2026)}\ }\BibitemShut {NoStop}%
\bibitem [{\citenamefont {Macke}\ \emph {et~al.}(2024)\citenamefont {Macke},
  \citenamefont {Timrov}, \citenamefont {Marzari},\ and\ \citenamefont
  {Ciacchi}}]{Macke2024}%
  \BibitemOpen
\bibfield  {journal} {  }\bibfield  {author} {\bibinfo {author} {\bibfnamefont
  {E.}~\bibnamefont {Macke}}, \bibinfo {author} {\bibfnamefont
  {I.}~\bibnamefont {Timrov}}, \bibinfo {author} {\bibfnamefont
  {N.}~\bibnamefont {Marzari}},\ and\ \bibinfo {author} {\bibfnamefont {L.~C.}\
  \bibnamefont {Ciacchi}},\ }\bibfield  {title} {\bibinfo {title}
  {Orbital-resolved {DFT+$U$} for molecules and solids},\ }\href
  {https://doi.org/10.1021/acs.jctc.3c01403} {\bibfield  {journal} {\bibinfo
  {journal} {J. Chem. Theory Comput.}\ }\textbf {\bibinfo {volume} {20}},\
  \bibinfo {pages} {4824} (\bibinfo {year} {2024})}\BibitemShut {NoStop}%
\bibitem [{\citenamefont {Zunger}\ \emph {et~al.}(1990)\citenamefont {Zunger},
  \citenamefont {Wei}, \citenamefont {Ferreira},\ and\ \citenamefont
  {Bernard}}]{Zunger1990}%
  \BibitemOpen
  \bibfield  {author} {\bibinfo {author} {\bibfnamefont {A.}~\bibnamefont
  {Zunger}}, \bibinfo {author} {\bibfnamefont {S.-H.}\ \bibnamefont {Wei}},
  \bibinfo {author} {\bibfnamefont {L.~G.}\ \bibnamefont {Ferreira}},\ and\
  \bibinfo {author} {\bibfnamefont {J.~E.}\ \bibnamefont {Bernard}},\
  }\bibfield  {title} {\bibinfo {title} {{Special quasirandom structures}},\
  }\href {https://doi.org/10.1103/PhysRevLett.65.353} {\bibfield  {journal}
  {\bibinfo  {journal} {Phys. Rev. Lett.}\ }\textbf {\bibinfo {volume} {65}},\
  \bibinfo {pages} {353} (\bibinfo {year} {1990})}\BibitemShut {NoStop}%
\bibitem [{\citenamefont {Alling}\ \emph {et~al.}(2010)\citenamefont {Alling},
  \citenamefont {Marten},\ and\ \citenamefont {Abrikosov}}]{Alling2010}%
  \BibitemOpen
  \bibfield  {author} {\bibinfo {author} {\bibfnamefont {B.}~\bibnamefont
  {Alling}}, \bibinfo {author} {\bibfnamefont {T.}~\bibnamefont {Marten}},\
  and\ \bibinfo {author} {\bibfnamefont {I.~A.}\ \bibnamefont {Abrikosov}},\
  }\bibfield  {title} {\bibinfo {title} {{Effect of magnetic disorder and
  strong electron correlations on the thermodynamics of CrN}},\ }\href
  {https://doi.org/10.1103/PhysRevB.82.184430} {\bibfield  {journal} {\bibinfo
  {journal} {Phys. Rev. B}\ }\textbf {\bibinfo {volume} {82}},\ \bibinfo
  {pages} {184430} (\bibinfo {year} {2010})}\BibitemShut {NoStop}%
\bibitem [{\citenamefont {Ponet}\ \emph {et~al.}(2024)\citenamefont {Ponet},
  \citenamefont {Lucente},\ and\ \citenamefont {Marzari}}]{Ponet2024}%
  \BibitemOpen
  \bibfield  {author} {\bibinfo {author} {\bibfnamefont {L.}~\bibnamefont
  {Ponet}}, \bibinfo {author} {\bibfnamefont {E.~D.}\ \bibnamefont {Lucente}},\
  and\ \bibinfo {author} {\bibfnamefont {N.}~\bibnamefont {Marzari}},\
  }\bibfield  {title} {\bibinfo {title} {The energy landscape of magnetic
  materials},\ }\href {https://doi.org/10.1038/s41524-024-01310-w} {\bibfield
  {journal} {\bibinfo  {journal} {npj Comput. Mater.}\ }\textbf {\bibinfo
  {volume} {10}},\ \bibinfo {pages} {151} (\bibinfo {year} {2024})}\BibitemShut
  {NoStop}%
\bibitem [{\citenamefont {Haddadi}\ \emph {et~al.}(2026)\citenamefont
  {Haddadi}, \citenamefont {Campi}, \citenamefont {{dos Santos}}, \citenamefont
  {Mounet}, \citenamefont {Ponet}, \citenamefont {Marzari},\ and\ \citenamefont
  {Gibertini}}]{Haddadi2026}%
  \BibitemOpen
  \bibfield  {author} {\bibinfo {author} {\bibfnamefont {F.}~\bibnamefont
  {Haddadi}}, \bibinfo {author} {\bibfnamefont {D.}~\bibnamefont {Campi}},
  \bibinfo {author} {\bibfnamefont {F.}~\bibnamefont {{dos Santos}}}, \bibinfo
  {author} {\bibfnamefont {N.}~\bibnamefont {Mounet}}, \bibinfo {author}
  {\bibfnamefont {L.}~\bibnamefont {Ponet}}, \bibinfo {author} {\bibfnamefont
  {N.}~\bibnamefont {Marzari}},\ and\ \bibinfo {author} {\bibfnamefont
  {M.}~\bibnamefont {Gibertini}},\ }\bibfield  {title} {\bibinfo {title}
  {{Exploring the Magnetic Landscape of Easily Exfoliable Two-Dimensional
  Materials}},\ }\href {https://doi.org/10.1021/acsnano.5c16067} {\bibfield
  {journal} {\bibinfo  {journal} {ACS Nano}\ }\textbf {\bibinfo {volume}
  {20}},\ \bibinfo {pages} {13528} (\bibinfo {year} {2026})}\BibitemShut
  {NoStop}%
\bibitem [{\citenamefont {Meredig}\ \emph {et~al.}(2010)\citenamefont
  {Meredig}, \citenamefont {Thompson}, \citenamefont {Hansen}, \citenamefont
  {Wolverton},\ and\ \citenamefont {van~de Walle}}]{Meredig:2010}%
  \BibitemOpen
  \bibfield  {author} {\bibinfo {author} {\bibfnamefont {B.}~\bibnamefont
  {Meredig}}, \bibinfo {author} {\bibfnamefont {A.}~\bibnamefont {Thompson}},
  \bibinfo {author} {\bibfnamefont {H.~A.}\ \bibnamefont {Hansen}}, \bibinfo
  {author} {\bibfnamefont {C.}~\bibnamefont {Wolverton}},\ and\ \bibinfo
  {author} {\bibfnamefont {A.}~\bibnamefont {van~de Walle}},\ }\bibfield
  {title} {\bibinfo {title} {{Method for locating low-energy solutions within
  \${\textbackslash}text\{{DFT}\}+{U}\$}},\ }\href
  {https://doi.org/10.1103/PhysRevB.82.195128} {\bibfield  {journal} {\bibinfo
  {journal} {Phys. Rev. B}\ }\textbf {\bibinfo {volume} {82}},\ \bibinfo
  {pages} {195128} (\bibinfo {year} {2010})}\BibitemShut {NoStop}%
\bibitem [{\citenamefont {Di~Lucente}\ \emph {et~al.}(2026)\citenamefont
  {Di~Lucente}, \citenamefont {Santos},\ and\ \citenamefont
  {Marzari}}]{DiLucente:2026}%
  \BibitemOpen
  \bibfield  {author} {\bibinfo {author} {\bibfnamefont {E.}~\bibnamefont
  {Di~Lucente}}, \bibinfo {author} {\bibfnamefont {F.~J.~d.}\ \bibnamefont
  {Santos}},\ and\ \bibinfo {author} {\bibfnamefont {N.}~\bibnamefont
  {Marzari}},\ }\bibfield  {title} {\bibinfo {title} {{Spin fluctuations steer
  the electronic behavior in the ${\mathrm{FeSb}}_{3}$ skutterudite}},\ }\href
  {https://doi.org/10.1103/lyy4-cmf6} {\bibfield  {journal} {\bibinfo
  {journal} {Phys. Rev. Res.}\ }\textbf {\bibinfo {volume} {8}},\ \bibinfo
  {pages} {013174} (\bibinfo {year} {2026})}\BibitemShut {NoStop}%
\bibitem [{\citenamefont {Lee}\ and\ \citenamefont {Pickett}(2005)}]{Lee2005}%
  \BibitemOpen
  \bibfield  {author} {\bibinfo {author} {\bibfnamefont {K.-W.}\ \bibnamefont
  {Lee}}\ and\ \bibinfo {author} {\bibfnamefont {W.}~\bibnamefont {Pickett}},\
  }\bibfield  {title} {\bibinfo {title} {{Na$_x$CoO$_2$ in the $x \rightarrow
  0$ regime: Coupling of structure and correlation effects}},\ }\href
  {https://doi.org/10.1103/PhysRevB.72.115110} {\bibfield  {journal} {\bibinfo
  {journal} {Phys. Rev. B}\ }\textbf {\bibinfo {volume} {72}},\ \bibinfo
  {pages} {115110} (\bibinfo {year} {2005})}\BibitemShut {NoStop}%
\bibitem [{\citenamefont {Iv\'ady}\ \emph {et~al.}(2014)\citenamefont
  {Iv\'ady}, \citenamefont {Armiento}, \citenamefont {Sz\'asz}, \citenamefont
  {Janz\'en}, \citenamefont {Gali},\ and\ \citenamefont
  {Abrikosov}}]{Ivady:2014}%
  \BibitemOpen
  \bibfield  {author} {\bibinfo {author} {\bibfnamefont {V.}~\bibnamefont
  {Iv\'ady}}, \bibinfo {author} {\bibfnamefont {R.}~\bibnamefont {Armiento}},
  \bibinfo {author} {\bibfnamefont {K.}~\bibnamefont {Sz\'asz}}, \bibinfo
  {author} {\bibfnamefont {E.}~\bibnamefont {Janz\'en}}, \bibinfo {author}
  {\bibfnamefont {A.}~\bibnamefont {Gali}},\ and\ \bibinfo {author}
  {\bibfnamefont {I.~A.}\ \bibnamefont {Abrikosov}},\ }\bibfield  {title}
  {\bibinfo {title} {{Theoretical unification of hybrid-DFT and $\text{DFT} +
  U$ methods for the treatment of localized orbitals}},\ }\href
  {https://doi.org/10.1103/PhysRevB.90.035146} {\bibfield  {journal} {\bibinfo
  {journal} {Phys. Rev. B}\ }\textbf {\bibinfo {volume} {90}},\ \bibinfo
  {pages} {035146} (\bibinfo {year} {2014})}\BibitemShut {NoStop}%
\bibitem [{\citenamefont {Chiarotti}\ \emph {et~al.}(2024)\citenamefont
  {Chiarotti}, \citenamefont {Ferretti},\ and\ \citenamefont
  {Marzari}}]{Chiarotti2024}%
  \BibitemOpen
  \bibfield  {author} {\bibinfo {author} {\bibfnamefont {T.}~\bibnamefont
  {Chiarotti}}, \bibinfo {author} {\bibfnamefont {A.}~\bibnamefont
  {Ferretti}},\ and\ \bibinfo {author} {\bibfnamefont {N.}~\bibnamefont
  {Marzari}},\ }\bibfield  {title} {\bibinfo {title} {{Energies and spectra of
  solids from the algorithmic inversion of dynamical Hubbard functionals}},\
  }\href {https://doi.org/10.1103/PhysRevResearch.6.L032023} {\bibfield
  {journal} {\bibinfo  {journal} {Phys. Rev. Res.}\ }\textbf {\bibinfo {volume}
  {6}},\ \bibinfo {pages} {L032023} (\bibinfo {year} {2024})}\BibitemShut
  {NoStop}%
\bibitem [{\citenamefont {Mahajan}\ \emph {et~al.}(2021)\citenamefont
  {Mahajan}, \citenamefont {Timrov}, \citenamefont {Marzari},\ and\
  \citenamefont {Kashyap}}]{Mahajan2021}%
  \BibitemOpen
  \bibfield  {author} {\bibinfo {author} {\bibfnamefont {R.}~\bibnamefont
  {Mahajan}}, \bibinfo {author} {\bibfnamefont {I.}~\bibnamefont {Timrov}},
  \bibinfo {author} {\bibfnamefont {N.}~\bibnamefont {Marzari}},\ and\ \bibinfo
  {author} {\bibfnamefont {A.}~\bibnamefont {Kashyap}},\ }\bibfield  {title}
  {\bibinfo {title} {Importance of intersite hubbard interactions in
  {$\beta$-MnO$_2$}: A first-principles {DFT+$U$+$V$} study},\ }\href
  {https://doi.org/10.1103/PhysRevMaterials.5.104402} {\bibfield  {journal}
  {\bibinfo  {journal} {Phys. Rev. Mater.}\ }\textbf {\bibinfo {volume} {5}},\
  \bibinfo {pages} {104402} (\bibinfo {year} {2021})}\BibitemShut {NoStop}%
\bibitem [{\citenamefont {Mahajan}\ \emph {et~al.}(2022)\citenamefont
  {Mahajan}, \citenamefont {Kashyap},\ and\ \citenamefont
  {Timrov}}]{Mahajan2022}%
  \BibitemOpen
  \bibfield  {author} {\bibinfo {author} {\bibfnamefont {R.}~\bibnamefont
  {Mahajan}}, \bibinfo {author} {\bibfnamefont {A.}~\bibnamefont {Kashyap}},\
  and\ \bibinfo {author} {\bibfnamefont {I.}~\bibnamefont {Timrov}},\
  }\bibfield  {title} {\bibinfo {title} {Pivotal role of intersite hubbard
  interactions in {Fe}-doped {$\alpha$-MnO$_2$}},\ }\href
  {https://doi.org/10.1021/acs.jpcc.2c04767} {\bibfield  {journal} {\bibinfo
  {journal} {J. Phys. Chem. C}\ }\textbf {\bibinfo {volume} {126}},\ \bibinfo
  {pages} {14353} (\bibinfo {year} {2022})}\BibitemShut {NoStop}%
\bibitem [{\citenamefont {Gebreyesus}\ \emph {et~al.}(2023)\citenamefont
  {Gebreyesus}, \citenamefont {Bastonero}, \citenamefont {Kotiuga},
  \citenamefont {Marzari},\ and\ \citenamefont {Timrov}}]{Gebreyesus2023}%
  \BibitemOpen
  \bibfield  {author} {\bibinfo {author} {\bibfnamefont {G.}~\bibnamefont
  {Gebreyesus}}, \bibinfo {author} {\bibfnamefont {L.}~\bibnamefont
  {Bastonero}}, \bibinfo {author} {\bibfnamefont {M.}~\bibnamefont {Kotiuga}},
  \bibinfo {author} {\bibfnamefont {N.}~\bibnamefont {Marzari}},\ and\ \bibinfo
  {author} {\bibfnamefont {I.}~\bibnamefont {Timrov}},\ }\bibfield  {title}
  {\bibinfo {title} {Understanding the role of hubbard corrections in the
  rhombohedral phase of {BaTiO$_3$}},\ }\href
  {https://doi.org/10.1103/PhysRevB.108.235171} {\bibfield  {journal} {\bibinfo
   {journal} {Phys. Rev. B}\ }\textbf {\bibinfo {volume} {108}},\ \bibinfo
  {pages} {235171} (\bibinfo {year} {2023})}\BibitemShut {NoStop}%
\bibitem [{\citenamefont {Binci}\ \emph {et~al.}(2023)\citenamefont {Binci},
  \citenamefont {Kotiuga}, \citenamefont {Timrov},\ and\ \citenamefont
  {Marzari}}]{Binci2023}%
  \BibitemOpen
  \bibfield  {author} {\bibinfo {author} {\bibfnamefont {L.}~\bibnamefont
  {Binci}}, \bibinfo {author} {\bibfnamefont {M.}~\bibnamefont {Kotiuga}},
  \bibinfo {author} {\bibfnamefont {I.}~\bibnamefont {Timrov}},\ and\ \bibinfo
  {author} {\bibfnamefont {N.}~\bibnamefont {Marzari}},\ }\bibfield  {title}
  {\bibinfo {title} {Hybridization driving distortions and multiferroicity in
  rare-earth nickelates},\ }\href
  {https://doi.org/10.1103/PhysRevResearch.5.033146} {\bibfield  {journal}
  {\bibinfo  {journal} {Phys. Rev. Research}\ }\textbf {\bibinfo {volume}
  {5}},\ \bibinfo {pages} {033146} (\bibinfo {year} {2023})}\BibitemShut
  {NoStop}%
\bibitem [{\citenamefont {Gelin}\ \emph {et~al.}(2024)\citenamefont {Gelin},
  \citenamefont {Kirchner-Hall}, \citenamefont {Katzbaer}, \citenamefont
  {Theibault}, \citenamefont {Xiong}, \citenamefont {Zhao}, \citenamefont
  {Khan}, \citenamefont {Andrewlavage}, \citenamefont {Orbe}, \citenamefont
  {Baksa}, \citenamefont {Cococcioni}, \citenamefont {Timrov}, \citenamefont
  {Campbell}, \citenamefont {Abruna}, \citenamefont {Schaak},\ and\
  \citenamefont {Dabo}}]{Gelin2024}%
  \BibitemOpen
  \bibfield  {author} {\bibinfo {author} {\bibfnamefont {S.}~\bibnamefont
  {Gelin}}, \bibinfo {author} {\bibfnamefont {N.~E.}\ \bibnamefont
  {Kirchner-Hall}}, \bibinfo {author} {\bibfnamefont {R.~R.}\ \bibnamefont
  {Katzbaer}}, \bibinfo {author} {\bibfnamefont {M.~J.}\ \bibnamefont
  {Theibault}}, \bibinfo {author} {\bibfnamefont {Y.}~\bibnamefont {Xiong}},
  \bibinfo {author} {\bibfnamefont {W.}~\bibnamefont {Zhao}}, \bibinfo {author}
  {\bibfnamefont {M.~M.}\ \bibnamefont {Khan}}, \bibinfo {author}
  {\bibfnamefont {E.}~\bibnamefont {Andrewlavage}}, \bibinfo {author}
  {\bibfnamefont {P.}~\bibnamefont {Orbe}}, \bibinfo {author} {\bibfnamefont
  {S.~M.}\ \bibnamefont {Baksa}}, \bibinfo {author} {\bibfnamefont
  {M.}~\bibnamefont {Cococcioni}}, \bibinfo {author} {\bibfnamefont
  {I.}~\bibnamefont {Timrov}}, \bibinfo {author} {\bibfnamefont
  {Q.}~\bibnamefont {Campbell}}, \bibinfo {author} {\bibfnamefont
  {H.}~\bibnamefont {Abruna}}, \bibinfo {author} {\bibfnamefont {R.~E.}\
  \bibnamefont {Schaak}},\ and\ \bibinfo {author} {\bibfnamefont
  {I.}~\bibnamefont {Dabo}},\ }\bibfield  {title} {\bibinfo {title} {Ternary
  oxides of {$s-$} and {$p-$}block metals for photocatalytic solar-to-hydrogen
  conversion},\ }\href {https://doi.org/10.1103/PRXEnergy.3.013007} {\bibfield
  {journal} {\bibinfo  {journal} {PRX Energy}\ }\textbf {\bibinfo {volume}
  {3}},\ \bibinfo {pages} {013007} (\bibinfo {year} {2024})}\BibitemShut
  {NoStop}%
\bibitem [{\citenamefont {Haddadi}\ \emph {et~al.}(2024)\citenamefont
  {Haddadi}, \citenamefont {Linscott}, \citenamefont {Timrov}, \citenamefont
  {Marzari},\ and\ \citenamefont {Gibertini}}]{Haddadi2024}%
  \BibitemOpen
  \bibfield  {author} {\bibinfo {author} {\bibfnamefont {F.}~\bibnamefont
  {Haddadi}}, \bibinfo {author} {\bibfnamefont {E.}~\bibnamefont {Linscott}},
  \bibinfo {author} {\bibfnamefont {I.}~\bibnamefont {Timrov}}, \bibinfo
  {author} {\bibfnamefont {N.}~\bibnamefont {Marzari}},\ and\ \bibinfo {author}
  {\bibfnamefont {M.}~\bibnamefont {Gibertini}},\ }\bibfield  {title} {\bibinfo
  {title} {On-site and inter-site hubbard corrections in magnetic monolayers:
  The case of {FePS$_3$} and {CrI$_3$}},\ }\href
  {https://doi.org/10.1103/PhysRevMaterials.8.014007} {\bibfield  {journal}
  {\bibinfo  {journal} {Phys. Rev. Mater.}\ }\textbf {\bibinfo {volume} {8}},\
  \bibinfo {pages} {014007} (\bibinfo {year} {2024})}\BibitemShut {NoStop}%
\bibitem [{\citenamefont {Bonfà}\ \emph {et~al.}(2024)\citenamefont {Bonfà},
  \citenamefont {Onuorah}, \citenamefont {Lang}, \citenamefont {Timrov},
  \citenamefont {Monacelli}, \citenamefont {Wang}, \citenamefont {Sun},
  \citenamefont {Petracic}, \citenamefont {Pizzi}, \citenamefont {Marzari},
  \citenamefont {Blundell},\ and\ \citenamefont {Renzi}}]{Bonfa2024}%
  \BibitemOpen
  \bibfield  {author} {\bibinfo {author} {\bibfnamefont {P.}~\bibnamefont
  {Bonfà}}, \bibinfo {author} {\bibfnamefont {I.~J.}\ \bibnamefont {Onuorah}},
  \bibinfo {author} {\bibfnamefont {F.}~\bibnamefont {Lang}}, \bibinfo {author}
  {\bibfnamefont {I.}~\bibnamefont {Timrov}}, \bibinfo {author} {\bibfnamefont
  {L.}~\bibnamefont {Monacelli}}, \bibinfo {author} {\bibfnamefont
  {C.}~\bibnamefont {Wang}}, \bibinfo {author} {\bibfnamefont {X.}~\bibnamefont
  {Sun}}, \bibinfo {author} {\bibfnamefont {O.}~\bibnamefont {Petracic}},
  \bibinfo {author} {\bibfnamefont {G.}~\bibnamefont {Pizzi}}, \bibinfo
  {author} {\bibfnamefont {N.}~\bibnamefont {Marzari}}, \bibinfo {author}
  {\bibfnamefont {S.~J.}\ \bibnamefont {Blundell}},\ and\ \bibinfo {author}
  {\bibfnamefont {R.~D.}\ \bibnamefont {Renzi}},\ }\bibfield  {title} {\bibinfo
  {title} {Magnetostriction-driven muon localisation in an antiferromagnetic
  oxide},\ }\href {https://doi.org/10.1103/PhysRevLett.132.046701} {\bibfield
  {journal} {\bibinfo  {journal} {Phys. Rev. Lett.}\ }\textbf {\bibinfo
  {volume} {132}},\ \bibinfo {pages} {046701} (\bibinfo {year}
  {2024})}\BibitemShut {NoStop}%
\bibitem [{\citenamefont {Uhrin}\ \emph {et~al.}(2025)\citenamefont {Uhrin},
  \citenamefont {Zadoks}, \citenamefont {Binci}, \citenamefont {Marzari},\ and\
  \citenamefont {Timrov}}]{Uhrin2025}%
  \BibitemOpen
  \bibfield  {author} {\bibinfo {author} {\bibfnamefont {M.}~\bibnamefont
  {Uhrin}}, \bibinfo {author} {\bibfnamefont {A.}~\bibnamefont {Zadoks}},
  \bibinfo {author} {\bibfnamefont {L.}~\bibnamefont {Binci}}, \bibinfo
  {author} {\bibfnamefont {N.}~\bibnamefont {Marzari}},\ and\ \bibinfo {author}
  {\bibfnamefont {I.}~\bibnamefont {Timrov}},\ }\bibfield  {title} {\bibinfo
  {title} {Machine learning hubbard parameters with equivariant neural
  networks},\ }\href {https://doi.org/10.1038/s41524-024-01501-5} {\bibfield
  {journal} {\bibinfo  {journal} {npj Comput. Mater.}\ }\textbf {\bibinfo
  {volume} {11}},\ \bibinfo {pages} {19} (\bibinfo {year} {2025})}\BibitemShut
  {NoStop}%
\bibitem [{\citenamefont {Bastonero}\ \emph {et~al.}(2025)\citenamefont
  {Bastonero}, \citenamefont {Malica}, \citenamefont {Macke}, \citenamefont
  {Bercx}, \citenamefont {Huber}, \citenamefont {Timrov},\ and\ \citenamefont
  {Marzari}}]{Bastonero2025}%
  \BibitemOpen
  \bibfield  {author} {\bibinfo {author} {\bibfnamefont {L.}~\bibnamefont
  {Bastonero}}, \bibinfo {author} {\bibfnamefont {C.}~\bibnamefont {Malica}},
  \bibinfo {author} {\bibfnamefont {E.}~\bibnamefont {Macke}}, \bibinfo
  {author} {\bibfnamefont {M.}~\bibnamefont {Bercx}}, \bibinfo {author}
  {\bibfnamefont {S.}~\bibnamefont {Huber}}, \bibinfo {author} {\bibfnamefont
  {I.}~\bibnamefont {Timrov}},\ and\ \bibinfo {author} {\bibfnamefont
  {N.}~\bibnamefont {Marzari}},\ }\bibfield  {title} {\bibinfo {title}
  {First-principles hubbard parameters with automated and reproducible
  workflows},\ }\href {https://doi.org/10.1038/s41524-025-01685-4} {\bibfield
  {journal} {\bibinfo  {journal} {npj Comput. Mater.}\ }\textbf {\bibinfo
  {volume} {11}},\ \bibinfo {pages} {183} (\bibinfo {year} {2025})}\BibitemShut
  {NoStop}%
\bibitem [{\citenamefont {Chang}\ \emph {et~al.}(2025)\citenamefont {Chang},
  \citenamefont {Timrov}, \citenamefont {Park}, \citenamefont {Zhou},
  \citenamefont {Marzari},\ and\ \citenamefont {Bernardi}}]{Chang2025}%
  \BibitemOpen
  \bibfield  {author} {\bibinfo {author} {\bibfnamefont {B.~K.}\ \bibnamefont
  {Chang}}, \bibinfo {author} {\bibfnamefont {I.}~\bibnamefont {Timrov}},
  \bibinfo {author} {\bibfnamefont {J.}~\bibnamefont {Park}}, \bibinfo {author}
  {\bibfnamefont {J.}~\bibnamefont {Zhou}}, \bibinfo {author} {\bibfnamefont
  {N.}~\bibnamefont {Marzari}},\ and\ \bibinfo {author} {\bibfnamefont
  {M.}~\bibnamefont {Bernardi}},\ }\bibfield  {title} {\bibinfo {title}
  {First-principles electron-phonon interactions and polarons in the parent
  cuprate {La$_2$CuO$_4$}},\ }\href
  {https://doi.org/10.1103/PhysRevResearch.7.L012073} {\bibfield  {journal}
  {\bibinfo  {journal} {Phys. Rev. Research}\ }\textbf {\bibinfo {volume}
  {7}},\ \bibinfo {pages} {L012073} (\bibinfo {year} {2025})}\BibitemShut
  {NoStop}%
\bibitem [{\citenamefont {Binci}\ \emph {et~al.}(2025)\citenamefont {Binci},
  \citenamefont {Marzari},\ and\ \citenamefont {Timrov}}]{Binci2025}%
  \BibitemOpen
  \bibfield  {author} {\bibinfo {author} {\bibfnamefont {L.}~\bibnamefont
  {Binci}}, \bibinfo {author} {\bibfnamefont {N.}~\bibnamefont {Marzari}},\
  and\ \bibinfo {author} {\bibfnamefont {I.}~\bibnamefont {Timrov}},\
  }\bibfield  {title} {\bibinfo {title} {Magnons from time-dependent
  density-functional perturbation theory and nonempirical hubbard
  functionals},\ }\href {https://doi.org/10.1038/s41524-025-01570-0} {\bibfield
   {journal} {\bibinfo  {journal} {npj Comput. Mater.}\ }\textbf {\bibinfo
  {volume} {11}},\ \bibinfo {pages} {100} (\bibinfo {year} {2025})}\BibitemShut
  {NoStop}%
\bibitem [{\citenamefont {dos Santos}\ \emph {et~al.}(2025)\citenamefont {dos
  Santos}, \citenamefont {Binci}, \citenamefont {Menichetti}, \citenamefont
  {Mahajan}, \citenamefont {Marzari},\ and\ \citenamefont
  {Timrov}}]{dosSantos2025}%
  \BibitemOpen
  \bibfield  {author} {\bibinfo {author} {\bibfnamefont {F.~J.}\ \bibnamefont
  {dos Santos}}, \bibinfo {author} {\bibfnamefont {L.}~\bibnamefont {Binci}},
  \bibinfo {author} {\bibfnamefont {G.}~\bibnamefont {Menichetti}}, \bibinfo
  {author} {\bibfnamefont {R.}~\bibnamefont {Mahajan}}, \bibinfo {author}
  {\bibfnamefont {N.}~\bibnamefont {Marzari}},\ and\ \bibinfo {author}
  {\bibfnamefont {I.}~\bibnamefont {Timrov}},\ }\bibfield  {title} {\bibinfo
  {title} {{Comparative study of magnetic exchange parameters and magnon
  dispersions in {NiO} and {MnO} from first principles}},\ }\href
  {https://doi.org/10.1103/gtxm-6vtg} {\bibfield  {journal} {\bibinfo
  {journal} {Phys. Rev. B}\ }\textbf {\bibinfo {volume} {113}},\ \bibinfo
  {pages} {024427} (\bibinfo {year} {2025})}\BibitemShut {NoStop}%
\bibitem [{\citenamefont {Warda}\ \emph {et~al.}(2026)\citenamefont {Warda},
  \citenamefont {Macke}, \citenamefont {Timrov}, \citenamefont {Ciacchi},\ and\
  \citenamefont {Kowalski}}]{Warda2025}%
  \BibitemOpen
  \bibfield  {author} {\bibinfo {author} {\bibfnamefont {K.}~\bibnamefont
  {Warda}}, \bibinfo {author} {\bibfnamefont {E.}~\bibnamefont {Macke}},
  \bibinfo {author} {\bibfnamefont {I.}~\bibnamefont {Timrov}}, \bibinfo
  {author} {\bibfnamefont {L.~C.}\ \bibnamefont {Ciacchi}},\ and\ \bibinfo
  {author} {\bibfnamefont {P.~M.}\ \bibnamefont {Kowalski}},\ }\bibfield
  {title} {\bibinfo {title} {Getting the manifold right: The crucial role of
  orbital resolution in {DFT+$U$} for mixed $d$-$f$ electron compounds},\
  }\href {https://doi.org/10.1021/acs.jctc.5c01406} {\bibfield  {journal}
  {\bibinfo  {journal} {J. Chem. Theory Comput.}\ }\textbf {\bibinfo {volume}
  {22}},\ \bibinfo {pages} {1016} (\bibinfo {year} {2026})}\BibitemShut
  {NoStop}%
\bibitem [{\citenamefont {Giannozzi}\ \emph {et~al.}(2009)\citenamefont
  {Giannozzi}, \citenamefont {Baroni}, \citenamefont {Bonini}, \citenamefont
  {Calandra}, \citenamefont {Car}, \citenamefont {Cavazzoni}, \citenamefont
  {Ceresoli}, \citenamefont {Chiarotti}, \citenamefont {Cococcioni},
  \citenamefont {Dabo}, \citenamefont {Corso}, \citenamefont {de~Gironcoli},
  \citenamefont {Fabris}, \citenamefont {Fratesi}, \citenamefont {Gebauer},
  \citenamefont {Gerstmann}, \citenamefont {Gougoussis}, \citenamefont
  {Kokalj}, \citenamefont {Lazzeri}, \citenamefont {Martin-Samos},
  \citenamefont {Marzari}, \citenamefont {Mauri}, \citenamefont {Mazzarello},
  \citenamefont {Paolini}, \citenamefont {Pasquarello}, \citenamefont
  {Paulatto}, \citenamefont {Sbraccia}, \citenamefont {Scandolo}, \citenamefont
  {Sclauzero}, \citenamefont {Seitsonen}, \citenamefont {Smogunov},
  \citenamefont {Umari},\ and\ \citenamefont {Wentzcovitch}}]{Giannozzi2009}%
  \BibitemOpen
  \bibfield  {author} {\bibinfo {author} {\bibfnamefont {P.}~\bibnamefont
  {Giannozzi}}, \bibinfo {author} {\bibfnamefont {S.}~\bibnamefont {Baroni}},
  \bibinfo {author} {\bibfnamefont {N.}~\bibnamefont {Bonini}}, \bibinfo
  {author} {\bibfnamefont {M.}~\bibnamefont {Calandra}}, \bibinfo {author}
  {\bibfnamefont {R.}~\bibnamefont {Car}}, \bibinfo {author} {\bibfnamefont
  {C.}~\bibnamefont {Cavazzoni}}, \bibinfo {author} {\bibfnamefont
  {D.}~\bibnamefont {Ceresoli}}, \bibinfo {author} {\bibfnamefont {G.~L.}\
  \bibnamefont {Chiarotti}}, \bibinfo {author} {\bibfnamefont {M.}~\bibnamefont
  {Cococcioni}}, \bibinfo {author} {\bibfnamefont {I.}~\bibnamefont {Dabo}},
  \bibinfo {author} {\bibfnamefont {A.~D.}\ \bibnamefont {Corso}}, \bibinfo
  {author} {\bibfnamefont {S.}~\bibnamefont {de~Gironcoli}}, \bibinfo {author}
  {\bibfnamefont {S.}~\bibnamefont {Fabris}}, \bibinfo {author} {\bibfnamefont
  {G.}~\bibnamefont {Fratesi}}, \bibinfo {author} {\bibfnamefont
  {R.}~\bibnamefont {Gebauer}}, \bibinfo {author} {\bibfnamefont
  {U.}~\bibnamefont {Gerstmann}}, \bibinfo {author} {\bibfnamefont
  {C.}~\bibnamefont {Gougoussis}}, \bibinfo {author} {\bibfnamefont
  {A.}~\bibnamefont {Kokalj}}, \bibinfo {author} {\bibfnamefont
  {M.}~\bibnamefont {Lazzeri}}, \bibinfo {author} {\bibfnamefont
  {L.}~\bibnamefont {Martin-Samos}}, \bibinfo {author} {\bibfnamefont
  {N.}~\bibnamefont {Marzari}}, \bibinfo {author} {\bibfnamefont
  {F.}~\bibnamefont {Mauri}}, \bibinfo {author} {\bibfnamefont
  {R.}~\bibnamefont {Mazzarello}}, \bibinfo {author} {\bibfnamefont
  {S.}~\bibnamefont {Paolini}}, \bibinfo {author} {\bibfnamefont
  {A.}~\bibnamefont {Pasquarello}}, \bibinfo {author} {\bibfnamefont
  {L.}~\bibnamefont {Paulatto}}, \bibinfo {author} {\bibfnamefont
  {C.}~\bibnamefont {Sbraccia}}, \bibinfo {author} {\bibfnamefont
  {S.}~\bibnamefont {Scandolo}}, \bibinfo {author} {\bibfnamefont
  {G.}~\bibnamefont {Sclauzero}}, \bibinfo {author} {\bibfnamefont {A.~P.}\
  \bibnamefont {Seitsonen}}, \bibinfo {author} {\bibfnamefont {A.}~\bibnamefont
  {Smogunov}}, \bibinfo {author} {\bibfnamefont {P.}~\bibnamefont {Umari}},\
  and\ \bibinfo {author} {\bibfnamefont {R.~M.}\ \bibnamefont {Wentzcovitch}},\
  }\bibfield  {title} {\bibinfo {title} {{QUANTUM ESPRESSO}: A modular and
  open-source software project for quantum simulations of materials},\ }\href
  {https://doi.org/10.1088/0953-8984/21/39/395502} {\bibfield  {journal}
  {\bibinfo  {journal} {J. Phys. Condens. Matter}\ }\textbf {\bibinfo {volume}
  {21}},\ \bibinfo {pages} {395502} (\bibinfo {year} {2009})}\BibitemShut
  {NoStop}%
\bibitem [{\citenamefont {Giannozzi}\ \emph {et~al.}(2017)\citenamefont
  {Giannozzi}, \citenamefont {Andreussi}, \citenamefont {Brumme}, \citenamefont
  {Bunau}, \citenamefont {Nardelli}, \citenamefont {Calandra}, \citenamefont
  {Car}, \citenamefont {Cavazzoni}, \citenamefont {Ceresoli}, \citenamefont
  {Cococcioni}, \citenamefont {Colonna}, \citenamefont {Carnimeo},
  \citenamefont {Corso}, \citenamefont {de~Gironcoli}, \citenamefont {Delugas},
  \citenamefont {DiStasio}, \citenamefont {Ferretti}, \citenamefont {Floris},
  \citenamefont {Fratesi}, \citenamefont {Fugallo}, \citenamefont {Gebauer},
  \citenamefont {Gerstmann}, \citenamefont {Giustino}, \citenamefont {Gorni},
  \citenamefont {Jia}, \citenamefont {Kawamura}, \citenamefont {Ko},
  \citenamefont {Kokalj}, \citenamefont {Küçükbenli}, \citenamefont
  {Lazzeri}, \citenamefont {Marsili}, \citenamefont {Marzari}, \citenamefont
  {Mauri}, \citenamefont {Nguyen}, \citenamefont {Nguyen}, \citenamefont {de~la
  Roza}, \citenamefont {Paulatto}, \citenamefont {Poncé}, \citenamefont
  {Rocca}, \citenamefont {Sabatini}, \citenamefont {Santra}, \citenamefont
  {Schlipf}, \citenamefont {Seitsonen}, \citenamefont {Smogunov}, \citenamefont
  {Timrov}, \citenamefont {Thonhauser}, \citenamefont {Umari}, \citenamefont
  {Vast}, \citenamefont {Wu},\ and\ \citenamefont {Baroni}}]{Giannozzi2017}%
  \BibitemOpen
  \bibfield  {author} {\bibinfo {author} {\bibfnamefont {P.}~\bibnamefont
  {Giannozzi}}, \bibinfo {author} {\bibfnamefont {O.}~\bibnamefont
  {Andreussi}}, \bibinfo {author} {\bibfnamefont {T.}~\bibnamefont {Brumme}},
  \bibinfo {author} {\bibfnamefont {O.}~\bibnamefont {Bunau}}, \bibinfo
  {author} {\bibfnamefont {M.~B.}\ \bibnamefont {Nardelli}}, \bibinfo {author}
  {\bibfnamefont {M.}~\bibnamefont {Calandra}}, \bibinfo {author}
  {\bibfnamefont {R.}~\bibnamefont {Car}}, \bibinfo {author} {\bibfnamefont
  {C.}~\bibnamefont {Cavazzoni}}, \bibinfo {author} {\bibfnamefont
  {D.}~\bibnamefont {Ceresoli}}, \bibinfo {author} {\bibfnamefont
  {M.}~\bibnamefont {Cococcioni}}, \bibinfo {author} {\bibfnamefont
  {N.}~\bibnamefont {Colonna}}, \bibinfo {author} {\bibfnamefont
  {I.}~\bibnamefont {Carnimeo}}, \bibinfo {author} {\bibfnamefont {A.~D.}\
  \bibnamefont {Corso}}, \bibinfo {author} {\bibfnamefont {S.}~\bibnamefont
  {de~Gironcoli}}, \bibinfo {author} {\bibfnamefont {P.}~\bibnamefont
  {Delugas}}, \bibinfo {author} {\bibfnamefont {R.~A.}\ \bibnamefont
  {DiStasio}}, \bibinfo {author} {\bibfnamefont {A.}~\bibnamefont {Ferretti}},
  \bibinfo {author} {\bibfnamefont {A.}~\bibnamefont {Floris}}, \bibinfo
  {author} {\bibfnamefont {G.}~\bibnamefont {Fratesi}}, \bibinfo {author}
  {\bibfnamefont {G.}~\bibnamefont {Fugallo}}, \bibinfo {author} {\bibfnamefont
  {R.}~\bibnamefont {Gebauer}}, \bibinfo {author} {\bibfnamefont
  {U.}~\bibnamefont {Gerstmann}}, \bibinfo {author} {\bibfnamefont
  {F.}~\bibnamefont {Giustino}}, \bibinfo {author} {\bibfnamefont
  {T.}~\bibnamefont {Gorni}}, \bibinfo {author} {\bibfnamefont
  {J.}~\bibnamefont {Jia}}, \bibinfo {author} {\bibfnamefont {M.}~\bibnamefont
  {Kawamura}}, \bibinfo {author} {\bibfnamefont {H.-Y.}\ \bibnamefont {Ko}},
  \bibinfo {author} {\bibfnamefont {A.}~\bibnamefont {Kokalj}}, \bibinfo
  {author} {\bibfnamefont {E.}~\bibnamefont {Küçükbenli}}, \bibinfo {author}
  {\bibfnamefont {M.}~\bibnamefont {Lazzeri}}, \bibinfo {author} {\bibfnamefont
  {M.}~\bibnamefont {Marsili}}, \bibinfo {author} {\bibfnamefont
  {N.}~\bibnamefont {Marzari}}, \bibinfo {author} {\bibfnamefont
  {F.}~\bibnamefont {Mauri}}, \bibinfo {author} {\bibfnamefont {N.~L.}\
  \bibnamefont {Nguyen}}, \bibinfo {author} {\bibfnamefont {H.-V.}\
  \bibnamefont {Nguyen}}, \bibinfo {author} {\bibfnamefont {A.~O.}\
  \bibnamefont {de~la Roza}}, \bibinfo {author} {\bibfnamefont
  {L.}~\bibnamefont {Paulatto}}, \bibinfo {author} {\bibfnamefont
  {S.}~\bibnamefont {Poncé}}, \bibinfo {author} {\bibfnamefont
  {D.}~\bibnamefont {Rocca}}, \bibinfo {author} {\bibfnamefont
  {R.}~\bibnamefont {Sabatini}}, \bibinfo {author} {\bibfnamefont
  {B.}~\bibnamefont {Santra}}, \bibinfo {author} {\bibfnamefont
  {M.}~\bibnamefont {Schlipf}}, \bibinfo {author} {\bibfnamefont {A.~P.}\
  \bibnamefont {Seitsonen}}, \bibinfo {author} {\bibfnamefont {A.}~\bibnamefont
  {Smogunov}}, \bibinfo {author} {\bibfnamefont {I.}~\bibnamefont {Timrov}},
  \bibinfo {author} {\bibfnamefont {T.}~\bibnamefont {Thonhauser}}, \bibinfo
  {author} {\bibfnamefont {P.}~\bibnamefont {Umari}}, \bibinfo {author}
  {\bibfnamefont {N.}~\bibnamefont {Vast}}, \bibinfo {author} {\bibfnamefont
  {X.}~\bibnamefont {Wu}},\ and\ \bibinfo {author} {\bibfnamefont
  {S.}~\bibnamefont {Baroni}},\ }\bibfield  {title} {\bibinfo {title} {Advanced
  capabilities for materials modelling with {QUANTUM ESPRESSO}},\ }\href
  {https://doi.org/10.1088/1361-648x/aa8f79} {\bibfield  {journal} {\bibinfo
  {journal} {J. Phys. Condens. Matter}\ }\textbf {\bibinfo {volume} {29}},\
  \bibinfo {pages} {465901} (\bibinfo {year} {2017})}\BibitemShut {NoStop}%
\bibitem [{\citenamefont {Giannozzi}\ \emph {et~al.}(2020)\citenamefont
  {Giannozzi}, \citenamefont {Baseggio}, \citenamefont {Bonf\`a}, \citenamefont
  {Brunato}, \citenamefont {Car}, \citenamefont {Carnimeo}, \citenamefont
  {Cavazzoni}, \citenamefont {de~Gironcoli}, \citenamefont {Delugas},
  \citenamefont {Ruffino}, \citenamefont {Ferretti}, \citenamefont {Marzari},
  \citenamefont {Timrov}, \citenamefont {Urru},\ and\ \citenamefont
  {Baroni}}]{Giannozzi2020}%
  \BibitemOpen
  \bibfield  {author} {\bibinfo {author} {\bibfnamefont {P.}~\bibnamefont
  {Giannozzi}}, \bibinfo {author} {\bibfnamefont {O.}~\bibnamefont {Baseggio}},
  \bibinfo {author} {\bibfnamefont {P.}~\bibnamefont {Bonf\`a}}, \bibinfo
  {author} {\bibfnamefont {D.}~\bibnamefont {Brunato}}, \bibinfo {author}
  {\bibfnamefont {R.}~\bibnamefont {Car}}, \bibinfo {author} {\bibfnamefont
  {I.}~\bibnamefont {Carnimeo}}, \bibinfo {author} {\bibfnamefont
  {C.}~\bibnamefont {Cavazzoni}}, \bibinfo {author} {\bibfnamefont
  {S.}~\bibnamefont {de~Gironcoli}}, \bibinfo {author} {\bibfnamefont
  {P.}~\bibnamefont {Delugas}}, \bibinfo {author} {\bibfnamefont {F.~F.}\
  \bibnamefont {Ruffino}}, \bibinfo {author} {\bibfnamefont {A.}~\bibnamefont
  {Ferretti}}, \bibinfo {author} {\bibfnamefont {N.}~\bibnamefont {Marzari}},
  \bibinfo {author} {\bibfnamefont {I.}~\bibnamefont {Timrov}}, \bibinfo
  {author} {\bibfnamefont {A.}~\bibnamefont {Urru}},\ and\ \bibinfo {author}
  {\bibfnamefont {S.}~\bibnamefont {Baroni}},\ }\bibfield  {title} {\bibinfo
  {title} {{QUANTUM ESPRESSO} toward the exascale},\ }\href
  {https://doi.org/10.1063/5.0005082} {\bibfield  {journal} {\bibinfo
  {journal} {J. Chem. Phys.}\ }\textbf {\bibinfo {volume} {152}},\ \bibinfo
  {pages} {154105} (\bibinfo {year} {2020})}\BibitemShut {NoStop}%
\bibitem [{\citenamefont {Prandini}\ \emph {et~al.}(2018)\citenamefont
  {Prandini}, \citenamefont {Marrazzo}, \citenamefont {Castelli}, \citenamefont
  {Mounet},\ and\ \citenamefont {Marzari}}]{Prandini2018}%
  \BibitemOpen
  \bibfield  {author} {\bibinfo {author} {\bibfnamefont {G.}~\bibnamefont
  {Prandini}}, \bibinfo {author} {\bibfnamefont {A.}~\bibnamefont {Marrazzo}},
  \bibinfo {author} {\bibfnamefont {I.~E.}\ \bibnamefont {Castelli}}, \bibinfo
  {author} {\bibfnamefont {N.}~\bibnamefont {Mounet}},\ and\ \bibinfo {author}
  {\bibfnamefont {N.}~\bibnamefont {Marzari}},\ }\bibfield  {title} {\bibinfo
  {title} {Precision and efficiency in solid-state pseudopotential
  calculations},\ }\href {https://doi.org/10.1038/s41524-018-0127-2} {\bibfield
   {journal} {\bibinfo  {journal} {npj Comput. Mater.}\ }\textbf {\bibinfo
  {volume} {4}},\ \bibinfo {pages} {72} (\bibinfo {year} {2018})}\BibitemShut
  {NoStop}%
\bibitem [{\citenamefont {Perdew}\ \emph {et~al.}(2008)\citenamefont {Perdew},
  \citenamefont {Ruzsinszky}, \citenamefont {Csonka}, \citenamefont {Vydrov},
  \citenamefont {Scuseria}, \citenamefont {Constantin}, \citenamefont {Zhou},\
  and\ \citenamefont {Burke}}]{Perdew2008}%
  \BibitemOpen
  \bibfield  {author} {\bibinfo {author} {\bibfnamefont {J.~P.}\ \bibnamefont
  {Perdew}}, \bibinfo {author} {\bibfnamefont {A.}~\bibnamefont {Ruzsinszky}},
  \bibinfo {author} {\bibfnamefont {G.~I.}\ \bibnamefont {Csonka}}, \bibinfo
  {author} {\bibfnamefont {O.~A.}\ \bibnamefont {Vydrov}}, \bibinfo {author}
  {\bibfnamefont {G.~E.}\ \bibnamefont {Scuseria}}, \bibinfo {author}
  {\bibfnamefont {L.~A.}\ \bibnamefont {Constantin}}, \bibinfo {author}
  {\bibfnamefont {X.}~\bibnamefont {Zhou}},\ and\ \bibinfo {author}
  {\bibfnamefont {K.}~\bibnamefont {Burke}},\ }\bibfield  {title} {\bibinfo
  {title} {Restoring the density-gradient expansion for exchange in solids and
  surfaces},\ }\href {https://doi.org/10.1103/PhysRevLett.100.136406}
  {\bibfield  {journal} {\bibinfo  {journal} {Phys. Rev. Lett.}\ }\textbf
  {\bibinfo {volume} {100}},\ \bibinfo {pages} {136406} (\bibinfo {year}
  {2008})}\BibitemShut {NoStop}%
\bibitem [{\citenamefont {Fletcher}(1987)}]{Fletcher1987}%
  \BibitemOpen
  \bibfield  {author} {\bibinfo {author} {\bibfnamefont {R.}~\bibnamefont
  {Fletcher}},\ }\href@noop {} {\emph {\bibinfo {title} {Practical Methods of
  Optimization}}},\ \bibinfo {edition} {2nd}\ ed.\ (\bibinfo  {publisher} {John
  Wiley \& Sons},\ \bibinfo {address} {New York, NY, USA},\ \bibinfo {year}
  {1987})\BibitemShut {NoStop}%
\bibitem [{\citenamefont {Timrov}\ \emph
  {et~al.}(2022{\natexlab{b}})\citenamefont {Timrov}, \citenamefont {Marzari},\
  and\ \citenamefont {Cococcioni}}]{Timrov2022b}%
  \BibitemOpen
  \bibfield  {author} {\bibinfo {author} {\bibfnamefont {I.}~\bibnamefont
  {Timrov}}, \bibinfo {author} {\bibfnamefont {N.}~\bibnamefont {Marzari}},\
  and\ \bibinfo {author} {\bibfnamefont {M.}~\bibnamefont {Cococcioni}},\
  }\bibfield  {title} {\bibinfo {title} {{HP} – a code for the calculation of
  hubbard parameters using density-functional perturbation theory},\ }\href
  {https://doi.org/10.1016/j.cpc.2022.108455} {\bibfield  {journal} {\bibinfo
  {journal} {Comput. Phys. Commun.}\ }\textbf {\bibinfo {volume} {279}},\
  \bibinfo {pages} {108455} (\bibinfo {year} {2022}{\natexlab{b}})}\BibitemShut
  {NoStop}%
\bibitem [{\citenamefont {Pizzi}\ \emph {et~al.}(2020)\citenamefont {Pizzi},
  \citenamefont {Vitale}, \citenamefont {Arita}, \citenamefont {Bl\"ugel},
  \citenamefont {Freimuth}, \citenamefont {G\'eranton}, \citenamefont
  {Gibertini}, \citenamefont {Gresch}, \citenamefont {Johnson}, \citenamefont
  {Koretsune}, \citenamefont {Ibanez-Azpiroz}, \citenamefont {Lee},
  \citenamefont {Lihm}, \citenamefont {Marchand}, \citenamefont {Marrazzo},
  \citenamefont {Mokrousov}, \citenamefont {Mustafa}, \citenamefont {Nohara},
  \citenamefont {Nomura}, \citenamefont {Paulatto}, \citenamefont {Ponc\'e},
  \citenamefont {Ponweiser}, \citenamefont {Qiao}, \citenamefont {Th\"ole},
  \citenamefont {Tsirkin}, \citenamefont {Wierzbowska}, \citenamefont
  {Marzari}, \citenamefont {Vanderbilt}, \citenamefont {Souza}, \citenamefont
  {Mostofi},\ and\ \citenamefont {Yates}}]{Pizzi2020}%
  \BibitemOpen
  \bibfield  {author} {\bibinfo {author} {\bibfnamefont {G.}~\bibnamefont
  {Pizzi}}, \bibinfo {author} {\bibfnamefont {V.}~\bibnamefont {Vitale}},
  \bibinfo {author} {\bibfnamefont {R.}~\bibnamefont {Arita}}, \bibinfo
  {author} {\bibfnamefont {S.}~\bibnamefont {Bl\"ugel}}, \bibinfo {author}
  {\bibfnamefont {F.}~\bibnamefont {Freimuth}}, \bibinfo {author}
  {\bibfnamefont {G.}~\bibnamefont {G\'eranton}}, \bibinfo {author}
  {\bibfnamefont {M.}~\bibnamefont {Gibertini}}, \bibinfo {author}
  {\bibfnamefont {D.}~\bibnamefont {Gresch}}, \bibinfo {author} {\bibfnamefont
  {C.}~\bibnamefont {Johnson}}, \bibinfo {author} {\bibfnamefont
  {T.}~\bibnamefont {Koretsune}}, \bibinfo {author} {\bibfnamefont
  {J.}~\bibnamefont {Ibanez-Azpiroz}}, \bibinfo {author} {\bibfnamefont
  {H.}~\bibnamefont {Lee}}, \bibinfo {author} {\bibfnamefont {J.-M.}\
  \bibnamefont {Lihm}}, \bibinfo {author} {\bibfnamefont {D.}~\bibnamefont
  {Marchand}}, \bibinfo {author} {\bibfnamefont {A.}~\bibnamefont {Marrazzo}},
  \bibinfo {author} {\bibfnamefont {Y.}~\bibnamefont {Mokrousov}}, \bibinfo
  {author} {\bibfnamefont {J.}~\bibnamefont {Mustafa}}, \bibinfo {author}
  {\bibfnamefont {Y.}~\bibnamefont {Nohara}}, \bibinfo {author} {\bibfnamefont
  {Y.}~\bibnamefont {Nomura}}, \bibinfo {author} {\bibfnamefont
  {L.}~\bibnamefont {Paulatto}}, \bibinfo {author} {\bibfnamefont
  {S.}~\bibnamefont {Ponc\'e}}, \bibinfo {author} {\bibfnamefont
  {T.}~\bibnamefont {Ponweiser}}, \bibinfo {author} {\bibfnamefont
  {J.}~\bibnamefont {Qiao}}, \bibinfo {author} {\bibfnamefont {F.}~\bibnamefont
  {Th\"ole}}, \bibinfo {author} {\bibfnamefont {S.}~\bibnamefont {Tsirkin}},
  \bibinfo {author} {\bibfnamefont {M.}~\bibnamefont {Wierzbowska}}, \bibinfo
  {author} {\bibfnamefont {N.}~\bibnamefont {Marzari}}, \bibinfo {author}
  {\bibfnamefont {D.}~\bibnamefont {Vanderbilt}}, \bibinfo {author}
  {\bibfnamefont {I.}~\bibnamefont {Souza}}, \bibinfo {author} {\bibfnamefont
  {A.}~\bibnamefont {Mostofi}},\ and\ \bibinfo {author} {\bibfnamefont
  {J.}~\bibnamefont {Yates}},\ }\bibfield  {title} {\bibinfo {title}
  {{Wannier90 as a community code: new features and applications}},\ }\href
  {https://doi.org/10.1088/1361-648X/ab51ff} {\bibfield  {journal} {\bibinfo
  {journal} {J. Phys.: Condens. Matter}\ }\textbf {\bibinfo {volume} {32}},\
  \bibinfo {pages} {165902} (\bibinfo {year} {2020})}\BibitemShut {NoStop}%
\bibitem [{\citenamefont {Carta}\ \emph {et~al.}(2025)\citenamefont {Carta},
  \citenamefont {Timrov}, \citenamefont {Mlkvik}, \citenamefont {Hampel},\ and\
  \citenamefont {Ederer}}]{Carta2025}%
  \BibitemOpen
  \bibfield  {author} {\bibinfo {author} {\bibfnamefont {A.}~\bibnamefont
  {Carta}}, \bibinfo {author} {\bibfnamefont {I.}~\bibnamefont {Timrov}},
  \bibinfo {author} {\bibfnamefont {P.}~\bibnamefont {Mlkvik}}, \bibinfo
  {author} {\bibfnamefont {A.}~\bibnamefont {Hampel}},\ and\ \bibinfo {author}
  {\bibfnamefont {C.}~\bibnamefont {Ederer}},\ }\bibfield  {title} {\bibinfo
  {title} {{Explicit demonstration of the equivalence between DFT+U and the
  Hartree-Fock limit of DFT+DMFT}},\ }\href
  {https://doi.org/10.1103/PhysRevResearch.7.013289} {\bibfield  {journal}
  {\bibinfo  {journal} {Phys. Rev. Res.}\ }\textbf {\bibinfo {volume} {7}},\
  \bibinfo {pages} {013289} (\bibinfo {year} {2025})}\BibitemShut {NoStop}%
\bibitem [{\citenamefont {Gygi}\ and\ \citenamefont
  {Baldereschi}(1986)}]{Gygi1986}%
  \BibitemOpen
  \bibfield  {author} {\bibinfo {author} {\bibfnamefont {F.}~\bibnamefont
  {Gygi}}\ and\ \bibinfo {author} {\bibfnamefont {A.}~\bibnamefont
  {Baldereschi}},\ }\bibfield  {title} {\bibinfo {title} {Self-consistent
  hartree-fock and screened-exchange calculations in solids: Application to
  silicon},\ }\href {https://doi.org/10.1103/PhysRevB.34.4405} {\bibfield
  {journal} {\bibinfo  {journal} {Phys. Rev. B}\ }\textbf {\bibinfo {volume}
  {34}},\ \bibinfo {pages} {4405} (\bibinfo {year} {1986})}\BibitemShut
  {NoStop}%
\bibitem [{\citenamefont {van Setten}\ \emph {et~al.}(2018)\citenamefont {van
  Setten}, \citenamefont {Giantomassi}, \citenamefont {Bousquet}, \citenamefont
  {Verstraete}, \citenamefont {Hamann}, \citenamefont {Gonze},\ and\
  \citenamefont {Rignanese}}]{vanSetten2018}%
  \BibitemOpen
  \bibfield  {author} {\bibinfo {author} {\bibfnamefont {M.~J.}\ \bibnamefont
  {van Setten}}, \bibinfo {author} {\bibfnamefont {M.}~\bibnamefont
  {Giantomassi}}, \bibinfo {author} {\bibfnamefont {E.}~\bibnamefont
  {Bousquet}}, \bibinfo {author} {\bibfnamefont {M.~J.}\ \bibnamefont
  {Verstraete}}, \bibinfo {author} {\bibfnamefont {D.~R.}\ \bibnamefont
  {Hamann}}, \bibinfo {author} {\bibfnamefont {X.}~\bibnamefont {Gonze}},\ and\
  \bibinfo {author} {\bibfnamefont {G.-M.}\ \bibnamefont {Rignanese}},\
  }\bibfield  {title} {\bibinfo {title} {The \textsc{PseudoDojo}: Training and
  grading a 85 element optimized norm-conserving pseudopotential table},\
  }\href {https://doi.org/10.1016/j.cpc.2018.01.012} {\bibfield  {journal}
  {\bibinfo  {journal} {Comput. Phys. Commun.}\ }\textbf {\bibinfo {volume}
  {226}},\ \bibinfo {pages} {39} (\bibinfo {year} {2018})}\BibitemShut
  {NoStop}%
\bibitem [{\citenamefont {Pizzi}\ \emph {et~al.}(2016)\citenamefont {Pizzi},
  \citenamefont {Cepellotti}, \citenamefont {Sabatini}, \citenamefont
  {Marzari},\ and\ \citenamefont {Kozinsky}}]{Pizzi2016-aiida}%
  \BibitemOpen
  \bibfield  {author} {\bibinfo {author} {\bibfnamefont {G.}~\bibnamefont
  {Pizzi}}, \bibinfo {author} {\bibfnamefont {A.}~\bibnamefont {Cepellotti}},
  \bibinfo {author} {\bibfnamefont {R.}~\bibnamefont {Sabatini}}, \bibinfo
  {author} {\bibfnamefont {N.}~\bibnamefont {Marzari}},\ and\ \bibinfo {author}
  {\bibfnamefont {B.}~\bibnamefont {Kozinsky}},\ }\bibfield  {title} {\bibinfo
  {title} {{AiiDA: automated interactive infrastructure and database for
  computational science}},\ }\href@noop {} {\bibfield  {journal} {\bibinfo
  {journal} {Comput. Mater. Sci.}\ }\textbf {\bibinfo {volume} {111}},\
  \bibinfo {pages} {218} (\bibinfo {year} {2016})}\BibitemShut {NoStop}%
\bibitem [{\citenamefont {Wang}\ \emph {et~al.}(2026)\citenamefont {Wang},
  \citenamefont {Bainglass}, \citenamefont {Bonacci}, \citenamefont
  {Ortega-Guerrero}, \citenamefont {Bastonero}, \citenamefont {Bercx},
  \citenamefont {Bonfà}, \citenamefont {De~Renzi}, \citenamefont {Du},
  \citenamefont {Gillespie}, \citenamefont {Hernández-Bertrán}, \citenamefont
  {Hollas}, \citenamefont {Huber}, \citenamefont {Molinari}, \citenamefont
  {Onuorah}, \citenamefont {Paulish}, \citenamefont {Prezzi}, \citenamefont
  {Qiao}, \citenamefont {Reents}, \citenamefont {Sewell}, \citenamefont
  {Timrov}, \citenamefont {Yakutovich}, \citenamefont {Yu}, \citenamefont
  {Marzari}, \citenamefont {Pignedoli},\ and\ \citenamefont
  {Pizzi}}]{Wang2026}%
  \BibitemOpen
  \bibfield  {author} {\bibinfo {author} {\bibfnamefont {X.}~\bibnamefont
  {Wang}}, \bibinfo {author} {\bibfnamefont {E.}~\bibnamefont {Bainglass}},
  \bibinfo {author} {\bibfnamefont {M.}~\bibnamefont {Bonacci}}, \bibinfo
  {author} {\bibfnamefont {A.}~\bibnamefont {Ortega-Guerrero}}, \bibinfo
  {author} {\bibfnamefont {L.}~\bibnamefont {Bastonero}}, \bibinfo {author}
  {\bibfnamefont {M.}~\bibnamefont {Bercx}}, \bibinfo {author} {\bibfnamefont
  {P.}~\bibnamefont {Bonfà}}, \bibinfo {author} {\bibfnamefont
  {R.}~\bibnamefont {De~Renzi}}, \bibinfo {author} {\bibfnamefont
  {D.}~\bibnamefont {Du}}, \bibinfo {author} {\bibfnamefont {P.~N.~O.}\
  \bibnamefont {Gillespie}}, \bibinfo {author} {\bibfnamefont {M.~A.}\
  \bibnamefont {Hernández-Bertrán}}, \bibinfo {author} {\bibfnamefont
  {D.}~\bibnamefont {Hollas}}, \bibinfo {author} {\bibfnamefont {S.~P.}\
  \bibnamefont {Huber}}, \bibinfo {author} {\bibfnamefont {E.}~\bibnamefont
  {Molinari}}, \bibinfo {author} {\bibfnamefont {I.~J.}\ \bibnamefont
  {Onuorah}}, \bibinfo {author} {\bibfnamefont {N.}~\bibnamefont {Paulish}},
  \bibinfo {author} {\bibfnamefont {D.}~\bibnamefont {Prezzi}}, \bibinfo
  {author} {\bibfnamefont {J.}~\bibnamefont {Qiao}}, \bibinfo {author}
  {\bibfnamefont {T.}~\bibnamefont {Reents}}, \bibinfo {author} {\bibfnamefont
  {C.~J.}\ \bibnamefont {Sewell}}, \bibinfo {author} {\bibfnamefont
  {I.}~\bibnamefont {Timrov}}, \bibinfo {author} {\bibfnamefont {A.~V.}\
  \bibnamefont {Yakutovich}}, \bibinfo {author} {\bibfnamefont
  {J.}~\bibnamefont {Yu}}, \bibinfo {author} {\bibfnamefont {N.}~\bibnamefont
  {Marzari}}, \bibinfo {author} {\bibfnamefont {C.~A.}\ \bibnamefont
  {Pignedoli}},\ and\ \bibinfo {author} {\bibfnamefont {G.}~\bibnamefont
  {Pizzi}},\ }\bibfield  {title} {\bibinfo {title} {Making atomistic materials
  calculations accessible with the aiidalab quantum espresso app},\ }\bibfield
  {journal} {\bibinfo  {journal} {npj Computational Materials}\ }\textbf
  {\bibinfo {volume} {12}},\ \href {https://doi.org/10.1038/s41524-025-01936-4}
  {10.1038/s41524-025-01936-4} (\bibinfo {year} {2026})\BibitemShut {NoStop}%
\bibitem [{\citenamefont {Marzari}\ \emph {et~al.}(1999)\citenamefont
  {Marzari}, \citenamefont {Vanderbilt}, \citenamefont {Vita},\ and\
  \citenamefont {Payne}}]{Marzari1999}%
  \BibitemOpen
  \bibfield  {author} {\bibinfo {author} {\bibfnamefont {N.}~\bibnamefont
  {Marzari}}, \bibinfo {author} {\bibfnamefont {D.}~\bibnamefont {Vanderbilt}},
  \bibinfo {author} {\bibfnamefont {A.~D.}\ \bibnamefont {Vita}},\ and\
  \bibinfo {author} {\bibfnamefont {M.~C.}\ \bibnamefont {Payne}},\ }\bibfield
  {title} {\bibinfo {title} {Thermal contraction and disordering of the
  {Al(110)} surface},\ }\href {https://doi.org/10.1103/PhysRevLett.82.3296}
  {\bibfield  {journal} {\bibinfo  {journal} {Phys. Rev. Lett.}\ }\textbf
  {\bibinfo {volume} {82}},\ \bibinfo {pages} {3296} (\bibinfo {year}
  {1999})}\BibitemShut {NoStop}%
\end{thebibliography}

%

\end{document}